\documentclass[
  journal=pasa,
  manuscript=research-article,
  year=2020,
  volume=37,
]{cup-journal}
\usepackage{amsmath,amssymb,amsfonts}
\usepackage{gensymb}
\usepackage[nopatch]{microtype}
\usepackage{booktabs}

\PassOptionsToPackage{hyphens}{url}\usepackage{hyperref}
\definecolor{dodgerblue}{RGB}{30, 144, 255}
\definecolor{crimson}{RGB}{220, 20, 60}
\definecolor{darkerblue}{RGB}{0, 0, 139}
\definecolor{darkred}{RGB}{150,20,20}
\definecolor{cpink}{RGB}{243, 141, 252}

\hypersetup{colorlinks,citecolor=dodgerblue,linkcolor=blue,urlcolor=magenta}
\usepackage{xurl}  

\usepackage{booktabs} 
\usepackage{threeparttable,longtable}  
\usepackage{makecell}
\usepackage{cprotect}
\usepackage{longtable}
\usepackage{caption}
\usepackage{threeparttable}
\usepackage[htt]{hyphenat}
\DeclareCaptionFormat{cont}{#1 (cont.)#2#3\par}
\usepackage[skip=0.5ex,labelformat=simple]{subcaption}
\renewcommand\thesubfigure{(\roman{subfigure})}

\usepackage{pdflscape}  

\definecolor{dodgerblue}{RGB}{30, 144, 255}

\def\haloPilot{21}
\def\chaloPilot{six}
\def\relicPilot{11}
\def\relicPilotClusters{seven}
\def\crelicPilot{five}

\def\uPilot{12}
\def\haloNsimple{254}

\def\haloN{$\haloNsimple_{-16}^{+88}$}

\def\haloP{30\%}
\def\relicNsimple{85}

\def\relicN{$\relicNsimple_{-9}^{+57}$}
\def\relicP{10\%}

\def\percHR{32\%}

\title{Evolutionary Map of the Universe (EMU): a pilot search for diffuse, non-thermal radio emission in galaxy clusters with the Australian SKA Pathfinder}

\author{S.~W.~Duchesne}
\affiliation{CSIRO Space and Astronomy, PO Box 1130, Bentley WA 6102, Australia}
\email[S.~W.~Duchesne]{stefan.duchesne.astro@gmail.com}

\author{A.~Botteon}
\affiliation{INAF-IRA, via P. Gobetti 101, I-40129, Bologna, Italy}

\author{B.~S.~Koribalski}
\affiliation{CSIRO Space and Astronomy, P.O. Box 76, Epping, NSW 1710, Australia}
\alsoaffiliation{Western Sydney University, Locked Bag 1797, Penrith, NSW 2751, Australia}

\author{F.~Loi}
\affiliation{INAF–Osservatorio Astronomico di Cagliari, via della Scienza 3, Selargius, Italy}

\author{K.~Rajpurohit}
\affiliation{Harvard-Smithsonian Center for Astrophysics, 60 Garden Street, Cambridge, MA 02138, USA}

\author{C.~J.~Riseley}
\affiliation{Dipartimento di Fisica e Astronomia, Università degli Studi di Bologna, via P. Gobetti 93/2, 40129 Bologna, Italy}
\alsoaffiliation{INAF–Istituto di Radioastronomia, via P. Gobetti 101, 40129 Bologna, Italy}
\alsoaffiliation{CSIRO Space and Astronomy, PO Box 1130, Bentley WA 6102, Australia}

\author{L.~Rudnick}
\affiliation{Minnesota Institute for Astrophysics, University of Minnesota, 116 Church St. SE, Minneapolis, MN 55455, USA}

\author{T.~Vernstrom}
\affiliation{ICRAR, The University of Western Australia, 35 Stirling Hw, 6009 Crawley, Australia}
\alsoaffiliation{CSIRO Space and Astronomy, PO Box 1130, Bentley WA 6102, Australia}

\author{H.~Andernach}
\affiliation{Th\"uringer Landessternwarte, Sternwarte 5, D-07778 Tautenburg, Germany}
\alsoaffiliation{Permanent Address: Depto.\ de Astronom\'{i}a, Univ.\ de Guanajuato,
Callej\'on de Jalisco s/n, Guanajuato, C.P.\ 36023, GTO, Mexico}

\author{A.~M.~Hopkins}
\affiliation{School of Mathematical and Physical Sciences, 12 Wally’s Walk, Macquarie University, Sydney, NSW 2109, Australia}

\author{A.~D.~Kapinska}
\affiliation{National Radio Astronomy Observatory, PO Box 0, Socorro, NM87801, USA}

\author{R.~P.~Norris}
\affiliation{Western Sydney University, Locked Bag 1797, Penrith, NSW 2751, Australia}
\alsoaffiliation{CSIRO Space and Astronomy, P.O. Box 76, Epping, NSW 1710, Australia}

\author{T. Zafar}
\affiliation{School of Mathematical and Physical Sciences, 12 Wally’s Walk, Macquarie University, Sydney, NSW 2109, Australia}


\begin{document}

\begin{abstract}

Clusters of galaxies have been found to host Mpc-scale diffuse, non-thermal radio emission in the form of central radio halos and peripheral relics. Turbulence and shock-related processes in the intra-cluster medium are generally considered responsible for the emission, though details of these processes are still not clear. The low surface brightness makes detection of the emission a challenge, but with recent surveys with high-sensitivity radio telescopes we are beginning to build large samples of these sources. The Evolutionary Map of the Universe (EMU) is a Southern Sky survey being performed by the Australian SKA Pathfinder (ASKAP) over the next few years and is well-suited to detect and characterise such emission. To assess prospects of the full survey, we have performed a pilot search of diffuse sources in 71 clusters from the \emph{Planck} Sunyaev--Zeldovich (SZ) cluster catalogue (PSZ2) found in archival ASKAP observations. After re-imaging the archival data and performing both $(u,v)$-plane and image-plane angular scale filtering, we detect \haloPilot\ radio halos (12 for the first time, excluding an additional \chaloPilot\ candidates), \relicPilot\ relics (in \relicPilotClusters\ clusters, and six for the first time, excluding a further \crelicPilot\ candidate relics), along with \uPilot\ other, unclassified diffuse radio sources. From these detections, we predict the full EMU survey will uncover up to $\approx \haloNsimple$ radio halos and $\approx \relicNsimple$ radio relics in the 858 PSZ2 clusters that will be covered by EMU. The percentage of clusters found to host diffuse emission in this work is similar to the number reported in recent cluster surveys with the LOw Frequency ARray (LOFAR) Two-metre Sky Survey {(Botteon, et al. 2022a, A\&A, 660, A78)}, suggesting EMU will complement similar searches being performed in the Northern Sky and provide us with statistically significant samples of halos and relics at the completion of the full survey. This work presents the first step towards large samples of the diffuse radio sources in Southern Sky clusters with ASKAP and eventually the SKA. 
~\\








\end{abstract}

\section{Introduction}

\defcitealias{Botteon2022}{BSC22}

\subsection{Radio emission in galaxy clusters}
Diffuse, non-thermal radio emission has been observed in hundreds of galaxy clusters and is thought to be linked to the dynamics of the hot ($\approx 10^7$--$10^8$\,K), X-ray--emitting intra-cluster medium (ICM). These synchrotron radio sources are generated by the \textmu G-level \citep[e.g.][]{Clarke2001,Bruggen2012} cluster magnetic fields, fuelled by energy deposited via cluster mergers and accretion---see \citet{bj14} for an overview of the theoretical  frameworks describing the physical mechanisms responsible for the diffuse radio emission and \citet{vda+19} and \citet{Paul2023} for observational reviews. While there are a number of different generation mechanisms for diffuse radio sources, they share some common properties: the spectral properties of the sources tend to preclude singular energy/particle injection sites, requiring in situ (re-)acceleration mechanisms \citep[e.g.][]{Jaffe1977}. The observed spectra are steep, with spectral indices\footnote{We define the spectral index, $\alpha$, via $S_\nu \propto \nu^\alpha$, for a flux density $S_\nu$ at frequency $\nu$.} $\alpha \lesssim -1$ \citep[e.g.\ see the compilation by][]{Duchesne2021b}. {Due to their steep radio spectra, diffuse cluster sources are best detected at low frequencies.}


\emph{Radio halos} are found at the centres of some clusters. The size of these sources is typically of order $\approx 1$\,Mpc and they are found predominantly in merging and post-merger cluster systems \citep[e.g.][]{Cassano2010,Cassano2023}. The mechanisms powering radio halos are not completely understood, though it is likely related to turbulence in the ICM as a result of major mergers \citep[e.g.][]{bsfg01,p01,Cassano2007}. At the smaller scale ($\lesssim 500$\,kpc) `mini-halos' are found in relaxed cool-core clusters \citep[e.g.][]{Giacintucci2017,Giacintucci2019}. Sloshing in the core of the cluster likely powers the emission, with the observed radio properties otherwise being similar to normal radio halos. Halos with multiple components have also been observed \citep[e.g.\ in RX~J1720.1$+$2638;][]{Biava2021}. {Indeed, the once-clear division between mini-halo and halo is becoming blurred with the new generation of radio interferometers, with transitional `mini'-halos showing mixed characteristics typical of both mini-halos and halos \citep[e.g.][]{Riseley2022_MS1455,Riseley2023}. Diffuse emission is also being found at much larger scales than the traditional radio halo \citep[e.g.][]{Shweta2020,Rajpurohit2021,Vacca2022a,Vacca2022b,Botteon2022b,Bruno2023b} with so-called `megahalos' also featuring a change in their radio surface brightness profile hinting at a change in physical conditions with increasing distance from the cluster centre \citep{Cuciti2022}.} 

In the low-density cluster outskirts, elongated radio sources other than tailed radio galaxies are occasionally observed. These are typically referred to as \emph{radio relics}\footnote{Note that these sources are also sometimes referred to as `radio shocks'.}, and have been observed to be co-located with shocks detected  via X-ray emission \citep[e.g.][]{Finoguenov2010,Akamatsu2015,Urdampilleta2018,DiGennaro2019}. Because of the coincidence with shocks and observed morphology and spectra the physical mechanism generating relics is thought to relate to shock-acceleration processes \citep[e.g.][]{Ensslin1998,Hoeft2007,Kang2018}. When observed with high sensitivity radio telescopes, some relics have been detected with a diffuse component that physically extends and spectrally steepens towards the cluster centre \citep[e.g.\ in 1RXS J0603.3$+$4214;][]{vanWeeren2012a,Rajpurohit2018,Rajpurohit2020}. In some cases, multiple relics are observed in a single cluster (e.g.\ in Abell~3667; \citealt{rwhe97,mj-h,Hindson2014,riseley2015,deGasperin2022}, and Abell~3376; \citealt{bdnp06,Kale2012}) and complex merging systems have been observed to host both radio relics and halos (e.g.\ in Abell~2744; \citealt{Pearce2017,Rajpurohit2021}, and Abell~3266; \citealt{Duchesne2022,Riseley2022}). The exact shock (re-)acceleration mechanism is still being investigated, and larger numbers of sources and highly-detailed multi-wavelength analyses are required to understand these sources further. 

Finally, other, smaller-scale diffuse emission is seen in some clusters \citep[e.g.][]{Slee2001,Hodgson2021}. These sources include steep-spectrum fossil plasmas left over from past episodes of active galactic nuclei (AGN). These sources may be simply ageing through synchrotron and inverse-Compton losses, or in some cases may be re-energised by mechanical processes in the ICM \citep[e.g.\ adiabatic compression due to shocks;][]{eg01}. Revived fossil plasmas have been observed connected to active radio galaxies \citep[e.g.][]{vanWeeren2017,deGasperin2017a}. Such sources provide possible links to the aforementioned radio relics and highlight a fossil electron population that may provide mildly-relativistic particles for shock--re-acceleration processes \citep[e.g.][]{Vazza2021}. 

\subsection{Surveys with modern radio telescopes}
With the new, sensitive radio interferometers at low frequencies ($\lesssim 1$\,GHz), there have been a number of radio surveys of clusters as well as detections of heretofore unseen types of emission. Below $231$\,MHz, work with the Murchison Widefield Array \citep[MWA;][]{tgb+13,wtt+18} has had its large fractional bandwidth leveraged to explore spectral properties of these steep-spectrum radio sources \citep{Hindson2014,gdj+17,Giacintucci2020,Duchesne2020a,Hodgson2021,Duchesne2021b,Duchesne2021a,Duchesne2022}. In the Northern Hemisphere, the LOw Frequency ARray \citep[LOFAR;][]{lofar} and the recent second data release from the LOFAR Two-metre Sky Survey \citep[LoTSS-DR2;][]{lotss:dr2} is also being used for surveys of new diffuse cluster sources \citep[e.g.][]{vanWeeren2020,Hoang2022} with the latest data release providing the largest single sample of clusters hosting diffuse radio sources \citep[][hereinafter \citetalias{Botteon2022}]{Botteon2022} along with in-depth statistical analyses of the sources and the hosting clusters \citep{Bruno2023,Zhang2023,Cassano2023,Cuciti2023,Jones2023}.

Closer to $\approx 1$\,GHz MeerKAT is also producing images of unprecedented sensitivity and resolution of clusters as part of surveys \citep[e.g.][]{Knowles2021} including the MeerKAT Galaxy Cluster Legacy Survey \citep[MGCLS;][]{Knowles2022} which is enabling highly-informative studies of selected cluster systems and the constituent diffuse, non-thermal radio emission \citep[e.g.][]{Riseley2022_MS1455,Giacintucci2022,Sikhosana2023,Trehaeven2023,Riseley2023}. The upgraded Giant Metrewave Radio Telescope \citep[uGMRT;][]{ugmrt} is also producing in-depth multi-wavelength studies of galaxy clusters, enabling new detections \citep[e.g.][]{Schellenberger2022,Pandge2022,Lee2022,Kurahara2023} and providing much-needed bandwidth to investigate the wideband spectral properties of these sources \citep[e.g.][]{Rajpurohit2020,Rajpurohit2021,DiGennaro2021,Kale2022}.




\subsection{The Australian SKA Pathfinder}
The Australian SKA Pathfinder (ASKAP; \citealt{Hotan2021}) is a 36-antenna radio interferometer located on Inyarrimanha Ilgari Bundara, the CSIRO\footnote{Commonwealth Scientific and Industrial Research Organisation.} Murchison Radio-astronomy Observatory. ASKAP operates between 700--1800\,MHz with an instantaneous bandwidth of 288\,MHz and features 12-m dishes. The array has baselines ranging from 22\,m to 6\,km, providing sensitivity to angular scales up to $\approx 50$~arcmin and an angular resolution of $\approx 12^{\prime\prime}$ at 900\,MHz. ASKAP's primary purpose is all-sky radio surveys, including continuum \citep[e.g.][]{Norris2011a, Norris2021}, linear polarization \citep{possum1,Thomson2023}, spectral line work \citep[e.g.][]{Rhee2022:dingo,gaskap1,wallaby1,Allison2022:flash}, and transient/variability surveys \citep{craft1,craft2,Murphy2021}. The main technology that allows ASKAP surveying capability is its phased array feeds \citep[PAF;][]{Hotan2014,McConnell2016}. The PAF simultaneously forms 36 mostly-independent primary beams that are arranged in a regular footprint on the sky covering $\approx 30$\,deg$^{2}$ at 900\,MHz.

While ASKAP has completed the first two epochs of the shallow Rapid ASKAP Continuum Survey \citep[RACS;][]{racs1,racs-mid}, one of the major continuum surveys undertaken by ASKAP is the Evolutionary Map of the Universe \citep[EMU;][]{Norris2011a,Norris2021}. EMU is to cover the Southern Sky at 943\,MHz with a total integration time of 10\,h per pointing. With a standard image point-spread function (PSF) of $15^{\prime\prime} \times 15^{\prime\prime}$ the expected noise characteristics approach $\approx 30$\,\textmu Jy\,PSF$^{-1}$. The survey is to be completed over the next few years, with $\approx 15$\% of the survey currently observed. The main science goals of EMU are 
 to trace the evolution of star-forming galaxies and supermassive black holes, to explore  large-scale structure and cosmological parameters, to use radio sources to help understand clusters, to study Galactic continuum emission, and to explore an uncharted region of observational parameter space, and find new classes of objects.

Shallow RACS data products have already been used to help characterise diffuse radio emission in galaxy clusters \citep{Duchesne2021b,Duchesne2022}, but the deep ASKAP observations of EMU and other deep ASKAP surveys are providing many new detections---and a more in-depth characterisation---of such sources \citep[e.g.][]{Wilber2020,HyeongHan2020,Bruggen2020,Duchesne2020b,Duchesne2021a,Venturi2022,Riseley2022,Loi2023, Macgregor2023}. Completion of the EMU survey is expected to yield an additional large sample of radio halos and relics, complementing similar surveys being conducted with LOFAR. In this work we aim to explore the prospects of the full EMU survey in the context of uncovering radio halos and relics using similar, archival ASKAP observations. 

Where relevant, we assume a flat $\Lambda$ cold dark matter cosmology with $H_0 = 70$~km\,s$^{-1}$\,Mpc$^{-1}$, $\Omega_\text{M} = 0.3$, and $\Omega_\Lambda = 1-\Omega_\text{M}$.

\section{The galaxy cluster sample}

\begin{figure}[!t]
    \centering
    \includegraphics[width=1\linewidth]{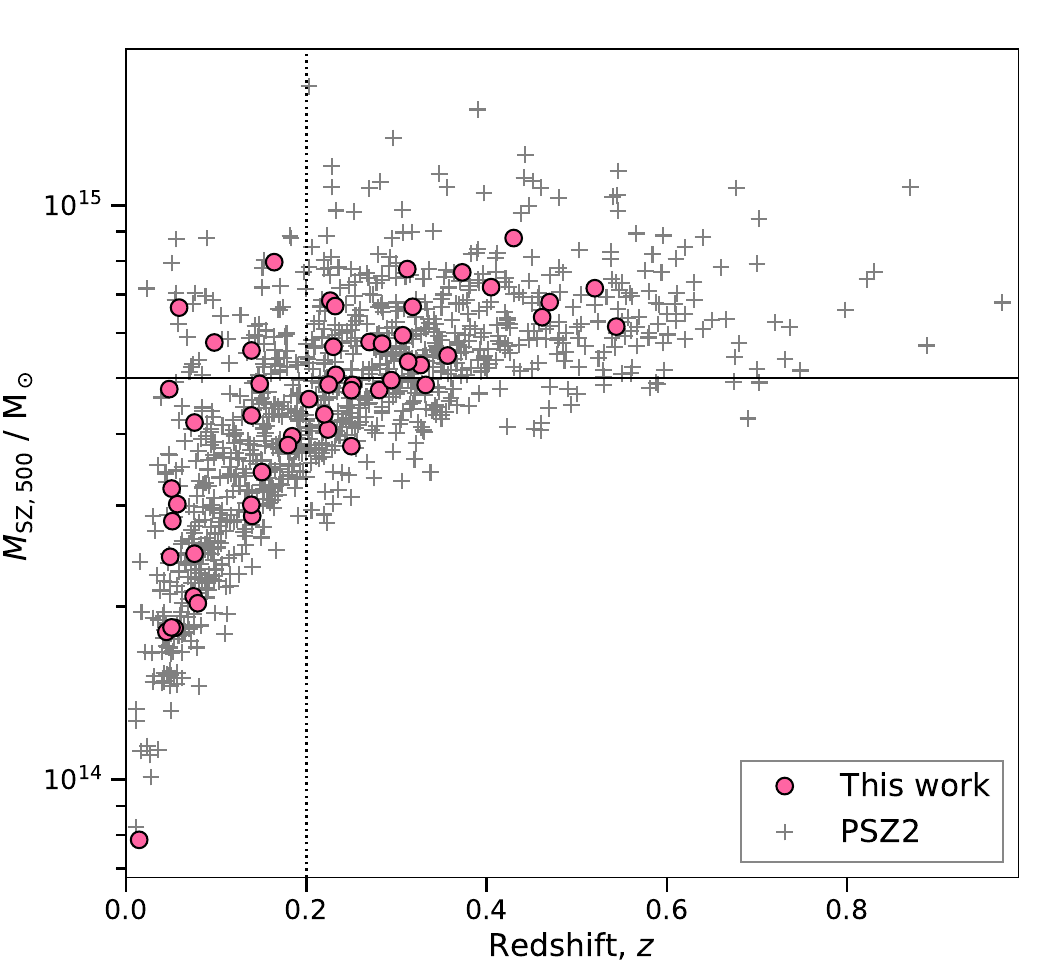}
    \caption{\label{fig:psz2} The distribution of cluster mass with redshift for the  PSZ2 catalogue (gray crosses) and the sample used in this work (pink circles) for clusters with reported redshifts. A vertical dotted line is drawn at $z=0.2$, the redshift we assume for clusters with no reported redshift. The horizontal line is drawn at $M_\text{SZ,500} = 5\times10^{14}$\,M$_\odot$.}
\end{figure}

\begin{figure*}[!t]
    \centering
    \includegraphics[width=1\linewidth]{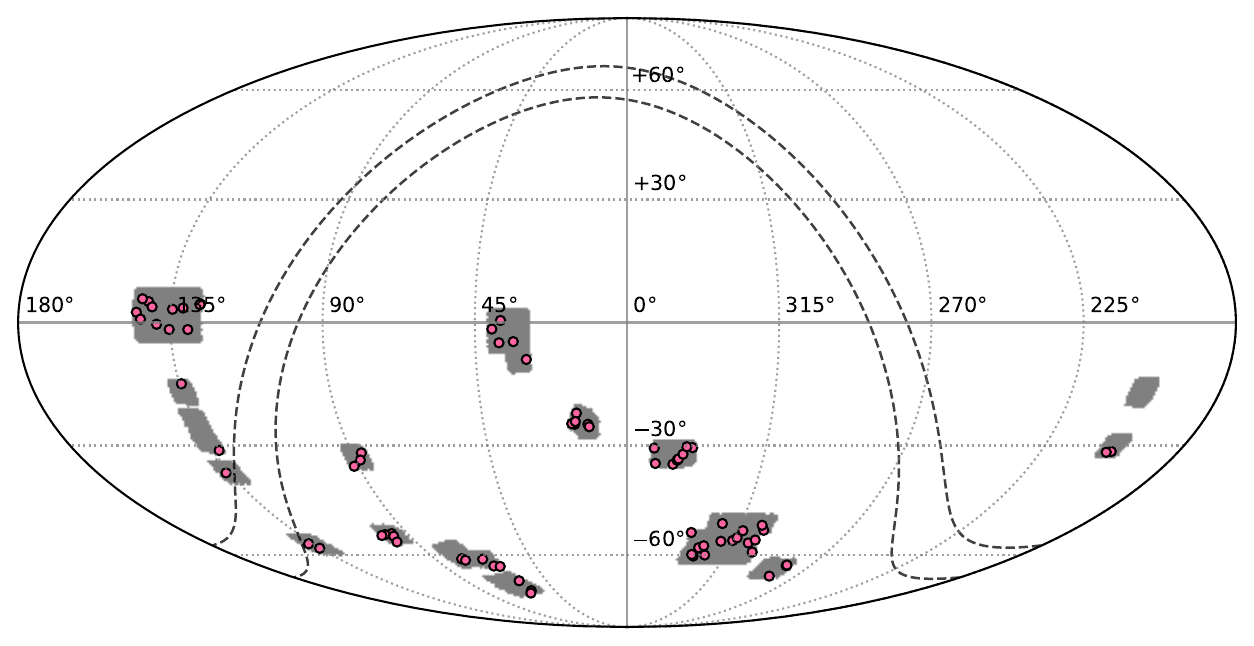}
    \caption{\label{fig:coverage} Equatorial map centered on $(\alpha_\text{J2000},\delta_\text{J2000}) = (0, 0)\degree$  showing the sky coverage of the ASKAP observations used in this work (shaded gray regions). The pink circles indicate the locations of the PSZ2 clusters used in this work. The black, dashed lines are drawn at Galactic latitudes $b \pm 5\degree$.}
\end{figure*}

\begin{table*}[!t]
    \centering
    \begin{threeparttable}
    \caption{\label{tab:observations} Archival ASKAP observations used in this work.}
    \begin{tabular}{c c c c c c c c c c}\toprule
         SBID & $\nu$ \tnote{a} & Coordinates \tnote{b} & Project  & Purpose & Footprint & Pitch & $\tau_\text{int}$ \tnote{c} & $\sigma_\text{rms}$ \tnote{d} & PSF \tnote{e}  \\
          & (MHz) & (J2000) & & & & (deg) & (h) & (\textmu Jy\,PSF$^{-1}$) & (arcsec $\times$ arcsec) \\\midrule
8132 & 888 & 22:46:14\,\,\,\,$-$32:15:35 & AS034 & Early Science & \texttt{square\_6x6} & 0.90 & 9.0 & 46.2 & $10.0 \times 8.5$ \\
8137 & 888 & 23:11:46\,\,\,\,$-$32:15:35 & AS034 & Early Science & \texttt{square\_6x6} & 0.90 & 10.0 & 44.4 & $10.1 \times 8.5$ \\
8275 & 1014 & 06:26:49\,\,\,\,$-$54:04:19 & AS034 & Early Science & \texttt{square\_6x6} & 0.90 & 10.0 & 28.2 & $11.2 \times 9.5$ \\
9287 & 944 & 21:00:00\,\,\,\,$-$51:07:06 & AS101 & EMU & \texttt{closepack36} & 0.90 & 10.0 & 34.3 & $14.8 \times 11.1$ \\
9325 & 944 & 20:34:17\,\,\,\,$-$60:19:18 & AS101 & EMU & \texttt{closepack36} & 0.90 & 10.0 & 35.3 & $12.5 \times 10.9$ \\
9351 & 944 & 20:42:00\,\,\,\,$-$55:43:29 & AS101 & EMU & \texttt{closepack36} & 0.90 & 10.0 & 38.5 & $12.0 \times 10.7$ \\
9410 & 944 & 21:15:26\,\,\,\,$-$60:19:18 & AS101 & EMU & \texttt{closepack36} & 0.90 & 10.0 & 32.6 & $11.8 \times 10.7$ \\
9434 & 944 & 21:32:44\,\,\,\,$-$51:07:06 & AS101 & EMU & \texttt{closepack36} & 0.90 & 10.0 & 30.1 & $12.1 \times 10.1$ \\
9437 & 944 & 20:27:16\,\,\,\,$-$51:07:06 & AS101 & EMU & \texttt{closepack36} & 0.90 & 10.0 & 35.3 & $12.0 \times 10.1$ \\
9442 & 944 & 21:18:00\,\,\,\,$-$55:43:29 & AS101 & EMU & \texttt{closepack36} & 0.90 & 10.0 & 33.3 & $14.0 \times 10.9$ \\
9501 & 944 & 21:56:34\,\,\,\,$-$60:19:18 & AS101 & EMU & \texttt{closepack36} & 0.90 & 10.0 & 31.9 & $12.1 \times 10.7$ \\
9596 & 944 & 05:56:19\,\,\,\,$-$33:09:10 & AS111 & GW follow-up & \texttt{closepack36} & 0.90 & 10.0 & 36.4 & $12.1 \times 9.7$ \\
10083 & 944 & 21:54:00\,\,\,\,$-$55:43:29 & AS101 & EMU & \texttt{closepack36} & 0.90 & 10.0 & 32.2 & $12.1 \times 9.9$ \\
10135 & 888 & 09:29:10\,\,\,\,$-$01:39:32 & AS112 & SWAG-X & \texttt{square\_6x6} & 1.05 & 8.6 & 45.1 & $13.3 \times 12.1$ \\
10475 & 888 & 08:38:45\,\,\,\,$+$04:38:27 & AS112 & SWAG-X & \texttt{square\_6x6} & 1.05 & 8.7 & 53.7 & $13.4 \times 12.4$ \\
10486 & 888 & 09:04:00\,\,\,\,$-$01:39:00 & AS112 & SWAG-X & \texttt{square\_6x6} & 1.05 & 8.6 & 57.7 & $13.2 \times 11.6$ \\
13570 & 944 & 00:58:00\,\,\,\,$-$23:45:00 & AS033 & Commissioning & \texttt{closepack36} & 0.90 & 10.0 & 37.0 & $12.6 \times 9.7$ \\
15191 & 944 & 00:50:38\,\,\,\,$-$25:16:57 & AS111 & GW follow-up & \texttt{closepack36} & 0.90 & 10.5 & 30.1 & $12.4 \times 9.9$ \\
20875 & 888 & 09:29:15\,\,\,\,$+$04:38:27 & AS112 & SWAG-X & \texttt{closepack36} & 1.05 & 8.0 & 46.4& $13.4 \times 12.5$ \\
20931 & 888 & 09:04:00\,\,\,\,$+$04:39:00 & AS112 & SWAG-X & \texttt{closepack36} & 1.05 & 8.0 & 48.1 & $13.5 \times 12.5$ \\
21021 & 888 & 08:38:50\,\,\,\,$-$01:39:32 & AS112 & SWAG-X & \texttt{closepack36} & 1.05 & 8.0 & 57.1 & $13.3 \times 11.9$ \\
25035 & 888 & 04:24:00\,\,\,\,$-$70:00:00 & AS113 & \emph{TESS} followup & \texttt{square\_6x6} & 1.05 & 13.0 & 34.9 & $13.9 \times 12.0$ \\
25077 & 888 & 05:08:00\,\,\,\,$-$60:00:00 & AS113 & \emph{TESS} followup & \texttt{square\_6x6} & 1.05 & 13.0 & 33.7 & $13.9 \times 11.4$ \\
28257 & 944 & 02:30:27\,\,\,\,$+$00:00:00 & AS101 & EMU & \texttt{closepack36} & 0.90 & 5.1 & 74.0& $18.0 \times 18.0$ \\
32235 & 944 & 19:08:00\,\,\,\,$-$64:30:00 & AS113 & \emph{TESS} followup & \texttt{closepack36} & 0.90 & 10.0 & 37.2 & $12.3 \times 11.2$ \\
33370 & 944 & 02:08:57\,\,\,\,$-$09:20:42 & AS101 & EMU & \texttt{closepack36} & 0.90 & 5.0 & 56.6 & $18.0 \times 18.0$ \\
33459 & 944 & 02:08:57\,\,\,\,$-$04:40:35 & AS101 & EMU & \texttt{closepack36} & 0.90 & 5.0 & 52.7 & $18.0 \times 18.0$ \\
33509 & 944 & 02:30:27\,\,\,\,$-$04:40:35 & AS101 & EMU & \texttt{closepack36} & 0.90 & 5.0 & 57.6 & $18.0 \times 18.0$ \\
34120 & 944 & 13:29:47\,\,\,\,$-$30:17:10 & AS103 & POSSUM & \texttt{closepack36} & 0.90 & 10.0 & 32.6 & $12.3 \times 9.9$ \\
41688 & 944 & 09:00:00\,\,\,\,$-$57:00:00 & AS113 & Pointing test & \texttt{closepack36} & 0.90 & 10.0 & 29.6 & $12.3 \times 10.7$ \\
41710 & 944 & 09:00:00\,\,\,\,$-$29:00:00 & AS113 & Pointing test & \texttt{closepack36} & 0.90 & 10.0 & 30.7 & $12.3 \times 10.1$ \\
41757 & 944 & 09:00:00\,\,\,\,$-$37:00:00 & AS113 & Pointing test & \texttt{closepack36} & 0.90 & 10.0 & 28.5 & $12.5 \times 10.3$ \\
41850 & 944 & 09:00:00\,\,\,\,$-$24:00:00 & AS113 & Pointing test & \texttt{closepack36} & 0.90 & 10.0 & 31.4 & $12.4 \times 10.2$ \\
41871 & 944 & 09:00:00\,\,\,\,$+$03:00:00 & AS113 & Pointing test & \texttt{closepack36} & 0.90 & 9.0 & 37.2  & $12.6 \times 11.8$ \\
41894 & 944 & 09:00:00\,\,\,\,$-$17:00:00 & AS113 & Pointing test & \texttt{closepack36} & 0.90 & 10.0 & 31.4 & $12.4 \times 10.5$ \\
         \bottomrule
    \end{tabular}
    \begin{tablenotes}[flushleft]
    {\footnotesize \item[a] Effective image frequency. \item[b] Coordinates at the centre of the tile. \item[c] Total integration time for the given observation. \item[d] Median rms noise over the archival, full-resolution image. \item[e] PSF as reported in the image metadata, though note for older images on CASDA the PSF may vary slightly over the image.}
    \end{tablenotes}
\end{threeparttable}
\end{table*}

\begin{table*}[t!]
    
    \begin{threeparttable}
    \caption{\label{tab:clusters} PSZ2 clusters covered by the archival ASKAP data shown in this work.}
    \begin{tabular}{cccccccc}\toprule
    Cluster & Other name & Coordinates \tnote{a} & $z$ & $z$ ref. \tnote{b} & $M_\text{SZ,500}$ & Source(s) \tnote{c} & Source(s) ref. \tnote{d} \\
     & & (hh:mm:ss, dd:mm:ss) & & & ($\times 10^{14}$~M$_\odot$) & & \\\midrule
PSZ2 G006.16$-$69.49 & - & 23:22:09\,\,\,\,$-$34:34:31 &0.23 & 3 & $3.73^{+0.50}_{-0.52}$ & - & -\\
PSZ2 G008.31$-$64.74 & Abell S1077 & 22:58:43\,\,\,\,$-$34:47:18 &0.312 & 1 & $7.75^{+0.41}_{-0.40}$ & R+R+cH & 1\\
PSZ2 G011.06$-$63.84 & Abell 3934 & 22:53:33\,\,\,\,$-$33:44:20 &0.224 & 1 & $4.07^{+0.47}_{-0.49}$ & cH & 1\\
PSZ2 G011.92$-$63.53 & Abell 3926 & 22:51:54\,\,\,\,$-$33:23:39 &0.24 & 4 & $4.36^{+0.46}_{-0.48}$ & H & 1\\
PSZ2 G014.72$-$62.49 & - & 22:46:32\,\,\,\,$-$32:12:04 &0.5 & 3 & $5.96^{+0.65}_{-0.71}$ & - & -\\
PSZ2 G017.25$-$70.71 & - & 23:24:41\,\,\,\,$-$30:39:04 &0.31 & 3 & $4.78^{+0.57}_{-0.64}$ & - & -\\
PSZ2 G018.18$-$60.00 & Abell 3889 & 22:34:51\,\,\,\,$-$30:32:31 &0.2515 & 1 & $4.87^{+0.45}_{-0.47}$ & - & -\\
PSZ2 G018.76$-$61.65 & - & 22:42:33\,\,\,\,$-$30:18:54 &0.24 & 3 & $4.33^{+0.52}_{-0.60}$ & - & -\\
PSZ2 G110.28$-$87.48 & - & 00:49:01\,\,\,\,$-$24:40:13 &0.52 & 1 & $7.17^{+0.59}_{-0.65}$ & H & 1\\
PSZ2 G149.63$-$84.19 & Abell 133 & 01:02:41\,\,\,\,$-$21:54:44 &0.0569 & 1 & $3.02^{+0.18}_{-0.16}$ & U+U & 2\\
PSZ2 G167.43$-$53.67 & - & 02:29:35\,\,\,\,$+$00:29:42 &- & - & - & - & -\\
PSZ2 G167.66$-$65.59 & - & 01:59:55\,\,\,\,$-$08:51:32 &0.405 & 1 & $7.20^{+0.61}_{-0.68}$ & - & -\\
PSZ2 G167.98$-$59.95 & Abell 329 & 02:14:44\,\,\,\,$-$04:35:08 &0.1393 & 1 & $4.31^{+0.38}_{-0.39}$ & - & -\\
PSZ2 G172.98$-$53.55 & Abell 370 & 02:39:52\,\,\,\,$-$01:34:05 &0.373 & 1 & $7.65^{+0.56}_{-0.57}$ & H & 3\\
PSZ2 G174.40$-$57.33 & Abell 362 & 02:31:43\,\,\,\,$-$04:51:40 &0.1843 & 1 & $3.96^{+0.49}_{-0.49}$ & - & -\\
PSZ2 G175.69$-$85.98 & Abell 141 & 01:05:30\,\,\,\,$-$24:39:17 &0.23 & 1 & $5.67^{+0.36}_{-0.40}$ & H & 4\\
PSZ2 G180.74$-$85.21 & - & 01:09:14\,\,\,\,$-$24:30:39 &- & - & - & - & -\\
PSZ2 G219.88$+$22.83 & Abell 664 & 08:25:08\,\,\,\,$+$04:27:23 &0.232813 & 1 & $5.07^{+0.44}_{-0.48}$ & cR & 1\\
PSZ2 G220.11$+$22.91 & - & 08:25:47\,\,\,\,$+$04:18:13 &0.2248 & 1 & $4.87^{+0.41}_{-0.45}$ & - & -\\
PSZ2 G223.47$+$26.85 & - & 08:45:29\,\,\,\,$+$03:28:31 &0.3269 & 1 & $5.27^{+0.65}_{-0.64}$ & H & 1\\
PSZ2 G225.48$+$29.41 & Abell 732 & 08:57:54\,\,\,\,$+$03:10:28 &0.203 & 1 & $4.60^{+0.45}_{-0.49}$ & H & 1\\
PSZ2 G227.59$+$22.98 & - & 08:39:26\,\,\,\,$-$01:40:44 &0.28085 & 1 & $4.77^{+0.66}_{-0.67}$ & cR+cR & 1\\
PSZ2 G227.89$+$36.58 & - & 09:26:52\,\,\,\,$+$05:00:35 &0.4616 & 1 & $6.39^{+0.67}_{-0.68}$ & cH & 1\\
PSZ2 G228.38$+$38.58 & - & 09:34:34\,\,\,\,$+$05:41:01 &0.543811 & 1 & $6.15^{+0.82}_{-0.92}$ & U & 1\\
PSZ2 G228.50$+$34.95 & - & 09:22:15\,\,\,\,$+$03:45:10 &0.2701 & 1 & $5.78^{+0.51}_{-0.58}$ & cH & 1\\
PSZ2 G230.73$+$27.70 & - & 09:01:29\,\,\,\,$-$01:39:30 &0.29435 & 1 & $4.96^{+0.57}_{-0.61}$ & - & -\\
PSZ2 G231.79$+$31.48 & Abell 776 & 09:16:14\,\,\,\,$-$00:24:42 &0.332405 & 1 & $4.87^{+0.63}_{-0.66}$ & R+H & 1\\
PSZ2 G232.84$+$38.13 & Abell 847 & 09:40:27\,\,\,\,$+$02:28:19 &0.1508 & 1 & $3.43^{+0.38}_{-0.40}$ & cR & 1\\
PSZ2 G233.68$+$36.14 & - & 09:35:17\,\,\,\,$+$00:49:06 &0.356823 & 1 & $5.48^{+0.65}_{-0.67}$ & R+R+H & 1\\
PSZ2 G236.92$-$26.65 & Abell 3364 & 05:47:36\,\,\,\,$-$31:52:23 &0.1483 & 1 & $4.89^{+0.29}_{-0.32}$ & - & 5\\
PSZ2 G239.27$-$26.01 & MACS J0553.4-3342 & 05:53:24\,\,\,\,$-$33:42:07 &0.43 & 1 & $8.77^{+0.44}_{-0.46}$ & H & 6,7\\
PSZ2 G241.79$-$24.01 & Abell 3378 & 06:05:53\,\,\,\,$-$35:18:33 &0.1392 & 1 & $5.59^{+0.20}_{-0.22}$ & U+U & 1\\
PSZ2 G241.98$+$19.56 & - & 08:57:49\,\,\,\,$-$14:44:31 &0.27 & 2 & $4.44^{+0.52}_{-0.60}$ & - & -\\
PSZ2 G254.52$+$08.27 & - & 08:51:30\,\,\,\,$-$31:15:52 &- & - & - & - & -\\
PSZ2 G260.80$+$06.71 & - & 09:05:12\,\,\,\,$-$37:01:30 &- & - & - & cR+cH & 1\\
PSZ2 G262.36$-$25.15 & Abell 3391 & 06:26:19\,\,\,\,$-$53:41:20 &0.0514 & 1 & $2.82^{+0.14}_{-0.13}$ & U & 8\\
PSZ2 G263.14$-$23.41 & Abell S592 & 06:38:54\,\,\,\,$-$53:58:53 &0.2266 & 1 & $6.83^{+0.34}_{-0.31}$ & H & 7\\
PSZ2 G263.19$-$25.19 & Abell 3395 & 06:27:09\,\,\,\,$-$54:25:48 &0.0506 & 1 & $3.21^{+0.18}_{-0.19}$ & U & 8\\
PSZ2 G263.68$-$22.55 & Abell 3404 & 06:45:29\,\,\,\,$-$54:13:46 &0.1644 & 1 & $7.96^{+0.23}_{-0.21}$ & H & 4\\
PSZ2 G265.21$-$24.83 & - & 06:32:17\,\,\,\,$-$56:08:51 &0.054 & 1 & $1.84^{+0.17}_{-0.17}$ & - & -\\[0.5em]\bottomrule
\end{tabular}
\begin{tablenotes}[flushleft]
{\footnotesize \item[] \textit{Notes.} \item[a] Coordinates reported in the PSZ2 catalogue. \item[b] Reference for the cluster redshift. 1: \citet{planck16}; 2: \citet{Aguado-Barahona2019}; 3: \citet{Maturi2019}; 4: \citet{Bleem2020}. \item[c] Detected sources, including: radio halo (any size, H), relic (R), miscellaneous diffuse emission (U), candidate object (c), `-' indicates no clear diffuse emission aside from active radio galaxies. \item[d] References for detections of radio emission. 1: this work; 2: \citet{Slee2001}; 3: \citet{Xie2020}; 4: \citet{Duchesne2021a}; 5: \citet{Knowles2022}; 6: \citet{Bonafede2012}; 7: \citet{Wilber2020}; 8: \citet{Bruggen2020}; 9: \citet{murphy99}; 10: \citet{Duchesne2022}; 11: \citet{Riseley2022}; 12: \citet{Venturi2022}; 13: \citet{Venturi2003}; 14: \citet{Duchesne2020b}; 15: \citet{HyeongHan2020}; 16: \citet{Loi2023}.}
\end{tablenotes}
\end{threeparttable}
\end{table*}
\begin{table*}[!t]
    
    \caption*{}
    \begin{tabular}{cccccccc}\toprule
    Cluster & Other name & Coordinates & $z$ & $z$ ref. & $M_\text{SZ,500}$  & Source(s) \tnote{c} & Radio ref. \tnote{d} \\
     & & (hh:mm:ss, dd:mm:ss) & & & ($\times 10^{14}$~M$_\odot$) & & \\\midrule
PSZ2 G270.63$-$35.67 & - & 05:09:55\,\,\,\,$-$61:17:15 &0.313 & 1 & $5.34^{+0.34}_{-0.34}$ & - & -\\
PSZ2 G271.28$-$36.11 & - & 05:05:34\,\,\,\,$-$61:44:28 &0.25 & 1 & $3.81^{+0.34}_{-0.38}$ & - & -\\
PSZ2 G272.08$-$40.16 & Abell 3266 & 04:31:14\,\,\,\,$-$61:24:25 &0.0589 & 1 & $6.64^{+0.11}_{-0.12}$ & R+U+H+U & 9,10,11\\
PSZ2 G275.24$-$40.42 & - & 04:22:09\,\,\,\,$-$63:35:16 &- & - & - & - & -\\
PSZ2 G275.73$-$06.12 & - & 09:08:04\,\,\,\,$-$56:41:42 &- & - & - & - & -\\
PSZ2 G276.09$-$41.53 & Abell 3230 & 04:10:39\,\,\,\,$-$63:44:28 &0.14 & 1 & $2.88^{+0.27}_{-0.27}$ & - & -\\
PSZ2 G276.14$-$07.68 & - & 09:01:23\,\,\,\,$-$58:01:50 &- & - & - & - & -\\
PSZ2 G282.32$-$40.15 & - & 04:00:53\,\,\,\,$-$68:33:05 &- & - & - & - & -\\
PSZ2 G286.28$-$38.36 & - & 03:59:13\,\,\,\,$-$72:05:50 &0.307 & 1 & $5.94^{+0.39}_{-0.40}$ & H & 1\\
PSZ2 G286.75$-$37.35 & - & 04:07:46\,\,\,\,$-$72:57:42 &0.47 & 1 & $6.79^{+0.49}_{-0.53}$ & H & 1\\
PSZ2 G311.98$+$30.71 & Abell 3558 & 13:27:58\,\,\,\,$-$31:30:45 &0.048 & 1 & $4.79^{+0.17}_{-0.17}$ & H & 12\\
PSZ2 G313.33$+$30.29 & Abell 3562 & 13:33:40\,\,\,\,$-$31:43:04 &0.049 & 1 & $2.44^{+0.21}_{-0.24}$ & H+B & 13,12\\
PSZ2 G328.58$-$25.25 & - & 18:56:04\,\,\,\,$-$66:56:26 &0.1797 & 1 & $3.82^{+0.44}_{-0.47}$ & - & -\\
PSZ2 G331.96$-$45.74 & Abell 3825 & 21:58:13\,\,\,\,$-$60:24:46 &0.075 & 1 & $2.08^{+0.24}_{-0.24}$ & - & -\\
PSZ2 G332.11$-$23.63 & - & 18:47:25\,\,\,\,$-$63:28:01 &- & - & - & - & -\\
PSZ2 G332.23$-$46.37 & Abell 3827 & 22:01:53\,\,\,\,$-$59:56:24 &0.098 & 1 & $5.77^{+0.18}_{-0.18}$ & H & 1\\
PSZ2 G332.29$-$23.57 & - & 18:47:12\,\,\,\,$-$63:17:23 &0.0146 & 1 & $0.79^{+0.10}_{-0.10}$ & - & -\\
PSZ2 G333.89$-$43.60 & SPT-CL J2138-6007 & 21:37:50\,\,\,\,$-$60:07:12 &0.318 & 1 & $6.66^{+0.47}_{-0.51}$ & H & 1\\
PSZ2 G335.58$-$46.44 & Abell 3822 & 21:54:07\,\,\,\,$-$57:51:47 &0.076 & 1 & $4.19^{+0.16}_{-0.18}$ & H+R & 1\\
PSZ2 G336.95$-$45.75 & Abell 3806 & 21:46:31\,\,\,\,$-$57:16:38 &0.076 & 1 & $2.47^{+0.21}_{-0.21}$ & - & -\\
PSZ2 G337.99$-$33.61 & - & 20:14:43\,\,\,\,$-$59:13:03 &- & - & - & - & -\\
PSZ2 G339.74$-$51.08 & - & 22:15:02\,\,\,\,$-$53:21:21 &- & - & - & - & -\\
PSZ2 G340.35$-$42.80 & - & 21:20:26\,\,\,\,$-$55:56:16 &- & - & - & - & -\\
PSZ2 G341.19$-$36.12 & Abell 3685 & 20:32:16\,\,\,\,$-$56:26:07 &0.284 & 1 & $5.74^{+0.56}_{-0.59}$ & R+R & 14\\
PSZ2 G341.44$-$40.19 & Abell 3732 & 21:01:01\,\,\,\,$-$55:43:00 &0.25 & 1 & $4.77^{+0.45}_{-0.46}$ & - & -\\
PSZ2 G342.33$-$34.93 & SPT-CL J2023-5535 & 20:23:25\,\,\,\,$-$55:34:30 &0.232 & 1 & $6.68^{+0.45}_{-0.43}$ & R+R+H & 15\\
PSZ2 G342.62$-$39.60 & Abell 3718 & 20:55:53\,\,\,\,$-$54:54:54 &0.139 & 1 & $3.01^{+0.33}_{-0.35}$ & U & 16\\
PSZ2 G345.38$-$39.32 & Abell 3716S & 20:52:13\,\,\,\,$-$52:50:12 &0.044831 & 1 & $1.81^{+0.18}_{-0.18}$ & - & -\\
PSZ2 G345.82$-$34.29 & Abell S861 & 20:18:49\,\,\,\,$-$52:42:55 &0.0505 & 1 & $1.84^{+0.21}_{-0.16}$ & - & -\\
PSZ2 G346.86$-$45.38 & Abell 3771 & 21:29:48\,\,\,\,$-$50:46:37 &0.0796 & 1 & $2.03^{+0.26}_{-0.27}$ & H & 1\\
PSZ2 G347.58$-$35.35 & Abell S871 & 20:25:49\,\,\,\,$-$51:16:38 &0.22 & 1 & $4.33^{+0.43}_{-0.45}$ & cH & 1\\[0.5em]
\bottomrule

\end{tabular}
\end{table*}

Following \citetalias{Botteon2022}, we select clusters reported in the second \emph{Planck} Sunyaev--Zeldovich (SZ) cluster catalogue \citep[PSZ2;][]{planck16}, which provides a selection of clusters across a range of redshifts with masses of $\approx 10^{14}$--$10^{15}$\,$\text{M}_\odot$. Figure~\ref{fig:psz2} shows the mass distribution of the full 1\,653 clusters in the PSZ2 catalogue as well as the subsample used in this work, as described below.

To select PSZ2 clusters to search, we obtain all archival ASKAP datasets with Stokes I total intensity images available, that fit the following criteria: \begin{enumerate}
    \item Observed in ASKAP Band 1 ($\lesssim 1$ GHz central frequency),
    \item Observed after all 36 antennas became operational (though ignoring flagged antennas),
    \item Observed for $\geq 5$~h, 
    \item Field direction not within Galactic latitudes $b\pm 5\degree$,
    \item Publicly `released' before October 1, 2022,
    \item Not a duplicated field---if fields were observed multiple times, only the `best' quality (i.e. lowest median rms noise) was selected.
\end{enumerate}

Other, particularly higher-frequency (Band 2), observations are available in the archive that could also be used, though we adhere to these criteria to ensure a similar dataset to the main EMU survey. During the commissioning and early science phase of ASKAP (prior to November 16, 2022, when full operations commenced), ASKAP had been performing numerous operational tests, including pilot surveys for the main ASKAP Survey Science Teams (SST). These SSTs cover a range of scientific goals, from neutral hydrogen absorption and emission studies of distant \citep{Rhee2022:dingo,Allison2022:flash} and nearby \citep{gaskap1,wallaby1} targets, to transient and variability studies of both Galactic and extra-Galactic objects \citep[e.g.][]{Murphy2021}, and studies of linear polarization of radio sources with POSSUM\footnote{The Polarisation Sky Survey of the Universe's Magnetism.} \citep{possum1}. Included in these SSTs are also the total intensity continuum surveys such as EMU, {covering most of the Southern Sky}, and the Survey with ASKAP of GAMA-09 + X-ray (SWAG-X; Moss et al., in prep.), overlapping with the GAMA-09\footnote{Galaxy And Mass Assembly at declination $-9^\circ$.} field. 

The chosen observation scheduling block IDs (SBIDs) and their details are recorded in Table~\ref{tab:observations}. A total of 36 SBIDs are available fitting the above criteria, however, two do not feature PSZ2 clusters. The total area covered is $\approx 1\,990$\,deg$^{2}$, which comprises 71 PSZ2 clusters ($\approx 0.036$ PSZ2 clusters per deg$^{2}$). Figure~\ref{fig:coverage} shows the distribution of clusters and selected observations across the sky, and Table~\ref{tab:clusters} summarises the 71 clusters. The selected SBIDs include observations from the EMU Pilot Survey \citep[][]{Norris2021}, Gravitational Wave follow-up observations, SWAG-X, \emph{TESS} follow-up observations \citep[][]{Rigney2022}, a POSSUM observation that is commensal with EMU, and various other EMU early science and commissioning observations \citep[e.g.][]{Bruggen2020,Gurkan2022:G23,Quici2021}. A handful of these SBIDs have already featured in work focused on diffuse emission in galaxy clusters \citep{Wilber2020,HyeongHan2020,Bruggen2020,Duchesne2021a,Duchesne2021b,Venturi2022,Riseley2022,Loi2023} {and for completeness in this work we report both previously detected sources and newly detected sources in the 71 PSZ2 clusters.} 

Of the 71 clusters selected for this work, 18 do not have redshifts reported in the original PSZ2 catalogue. {We find six of these have spectroscopic redshifts reported in the literature, and for these clusters we follow \citetalias{Botteon2022} and compute $M_\text{SZ,500}$ by interpolating the $M_\text{SZ}$--$z$ curves provided in the PSZ2 individual algorithm catalogues \citep{planck16b}.} One cluster, PSZ2~G167.43$-$53.67, has an angular separation $<10$~arcmin from 15 other catalogued clusters (and one group) with redshifts in the range $0.132 \lesssim z \lesssim 1.35$. Due to the ambiguity in an possible cross-match, we do not obtain a redshift for this cluster. Cluster redshifts are reported in Table~\ref{tab:clusters} along with the relevant redshift reference where available. 


\section{Data}
\subsection{Archival ASKAP observations}

\subsubsection{Re-imaging the ASKAP data}\label{sec:ddselfcal}

\begin{figure*}[t]
\centering
\includegraphics[width=1\linewidth]{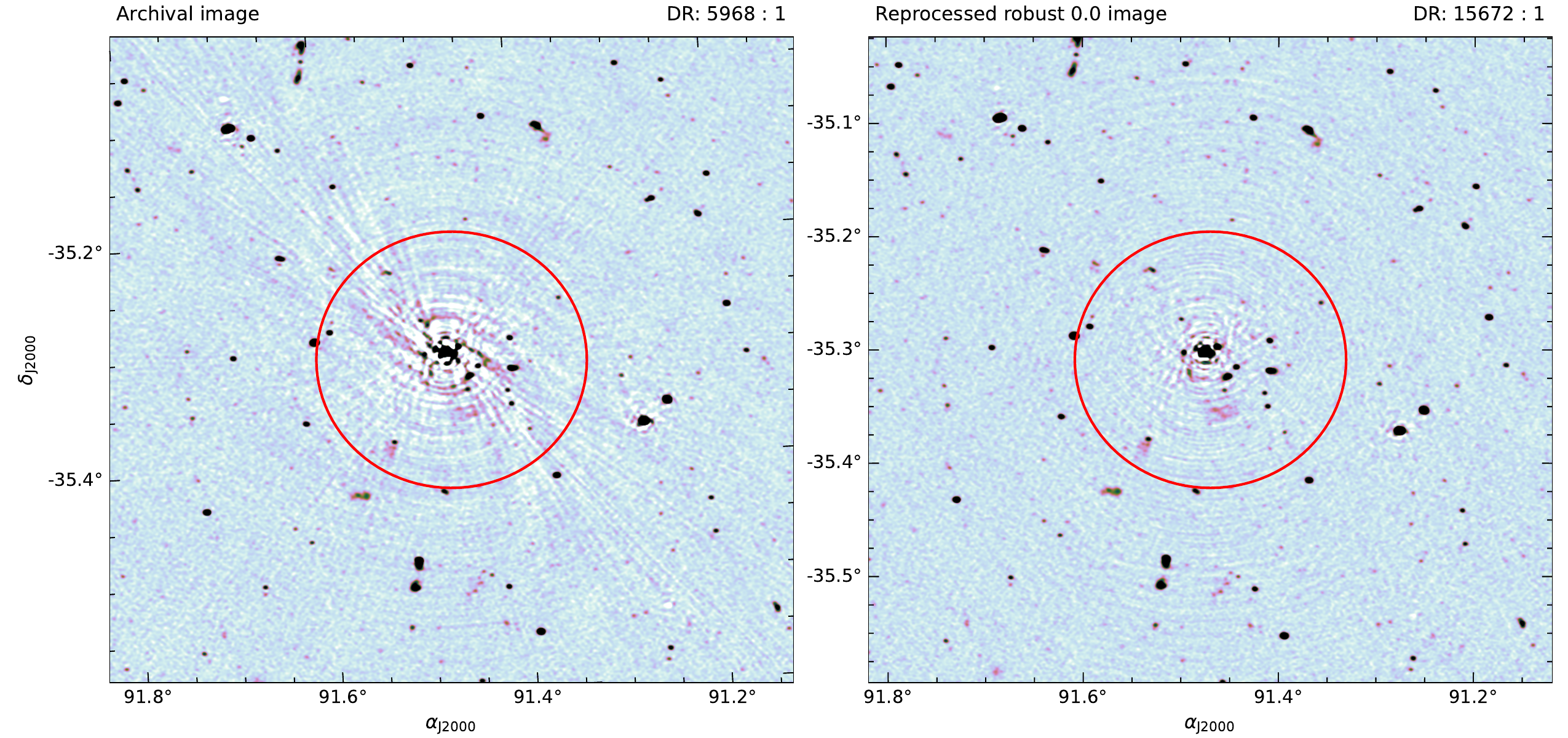}
\caption{\label{fig:sc} PSZ2~G241.79$-$24.01 in beam 5 of SB9596 in the archival image (\emph{left}) and the phase-rotated and self-calibrated robust 0.0 image (\emph{right}). The dynamic range for each image is shown in the top right of each panel. The red circle is centred on PSZ2~G241.79$-$24.01 and has a radius of 1\,Mpc at the cluster's redshift ($z=0.1392$). The linear colour scale is the same in each panel and shown in the range $[-150, 1000]$\,\textmu Jy\,beam$^{-1}$.}
\end{figure*}

As the archival ASKAP data have been processed at various stages of pipeline development and with a range of imaging settings (particularly image weighting), we opt to re-image all datasets containing the PSZ2 clusters from our sample. To avoid re-imaging PAF beams without significant sensitivity to clusters in our sample, we identify PAF beams that lie within 0.75~deg of a cluster from our sample. This results in 144 beams covering the 71 clusters, ranging from 1--4 beams per cluster. Each beam has its own visibility dataset and is re-imaged independently prior to co-addition/linear mosaicking for each target. Each beam dataset is retrieved from the CSIRO ASKAP Science Data Archive \citep[CASDA][]{casda,Huynh2020} and has already been self-calibrated as part of the usual \texttt{ASKAPsoft} data processing strategy. This direction-independent self-calibration procedure has remained reasonably unchanged over the course of processing the archival datasets and comprises two rounds of phase-only self-calibration. Our re-imaging process is in principle similar to the process described by \citet[][see also \citealt{Botteon2022}]{vanWeeren2020} used for LOFAR, though due to the higher frequency the direction-dependent effects caused by the ionosphere are not as problematic for the ASKAP data.

We stage each PAF beam dataset on the internal CSIRO supercomputer and use a bespoke processing pipeline \texttt{SASKAP}\footnote{\url{https://gitlab.com/Sunmish/saskap/-/tree/petrichor}.} for processing single ASKAP beams. We begin by creating large template images for each beam out to the first sidelobe with a Briggs \citep{db95} robust $+0.25$ image weighting. We use \texttt{WSClean}\footnote{\url{https://gitlab.com/aroffringa/wsclean/}.} \citep{wsclean1,wsclean2} for imaging, and make use of the multi-scale CLEAN algorithm for deconvolution and the \texttt{wgridder} algorithm \citep{wgridder1,wgridder2} for gridding/de-gridding. This template image provides a good model of the sky for primary beam modelling later on, and allows us to subtract the sky $>5$~Mpc from the cluster centre. For clusters without a redshift, we follow \citetalias{Botteon2022} and assume $z=0.2$ (here and for other redshift-dependent processing parameters described further on). For clusters with $z < 0.1$, we opt to reduce this size (cluster-dependent) to aid in processing. After subtracting the sky away from the cluster, we phase shift the individual beam datasets towards the direction of the target cluster and begin further self-calibration. This self-calibration process uses the \texttt{CASA}\footnote{\url{https://casa.nrao.edu/}.} \citep{casa2022} task \texttt{gaincal} with the CLEAN component model generated by \texttt{WSClean}, and performs two loops in most cases: (1) phase-only on 300\,s intervals, and (2) amplitude and phase on 60\,s intervals. Generally the amplitude self-calibration makes a significant improvement near bright sources, but in two cases failed. In the two failed cases, we simply turn this second stage off and rely on the phase-only self-calibration which yielded sufficient improvement for our purposes. 

Figure~\ref{fig:sc} shows a comparison of the archival data with the phase-shifted and self-calibrated data for  PSZ2~G241.79$-$24.01 in beam 5 of SB9596. The left panel shows the archival image, and the right panel shows a robust $0.0$ image after self-calibration. PSZ2~G241.79$-$24.01 is the most extreme example of the self-calibration improvements, as it features a $S_\text{887\,MHz} \approx 1$\,Jy source at its centre. The self-calibration reduces artefacts significantly enough to reveal heretofore unseen diffuse emission near the centre. These improvements are commensurate with the improvements seen using direction-dependent calibration and imaging software such as \texttt{killMS} \citep{Tasse2014,Smirnov2015} and \texttt{DDFacet} \citep[][see \citealt{Wilber2020,Bruggen2020,Riseley2022} for ASKAP examples]{2018A&A...611A..87T} and are functionally similar to a single facet in cases without bright, off-axis sources. 

Following the self-calibration, we create a range of images: \begin{enumerate}
    \item Uniform image (highest resolution),
    \item Robust 0.0 image (high resolution with sensitivity to extended structure),
    \item Robust $+0.25$ image (as above, but generally more suitable for extended emission, depending on ($u,v$) coverage),
    \item Robust $+0.25$ image, with Gaussian taper corresponding to 100~kpc (25--50~kpc if $z<0.09$),
    \item Robust $+0.25$ image, with Gaussian taper corresponding to 250~kpc (63--125~kpc if $z<0.09$).
\end{enumerate}
Note that this is similar to the  image set created by \citetalias{Botteon2022} for their work with the LoTSS-DR2, though we optimize the weighting and and tapering scales for the lower-resolution ASKAP data. Generally the first three images are to provide a range of reference images at high resolution while retaining sensitivity to extended sources, and the tapered, low-resolution maps provide better sensitivity to large-scale halos and relics. We then subtract compact emission from the visibility datasets by imaging with data with a $(u,v)$ cut to remove emission on physical scales $<250$~kpc (between 63--125\,kpc if $z<0.09$, depending on cluster). After subtraction of the compact emission model in the $(u,v)$ data, we re-image the residual datasets following the previous round of imaging, excluding the robust 0.0 image.



\subsubsection{Modelling the ASKAP primary beams}

\begin{figure}[t]
    \centering
    \includegraphics[width=1\linewidth]{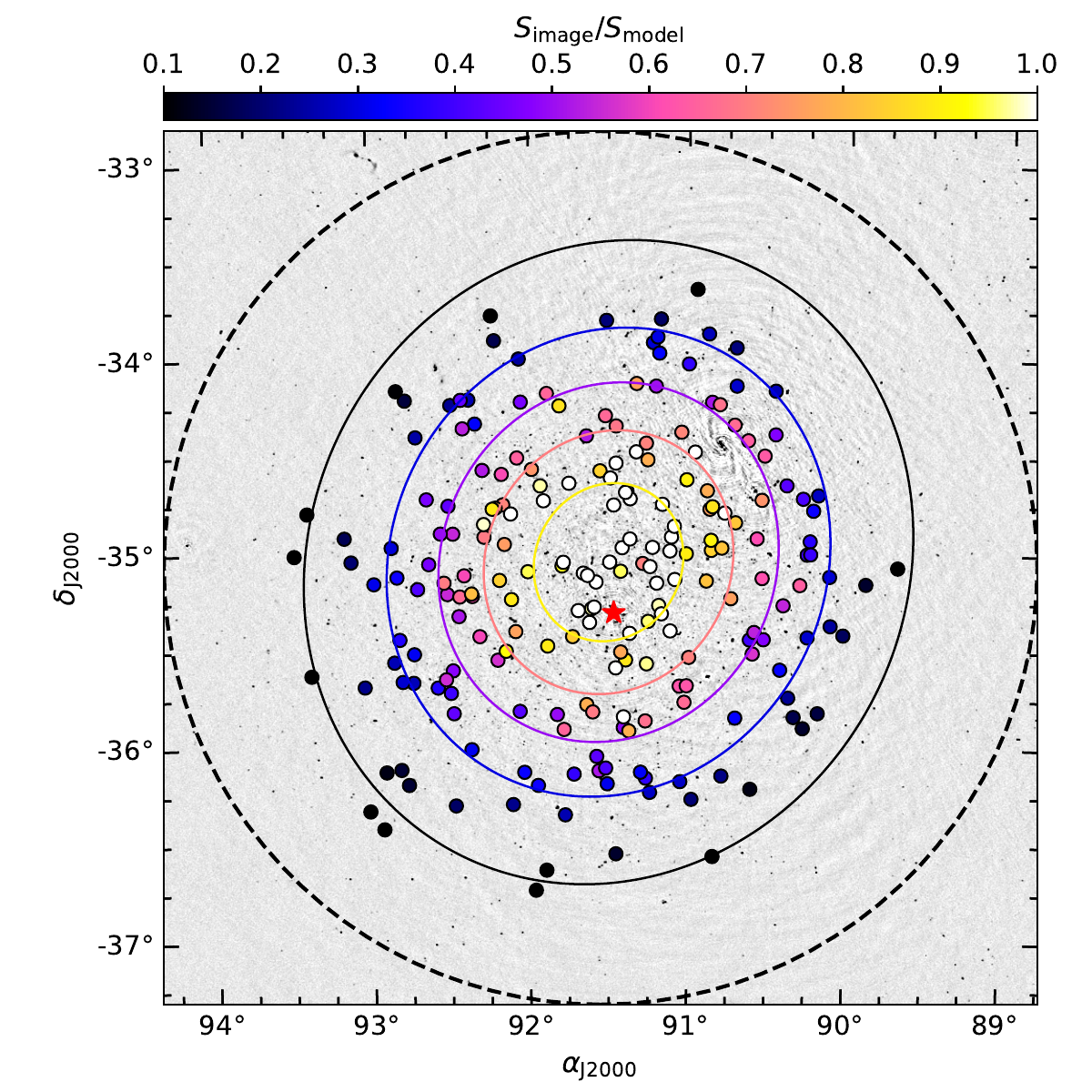}
    \caption{\label{fig:beamexample}Example beam 5 from SB9596---a corner beam in the \texttt{closepack36} footprint. The background is the template image prior to directional self-calibration and source subtraction, used to generate the beam model. Overlaid are the sources used in modelling, coloured by the ratio of the measured flux density to the model flux density ($S_\text{image}/S_\text{model}$). Also overlaid are contours from the model beam, in levels of [0.1, 0.3, 0.5, 0.7, 0.9]. The larger dashed, black circle indicates the 2.25-deg radius within which sources are selected. The red star indicates the location of PSZ2~G241.79$-$24.01.}
\end{figure}

The 36 primary beams of the PAFs are formed by adjusting weights to maximise SNR while observing the Sun \citep{Hotan2021}. This process is undertaken every 1--2 months, and can create primary beam responses that shift in position by $\approx$\,arcmin and change shape slightly. These changes can result in a factor of two difference to the response towards the beam edges \citep{racs-mid}. ASKAP now measures the primary beam responses via holographic measurements while observing PKS~J0408$-$6544. During most early science and commissioning, appropriate holographic measurements were not available and so a 2-D circular Gaussian model was assumed for primary beam correction and mosaicking\footnote{The first regularly scheduled holography observation was SB34422---at the end of 2021---and prior to that most observations did not have a holography observation that corresponded to the PAF beam-former weights used for that particular time period.}. This was found to be inadequate, particularly at the beam edges \citep{racs1}. While resulting per-beam brightness scale errors can average out in the centre of an ASKAP image (formed via linear mosaic of all PAF beams), tile edges and individual beam images will retain these significant errors.

For consistency we opt to measure an in-field primary beam response for all observations. We created a global sky model from the existing RACS source-lists. At present, RACS has completed two sub-surveys, one at 887.5 MHz \citep[RACS-low;][]{racs1,racs2} and the other at 1367.5 MHz \citep[RACS-mid;][]{racs-mid}. All imaging data products for RACS-low and RACS-mid are available through CASDA\footnote{Under the RACS DOI for catalogue data products: \url{https://doi.org/10.25919/1khs-c716}.}. The sky model is created by merging the existing per-observation source-lists from RACS-low and RACS-mid\footnote{While a 25~arcsec resolution catalogue is available for RACS-low \citep{racs2}, we opt to create our own catalogue for this work to retain the highest-possible resolution across the survey to match the RACS-mid data. {Similarly, all-sky RACS-mid catalogues are available \citep{racs-mid2}, but were not} available at the time of processing these data.}. These source-lists were created using the \texttt{selavy} source-finding software during the processing of the surveys, which decomposes grouped pixels (`sources') into 2-D Gaussian components. For this purpose, we use the `component' lists to represent individual sources. The source-lists are retrieved from CASDA for each survey. We merge the RACS-low and RACS-mid source-lists separately, removing duplicated sources in overlap regions. This duplicate removal simply matches sources within their respective reported angular size in each observation that comprises the overlap regions. If a source is detected in two or more source-lists based on this criterion, we take the source that has the smallest separation from its tile centre. This process results in 3\,313\,521 components for RACS-low and 3\,916\,193 components for RACS-mid.

With separate merged source-lists for RACS-low and RACS-mid, we perform a cross-match using \texttt{match\_catalogues}\footnote{Packaged as part of \texttt{flux\_warp} \citep{Duchesne2020a}.} accepting a maximum separation of 10\,arcsec and excluding sources if they have neighbours within 25\,arcsec. This yields 2\,088\,670 sources. We then calculated two-point spectral indices for every source, following \begin{equation}
    \alpha = \dfrac{\log_{10} \left( S_\text{RACS-low} / S_\text{RACS-mid} \right)}{\log_{10} \left( 887.5 / 1367.5 \right)} \, ,
\end{equation}
where $S_{\text{RACS-low}}$ and $S_\text{RACS-mid}$ are the RACS-low and RACS-mid integrated flux densities for the given source. We clip the catalogue where sources have $\alpha$ outside of the range $[-3, 2]$ and where sources have integrated flux densities $<10\,\sigma_\text{rms}$ in the respectively catalogues. This results in a final sky model with 1\,092\,183 sources with spectral indices. The sky model has a median spectral index of $\approx -0.84$.

The apparent brightness template image is used to create the primary beam response. We generate a source-list using the source-finder \texttt{PyBDSF}\footnote{\url{https://github.com/lofar-astron/PyBDSF}.} \citep{pybdsf} for each template beam image, and cross-match these per-beam source lists to the RACS sky model. We restrict the match to sources within 2.25\,deg of the beam centre to ensure we are not matching sources in the primary beam sidelobes (which are imaged). Additionally, we restrict the per-beam source-list to compact components, with integrated to peak flux density ratios of $<1.2$. We use \texttt{flux\_warp} to generate the primary beam model, by taking the sky model cross-match results, extrapolating to the relevant frequency, and fitting a 2-D elliptical Gaussian model to the ratio $S_\text{image} / S_\text{sky model}$. While \citet{racs-mid} found Zernike polynomial models represented the mid-band beams more accurately than 2-D Gaussian models, the fitted elliptical Gaussian is sufficiently accurate for the main lobe within 0.75~deg for the low-band ASKAP data. An example beam model is shown in Figure~\ref{fig:beamexample} along with the calibrator sources used in generating the model. We use these models for primary beam correction. For clusters with multiple beams, we form a linear mosaic of the beam images as in the usual \texttt{ASKAPSoft} processing, applying the primary beam responses and weighting the co-addition by the square of the primary beam response.

As an estimate of the uncertainty in the brightness scale, $\xi_\text{scale}$, of the mosaicked ASKAP images, we take the quadrature sum of standard deviations of the residuals ($\sigma_{b,\text{residual}}$) from the calibrator sources used in generating the individual 2-D Gaussian beam models. In addition to the beam model uncertainty, we also add the uncertainties from the RACS-low and RACS-mid brightness scales, which are 7\% \citep{racs1} and 6\% \citep{racs-mid} in this case, respectively. In total, the brightness scale uncertainty is then \begin{equation}\label{eq:brightnessunc}
\xi{_\text{scale}}^2 = 0.07^2 + 0.06^2 + \sum_{b}^{N_\text{beams}} {\sigma_{b,\text{residual}}}^2 \, .
\end{equation}
Only the residuals from calibrator sources with a model beam attenuation of $\geq 0.1$ are included as the images are clipped for attenuation $<0.1$.

\subsubsection{A comparison of the new and archival images}

\begin{figure*}[t!]
    \centering
    \includegraphics[width=1\linewidth]{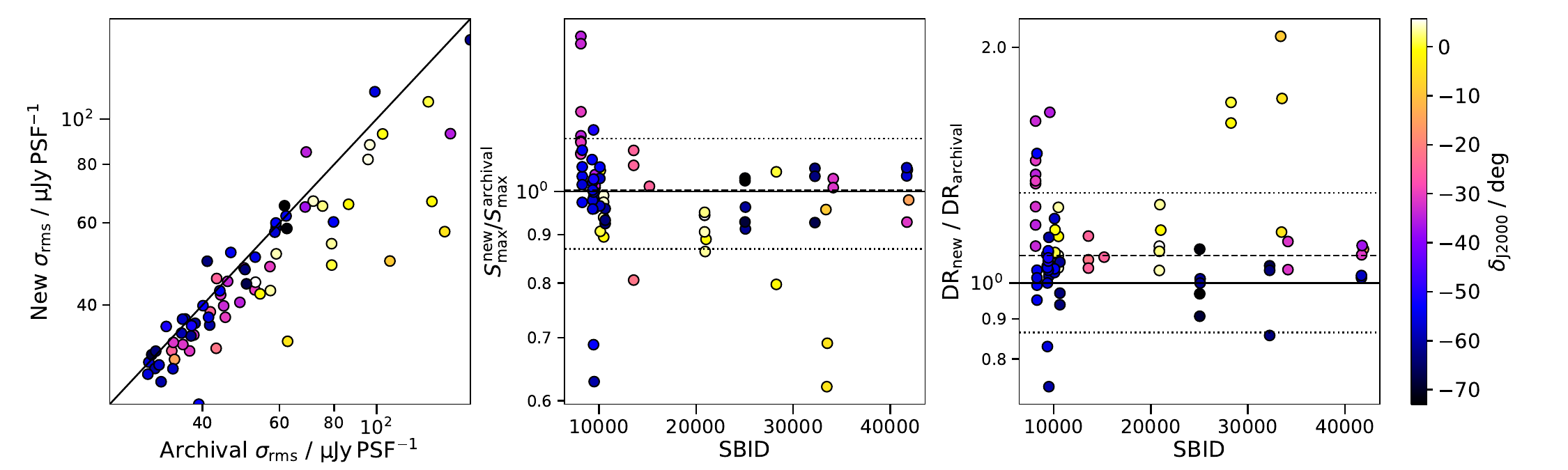}
    \caption{\label{fig:rms_comparison} {Comparison of the rms noise ($\sigma_\text{rms}$, \emph{left}), peak flux density ($S_\text{ms}$, as a function of SBID, \emph{centre}), and dynamic range (DR, as a function of SBID, \emph{right}) calculated within 1\,Mpc of cluster centres between the new re-processed, robust 0.0 images and the original archival images as they appear on CASDA. The points are coloured by the cluster declination. The solid black lines indicate equal values between the images.}}
\end{figure*}

{To compare the new, re-processed images with the archival images, we look at the rms noise ($\sigma_\text{rms}$) and peak flux density ($S_\text{max}$) within 1~Mpc of the cluster centres, and by extension the dynamic range ($\text{DR} = S_\text{max} / \sigma_\text{rms}$). For this comparison we use the re-processed robust 0.0 image as that image weighting is generally the closest match to the weighting used by \texttt{ASKAPsoft}. Figure~\ref{fig:rms_comparison} shows the comparison of the three quantities, with each cluster (and resulting image) coloured according to the cluster's declination. In general, there is a marginal improvement in the overall DR of the re-processed images (median $\text{DR}_\text{new}/\text{DR}_\text{archival} = 1.08_{-0.07}^{+0.17}$, with uncertainties drawn from the 16$^\text{th}$ and 84$^\text{th}$ percentiles of the distribution), though not all clusters see an improvement. Some of the largest improvements are in the equatorial fields, which is simply a combination of difference in image weighting (where robust 0.0 is not as close to the \texttt{ASKAPsoft} weighting) and some differences in the treatment of $w$-terms between \texttt{WSClean} (via \texttt{wgridder}) and the \texttt{ASKAPSoft} $w$-projection implementation\footnote{\url{https://www.atnf.csiro.au/computing/software/askapsoft/sdp/docs/current/calim/gridder.html}.}. In cases where the re-processed image has lower DR, this is generally the result of a difference of image PSF. There is also some variation to the brightness scales between re-processed and archival images due to the different primary beam models used, illustrated in the centre panel of Figure~\ref{fig:rms_comparison}, though there is general agreement with median $S_\text{max}^\text{new}/S_\text{max}^\text{archival} = 1.00_{-0.09}^{+0.07}$). For consistency we only use the re-processed images for analysis in this work.}  

{In Appendix~\ref{sec:app:images} we also summarise the rms noise (Table~\ref{tab:rms_properties}) and PSF (Table~\ref{tab:psf_properties}) of the five re-processed images (robust 0.0, robust $+0.25$, uniform, and the two tapered images) and archival images of each cluster. While the images used in this work are not directly output from the \texttt{ASKAPsoft} pipeline, we suggest they form an approximate representation of the images being produced for the main EMU survey that is currently underway.}

\subsubsection{Image-based angular scale filtering}\label{sec:filtering}

\begin{figure*}[p]
\centering
\begin{subfigure}[b]{1\linewidth}
\includegraphics[width=1\linewidth]{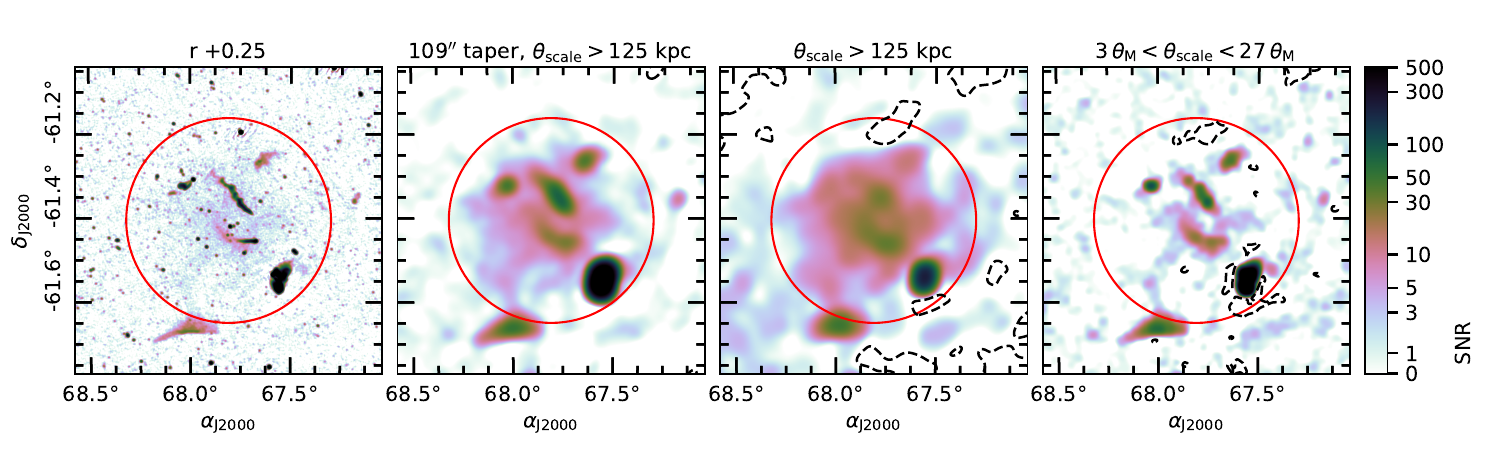}
\caption{\label{fig:filter:example1} PSZ2 G272.08$-$40.16 (Abell 3266).}
\end{subfigure}\\%
\begin{subfigure}[b]{1\linewidth}
\includegraphics[width=1\linewidth]{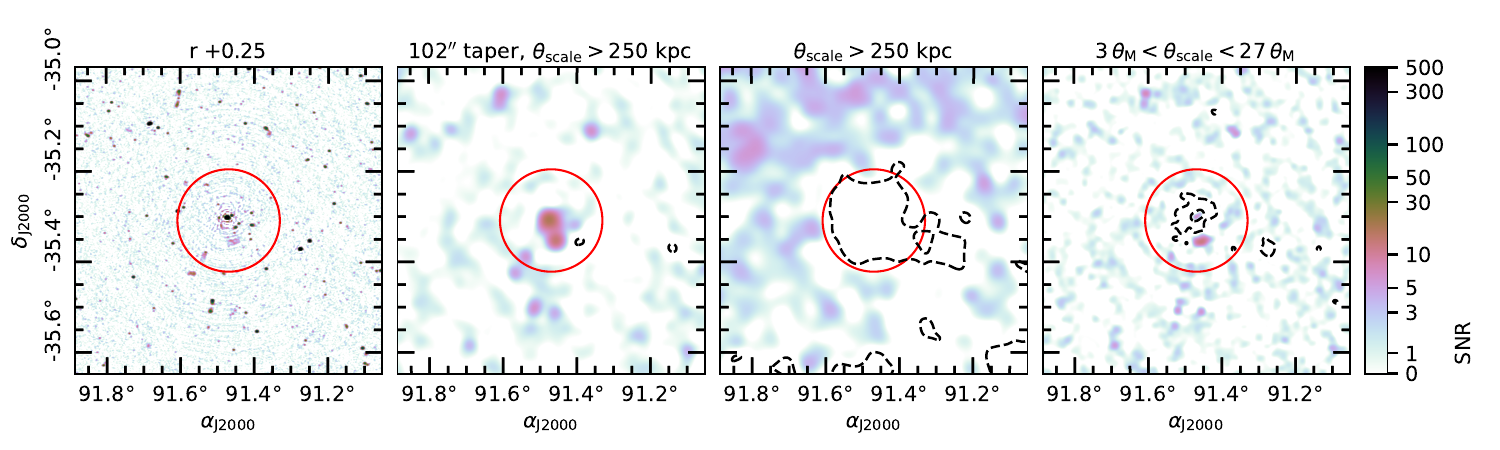}
\caption{\label{fig:filter:example2} PSZ2 G241.79$-$24.01 (Abell 3378).}
\end{subfigure}\\%
\begin{subfigure}[b]{1\linewidth}
\includegraphics[width=1\linewidth]{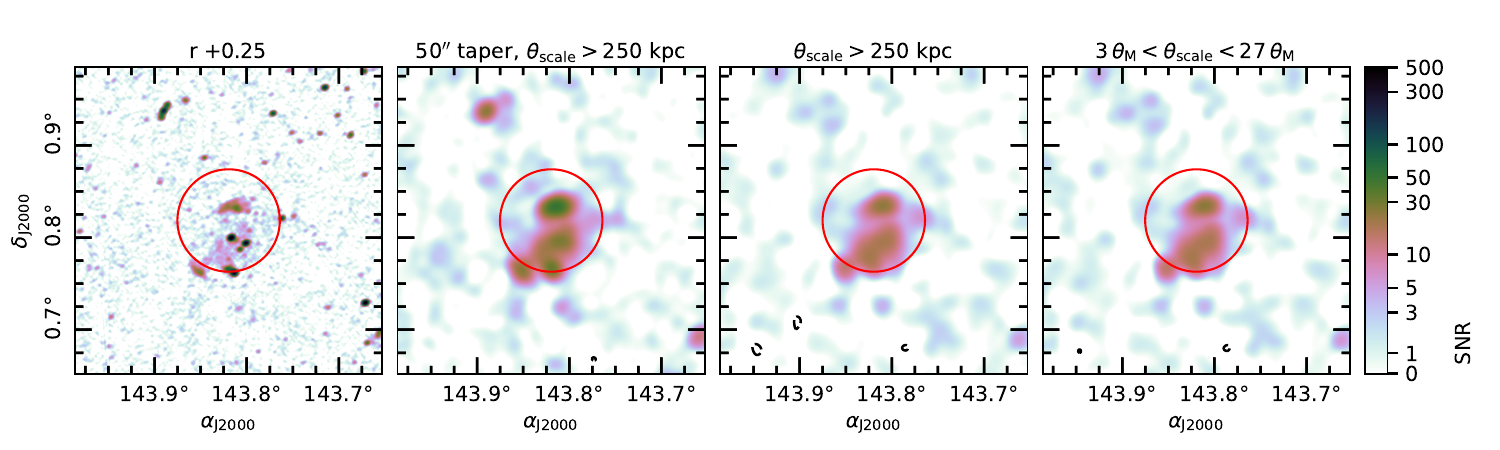}
\caption{\label{fig:filter:example3} PSZ2 G233.68$+$36.14.}
\end{subfigure}\\%
\caption{\label{fig:filter} Examples of angular scale filtering. \textit{Left.} Robust $+0.25$ reference image. \emph{Centre left.} $(u,v)$-filtered image, with corresponding taper applied during imaging. \emph{Centre right.} Image-based filtering using the same scale as the $(u,v)$ filtering. Note the image is convolved to the same resolution as the filter. \emph{Right.} Image-based filtering used for the EMU survey. Note that the image is convolved to the resolution of the lower filter. The red circles are centred on the cluster with a 1\,Mpc radius. Black, dashed contours are drawn on the filtered images at $-3\,\sigma_\text{rms}$.}
\end{figure*}

While not originally performed on these archival datasets, as part of the EMU processing pipeline, the main survey images go through an additional image-based angular scale filtering. This filtering is based on the multi-resolution filtering method described by \citet{Rudnick2002a}, and employs maximum and minimum sliding box filters at two angular scales to remove features in images that fall outside of the two chosen angular scales. We introduce a \texttt{python} implementation, \texttt{DiffuseFilter}\footnote{\url{https://gitlab.com/Sunmish/diffusefilter}.}. This implementation has a curated mode for filtering EMU images that removes angular scales outside of $3\,\theta_\text{M} \lesssim \theta_\text{scale} \lesssim 27\,\theta_\text{M}$, where $\theta_\text{M}$ is the {full width at half maximum of the} major axis of the image PSF. The smaller scale typically removes compact emission unassociated with diffuse cluster sources, and the larger scale is used to remove large-scale ripples. The ripples are generally a combination of undeconvolved sidelobes of off-axis extended (usually Galactic) emission, solar interference, and generally the poorer calibration of short baselines. 

Similar angular scale filtering has been used in previous cluster studies to identify diffuse radio sources within images with a large number of compact sources \citep[e.g.][]{Knowles2022,Venturi2022} and will be a feature of the upcoming EMU survey. For assessing expectations of the full EMU survey, we opt to generate these filtered maps for the robust $+0.25$ images alongside the $(u,v)$-plane subtraction method outlined earlier. We also create a separate filtered map similar to the $(u,v)$-plane subtraction method, removing similar scales only as a point of comparison. Figure~\ref{fig:filter} shows some examples of the different filter methods on a selection of clusters. {A comparison of the $(u,v)$-plane and image-plane filtering methods is presented in Section~\ref{sec:comparison_filtering}.}


\subsection{Optical and X-ray data}
We use optical data to inform positions of clusters and of any potential hosts to candidate diffuse radio sources. We typically only collect a single optical dataset per cluster, depending on availability and the sensitivity necessary.  The optical data include images from the Dark Energy Survey data release 2 \citep[DES DR2;][]{decam,des1,des:dr2}, the Pan-STARRS\footnote{Panoramic Survey Telescope And Rapid Response System} survey \citep[PS1;][]{tsl+12,cmm+16}, the Digitized Sky Survey (DSS2), and the Sloan Digitized Sky Survey data release 7 \citep[SDSS DR7;][]{sdssdr7}.

For associated X-ray observations, we query the \emph{Chandra}\footnote{\url{https://cxc.harvard.edu/cda/}.} and \emph{XMM}-Newton\footnote{\url{https://www.cosmos.esa.int/web/xmm-newton/xsa}.} online archives for observations of the PSZ2 clusters in our sample. {In total we find that 36 of the 71 PSZ2 clusters have existing \emph{XMM}-Newton observations, which we make use of, and we also make use of \emph{Chandra} observations for three additional clusters without \emph{XMM-Newton} data.} We used the standard pipeline data products, which are generally sufficient for this work. For the \emph{XMM}-Newton pipeline\footnote{\url{https://www.cosmos.esa.int/web/xmm-newton/pipeline}.} data products, we make use of the three-colour image generated from data taken by the European Photon Imaging Camera \citep[EPIC;][]{turner2001,struder2001}. For the \emph{Chandra} pipeline data products, we use images from the Advanced CCD Imaging Spectrometer (ACIS) produced through standard data processing by the \emph{Chandra} X-ray Centre\footnote{\url{https://cxc.cfa.harvard.edu/ciao/dictionary/sdp.html}.}. Optical and X-ray images are only used for qualitative analysis.

\begin{landscape}

\begin{table}[t!]
    \centering
    
    \resizebox{\textwidth}{!}{
    \begin{threeparttable}
    \caption{\label{tab:source:flux} Measured properties of the (candidate) diffuse radio sources detected in the PSZ2 clusters.}
    \begin{tabular}{l cc c c c c c c c ccc  c c c}\toprule
         Cluster & $z$ \tnote{a} & Source & New? & $\nu$ \tnote{b} & Taper \tnote{c} & $\xi_\text{scale}$ & Notes \tnote{d} & $S_\text{int}$ & $S_\text{sub}$ & Model & $S_\text{model}$ & Luminosity \tnote{e}  & LAS \tnote{f} & LLS \tnote{g} & Dist. \tnote{h} \\
                 &      &  & & (MHz) & (kpc) & & & (mJy) & (mJy) & & (mJy)       & $(\times 10^{24}$\,W\,Hz$^{-1}$) & (arcmin) & (kpc) & (kpc) \\[0.5em]\midrule
PSZ2 G008.31$-$64.74 &0.312 &R (SE) & \checkmark & 887.5 &- &0.22 &  - & $11.2 \pm 2.5$ &  - &  - & - & $3.74 \pm 0.30$ &  3.7 &  1000 &  2000 \\
PSZ2 G008.31$-$64.74 &0.312 &R (NW) & \checkmark & 887.5 &- &0.22 &  - & $11.2 \pm 2.6$ &  - &  - & - & $3.72 \pm 0.36$ &  4.6 &  1300 &  1200 \\
PSZ2 G008.31$-$64.74 &0.312 &cH & \checkmark & 887.5 &250 &0.22 &  - & $10.3 \pm 2.8$ & 1.5 & circle & $21.9 \pm 9.1$ & $7.3 \pm 3.1$ &  5.5 &  1500 &  - \\
PSZ2 G011.06$-$63.84 &0.224 &cH & \checkmark & 887.5 &100 &0.21 &  - & $5.4 \pm 1.4$ & 1.1 & skewed & $5.4 \pm 1.9$ & $0.84 \pm 0.30$ &  3.0 &  650 &  - \\
PSZ2 G011.92$-$63.53 &0.24 &H & \checkmark & 887.5 &100 &0.18 &  - & $7.7 \pm 1.6$ & 0.27 & skewed & $8.9 \pm 5.0$ & $1.62 \pm 0.90$ &  6.8 &  1500 &  - \\
PSZ2 G110.28$-$87.48 &0.52 &H & \checkmark & 943.5 &100 &0.29 &  - & $2.97 \pm 0.96$ & 0.16 & skewed & $3.4 \pm 1.2$ & $3.9 \pm 1.4$ &  2.7 &  1000 &  - \\
PSZ2 G149.63$-$84.19 &0.0569 &U (S) & $\times$ & 943.5 &- &0.25 &  - & $18.7 \pm 4.6$ &  - &  - & - & $0.1463 \pm 0.0028$ &  3.3 &  220 &  210 \\
PSZ2 G149.63$-$84.19 &0.0569 &U (N) & $\times$ & 943.5 &- &0.25 &  conf. & $337 \pm 89$ &  - &  - & - & $2.818 \pm 0.031$ &  1.7 &  110 &  150 \\
PSZ2 G172.98$-$53.55 &0.373 &H & $\times$ & 943.5 &100 &0.25 &  conf. & $5.6 \pm 1.7$ & 0.63 & skewed & $6.9 \pm 3.0$ & $3.5 \pm 1.6$ &  3.2 &  1000 &  - \\
PSZ2 G175.69$-$85.98 &0.23 &H & $\times$ & 943.5 &100 &0.33 &  - & $22.6 \pm 8.3$ & 1.8 & skewed & $22.9 \pm 8.7$ & $3.8 \pm 1.4$ &  7.1 &  1600 &  - \\
PSZ2 G219.88$+$22.83 &0.232813 &cR & \checkmark & 887.5 &100 &0.18 &  - & $14.5 \pm 2.9$ & 0.66 &  - & - & $2.58 \pm 0.19$ &  5.6 &  1200 &  2500 \\
PSZ2 G223.47$+$26.85 &0.3269 &H & \checkmark & 887.5 &250 &0.14 &  - & $3.24 \pm 0.69$ & 0.25 & skewed & $3.7 \pm 2.0$ & $1.38 \pm 0.74$ &  3.8 &  1100 &  - \\
PSZ2 G225.48$+$29.41 &0.203 &H & \checkmark & 943.5 &100 &0.27 &  conf. & $8.7 \pm 2.9$ & 1.6 & skewed & $5.3 \pm 2.1$ & $0.65 \pm 0.26$ &  5.1 &  1000 &  - \\
PSZ2 G227.59$+$22.98 &0.28085 &cR (NW) & \checkmark & 887.5 &100 &0.24 &  - & $2.88 \pm 0.76$ &  - &  - & - & $0.750 \pm 0.087$ &  2.2 &  570 &  1200 \\
PSZ2 G227.59$+$22.98 &0.28085 &cR (SE) & \checkmark & 887.5 &100 &0.24 &  - & $1.42 \pm 0.40$ &  - &  - & - & $0.369 \pm 0.055$ &  1.9 &  480 &  1500 \\
PSZ2 G227.89$+$36.58 &0.4616 &cH & \checkmark & 887.5 &100 &0.21 &  conf. & $1.30 \pm 0.33$ &  - &  - & - & $1.11 \pm 0.19$ &  1.6 &  560 &  - \\
PSZ2 G228.38$+$38.58 &0.543811 &U & \checkmark & 887.5 &- &0.18 &  - & $1.26 \pm 0.28$ &  - &  - & - & $1.61 \pm 0.26$ &  1.5 &  570 &  620 \\
PSZ2 G228.50$+$34.95 &0.2701 &cH & \checkmark & 887.5 &100 &0.20 &  - & $1.77 \pm 0.46$ & 0.036 & circle & $8.8 \pm 2.8$ & $2.10 \pm 0.67$ &  2.6 &  640 &  - \\
PSZ2 G231.79$+$31.48 &0.332405 &R & \checkmark & 887.5 &- &0.23 &  - & $16.1 \pm 4.3$ & 2.4 &  - & - & $7.28 \pm 0.46$ &  3.3 &  930 &  790 \\
PSZ2 G231.79$+$31.48 &0.332405 &H & \checkmark & 887.5 &250 &0.23 &  conf. & $7.8 \pm 5.8$ & 1.1 & ellipse & $5.1 \pm 2.0$ & $1.99 \pm 0.77$ &  6.0 &  1700 &  - \\
PSZ2 G232.84$+$38.13 &0.1508 &cR & \checkmark & 887.5 &100 &0.18 &  conf. & $26.2 \pm 4.9$ &  - &  - & - & $1.722 \pm 0.068$ &  6.0 &  940 &  580 \\
PSZ2 G232.84$+$38.13 &0.1508 &U & \checkmark & 887.5 &- &0.18 &  - & $9.8 \pm 1.8$ &  - &  - & - & $0.622 \pm 0.023$ &  1.4 &  230 &  74 \\
PSZ2 G233.68$+$36.14 &0.356823 &R (N) & \checkmark & 887.5 &- &0.18 &  - & $15.2 \pm 2.8$ &  - &  - & - & $6.97 \pm 0.46$ &  2.9 &  870 &  290 \\
PSZ2 G233.68$+$36.14 &0.356823 &R (S) & \checkmark & 887.5 &- &0.18 &  - & $6.9 \pm 1.3$ &  - &  - & - & $3.17 \pm 0.25$ &  2.2 &  650 &  1100 \\
PSZ2 G233.68$+$36.14 &0.356823 &H & \checkmark & 887.5 &- &0.18 &  conf. & $19.8 \pm 4.3$ & 1.9 & skewed & $19.7 \pm 5.5$ & $9.0 \pm 2.6$ &  4.6 &  1400 &  - \\
PSZ2 G239.27$-$26.01 &0.43 &H & $\times$ & 943.5 &100 &0.21 &  - & $15.4 \pm 3.4$ & 0.72 & skewed & $14.3 \pm 3.2$ & $10.3 \pm 2.4$ &  4.5 &  1500 &  - \\
PSZ2 G241.79$-$24.01 &0.1392 &U (S) & \checkmark & 943.5 &100 &0.17 &  - & $6.4 \pm 1.2$ & 0.30 &  - & - & $0.360 \pm 0.024$ &  2.7 &  390 &  390 \\
PSZ2 G241.79$-$24.01 &0.1392 &U (central) & \checkmark & 943.5 &100 &0.17 &  - & $9.8 \pm 2.2$ & 2.2 &  - & - & $0.651 \pm 0.033$ &  4.1 &  600 &  37 \\

\bottomrule

    \end{tabular}
    \begin{tablenotes}[flushleft]
    {\footnotesize \item[a] `*' indicates assumed redshift. \item[b] Image frequency. \item[c] Image taper used. `-' if the standard robust $+0.25$ image is used (see Section~\ref{sec:ddselfcal} for details). \item[d] `conf.' refers to sources that are blended with either residual emission after subtraction or with other diffuse sources. \item[e] At the image frequency, assuming $\alpha = -1.2 \pm 0.2$ for all sources. \item[f] Largest deconvolved angular scale (or extent) within $2\,\sigma$ contours. \item[g] Largest deconvolved linear scale (or extent) from the LAS at the reported redshift. \item[h] Linear projected distance from the PSZ2 cluster centre.}
    \end{tablenotes}
\end{threeparttable}
    }
\end{table}

\end{landscape}
\begin{landscape}
    
\begin{table}[!t]
    \centering
    \resizebox{\textwidth}{!}{
    \caption*{}
    \begin{tabular}{l cc c c c c c c c ccc  c c c}\toprule
         Cluster & $z$ \tnote{a} & Source & New? & $\nu$ \tnote{b} & Taper \tnote{c} & $\xi_\text{scale}$ & Notes \tnote{d} & $S_\text{int}$ & $S_\text{sub}$ & Model & $S_\text{model}$ & Luminosity \tnote{e}  & LAS \tnote{f} & LLS \tnote{g} & Dist. \tnote{h} \\
                 &      &  & & (MHz) & (kpc) & & & (mJy) & (mJy) & & (mJy)       & $(\times 10^{24}$\,W\,Hz$^{-1}$) & (arcmin) & (kpc) & (kpc) \\[0.5em]\midrule

PSZ2 G260.80$+$06.71 &0.2* &cR & \checkmark & 943.5 &100 &0.26 &  - & $64 \pm 17$ &  - &  - & - & - & 5.7 & - &  - \\
PSZ2 G260.80$+$06.71 &0.2* &cH & \checkmark & 943.5 &100 &0.26 &  - & $11.6 \pm 3.3$ & 0.59 & skewed & $10.7 \pm 5.2$ & - & 7.2 & - &  - \\
PSZ2 G262.36$-$25.15 &0.0514 &U & $\times$ & 1013.5 &- &0.30 &  - & $7.6 \pm 2.3$ &  - &  - & - & $0.0479 \pm 0.0026$ &  2.2 &  130 &  730 \\
PSZ2 G263.14$-$23.41 &0.2266 &H & $\times$ & 1013.5 &100 &0.26 &  - & $13.9 \pm 4.5$ & 2.8 & skewed & $12.5 \pm 4.9$ & $1.97 \pm 0.78$ &  4.4 &  970 &  - \\
PSZ2 G263.19$-$25.19 &0.0506 &U & $\times$ & 1013.5 &- &0.20 &  conf. & $345 \pm 69$ &  - &  - & - & $2.109 \pm 0.024$ &  8.7 &  510 &  640 \\
PSZ2 G263.68$-$22.55 &0.1644 &H & $\times$ & 1013.5 &250 &0.41 &  - & $28 \pm 14$ & 4.0 & skewed & $30 \pm 15$ & $2.3 \pm 1.2$ &  8.4 &  1400 &  - \\
PSZ2 G272.08$-$40.16 &0.0589 &R & $\times$ & 943.5 &50 &0.19 &  - & $74 \pm 14$ &  - &  - & - & $0.625 \pm 0.012$ &  10. &  700 &  1100 \\
PSZ2 G272.08$-$40.16 &0.0589 &U (N) & $\times$ & 943.5 &50 &0.19 &  - & $33.8 \pm 6.6$ &  - &  - & - & $0.2843 \pm 0.0098$ &  6.4 &  440 &  670 \\
PSZ2 G272.08$-$40.16 &0.0589 &H & $\times$ & 943.5 &50 &0.19 &  conf. & $256 \pm 58$ & 28 & skewed & $161 \pm 45$ & $1.35 \pm 0.38$ &  24 &  1600 &  - \\
PSZ2 G272.08$-$40.16 &0.0589 &U (W) & $\times$ & 943.5 &- &0.19 &  conf. & $3.40 \pm 0.74$ &  - &  - & - & $0.0286 \pm 0.0028$ &  2.2 &  150 &  870 \\
PSZ2 G286.28$-$38.36 &0.307 &H & \checkmark & 887.5 &100 &0.20 &  - & $11.8 \pm 2.8$ & 1.8 & skewed & $12.8 \pm 3.9$ & $4.1 \pm 1.3$ &  5.5 &  1500 &  - \\
PSZ2 G286.75$-$37.35 &0.47 &H & \checkmark & 887.5 &100 &0.14 &  - & $8.3 \pm 1.4$ & 0.57 & circle & $8.9 \pm 1.9$ & $7.9 \pm 1.8$ &  3.8 &  1300 &  - \\
PSZ2 G311.98$+$30.71 &0.048 &H & $\times$ & 943.5 &100 &0.20 &  - & $30.2 \pm 7.7$ & 4.9 & skewed & $44 \pm 15$ & $0.242 \pm 0.082$ &  11 &  600 &  - \\
PSZ2 G313.33$+$30.29 &0.049 &H & $\times$ & 943.5 &100 &0.26 &  - & $49 \pm 16$ & 8.8 & skewed & $70 \pm 27$ & $0.40 \pm 0.15$ &  12 &  700 &  - \\
PSZ2 G313.33$+$30.29 &0.049 &U & $\times$ & 943.5 &100 &0.26 &  - & $35 \pm 10$ & 1.8 &  - & - & $0.215 \pm 0.014$ &  8.1 &  470 &  1100 \\
PSZ2 G313.33$+$30.29 &0.049 &Bridge & $\times$ & 943.5 &100 &0.26 &  conf. & $73 \pm 45$ &  - &  - & - & $0.977 \pm 0.044$ &  28 &  1600 &  - \\
PSZ2 G332.23$-$46.37 &0.098 &H & \checkmark & 943.5 &100 &0.13 &  - & $8.1 \pm 1.3$ & 1.2 & skewed & $9.7 \pm 2.3$ & $0.241 \pm 0.058$ &  5.7 &  620 &  - \\
PSZ2 G333.89$-$43.60 &0.318 &H & \checkmark & 943.5 &100 &0.22 &  conf. & $4.4 \pm 1.2$ & 0.57 & skewed & $4.4 \pm 1.7$ & $1.52 \pm 0.60$ &  3.5 &  980 &  - \\
PSZ2 G335.58$-$46.44 &0.076 &H & \checkmark & 943.5 &100 &0.19 &  - & $38.1 \pm 8.1$ & 2.9 & skewed & $55 \pm 13$ & $0.79 \pm 0.19$ &  12 &  1100 &  - \\
PSZ2 G335.58$-$46.44 &0.076 &R & \checkmark & 943.5 &- &0.19 &  - & $17.7 \pm 3.4$ &  - &  - & - & $0.2548 \pm 0.0071$ &  4.2 &  360 &  470 \\
PSZ2 G341.19$-$36.12 &0.284 &R (NW) & $\times$ & 943.5 &- &0.21 &  conf. & $3.48 \pm 0.98$ & 0.22 &  - & - & $1.200 \pm 0.098$ &  3.0 &  760 &  450 \\
PSZ2 G341.19$-$36.12 &0.284 &R (SE) & $\times$ & 943.5 &- &0.21 &  conf. & $7.3 \pm 1.8$ & 0.42 &  - & - & $2.22 \pm 0.15$ &  3.8 &  960 &  1100 \\
PSZ2 G342.33$-$34.93 &0.232 &R (W) & $\times$ & 943.5 &- &0.21 &  - & $15.8 \pm 3.3$ &  - &  - & - & $2.64 \pm 0.13$ &  2.9 &  640 &  640 \\
PSZ2 G342.33$-$34.93 &0.232 &R (E) & $\times$ & 943.5 &100 &0.21 &  - & $3.22 \pm 0.74$ &  - &  - & - & $0.539 \pm 0.055$ &  2.8 &  630 &  780 \\
PSZ2 G342.33$-$34.93 &0.232 &H & $\times$ & 943.5 &100 &0.21 &  conf. & $32.8 \pm 7.7$ & 4.0 & skewed & $53 \pm 14$ & $8.9 \pm 2.4$ &  5.7 &  1300 &  - \\
PSZ2 G342.62$-$39.60 &0.139 &U & $\times$ & 943.5 &- &0.26 &  conf. & $12.5 \pm 4.0$ & 2.2 &  - & - & $0.818 \pm 0.039$ &  4.1 &  600 &  150 \\
PSZ2 G346.86$-$45.38 &0.0796 &H & \checkmark & 943.5 &100 &0.16 &  - & $7.6 \pm 1.6$ & 1.4 & skewed & $7.2 \pm 2.2$ & $0.115 \pm 0.035$ &  4.8 &  430 &  - \\
PSZ2 G347.58$-$35.35 &0.22 &cH & \checkmark & 943.5 &- &0.24 &  - & $4.0 \pm 1.0$ & 0.0027 & skewed & $6.1 \pm 1.5$ & $0.90 \pm 0.23$ &  2.4 &  520 &  - \\

\bottomrule
\end{tabular}}
\end{table}

\end{landscape}

\section{Survey results}

\begin{figure*}[p]
    \centering
    \includegraphics[width=1\linewidth]{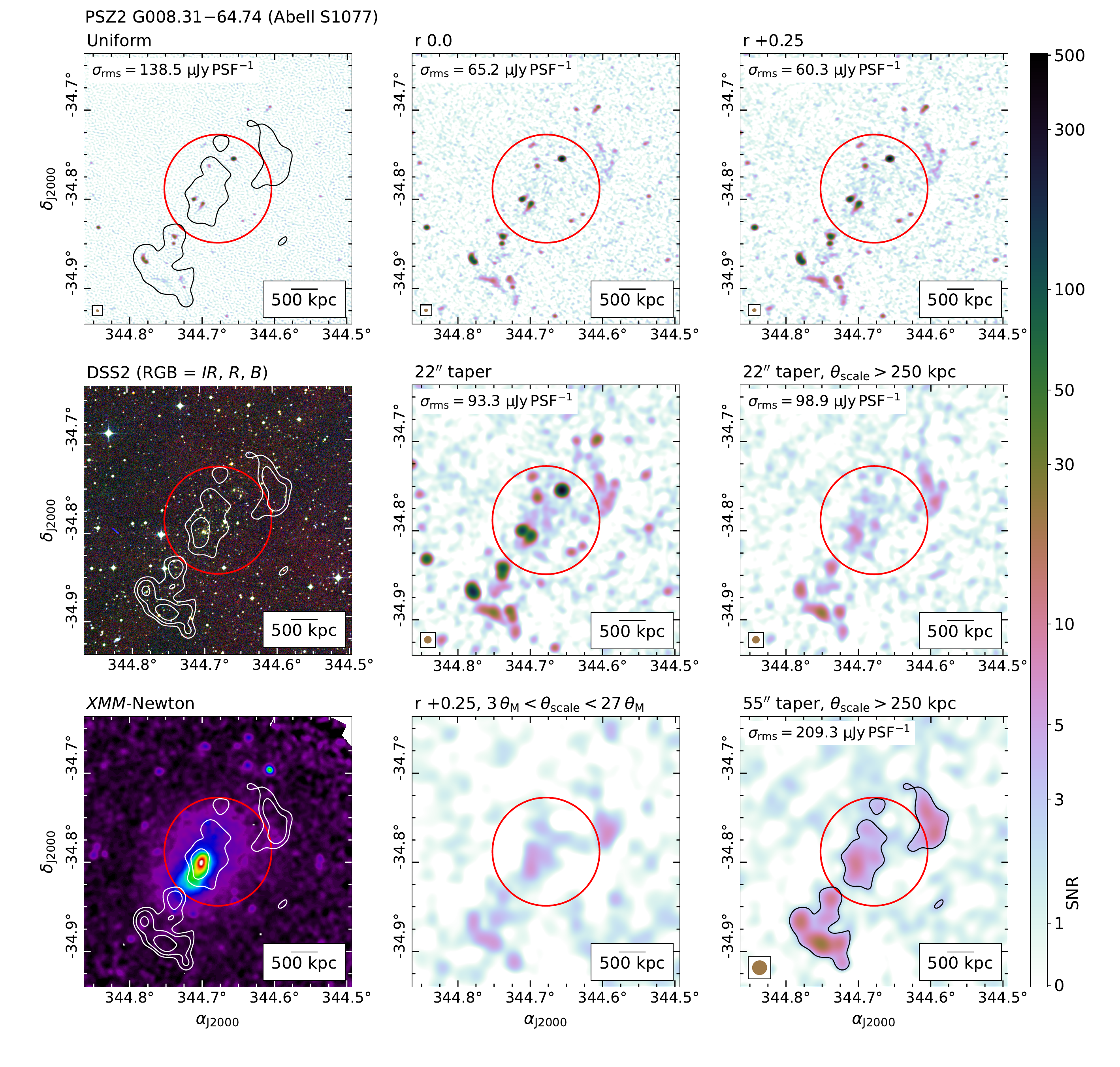}
    \caption{\label{fig:example} Example set of images used for identification of sources in PSZ2~G008.31$-$64.74. The colour scales in all radio images are linear between the range $[0, 3\,\sigma_\text{rms}]$ and logarithmic in the range $(3\,\sigma_\text{rms}, 500\,\sigma_\text{rms}]$. The white and black contours are of the bottom right image, and are drawn at $[3, 6, 12, 24, 48]\times\sigma_\text{rms}$ in the optical and X-ray panels and at $3\,\sigma_\text{rms}$ in other panels. The solid circle is centred on the reported PSZ2 coordinates and has a 1~Mpc radius at the cluster redshift. Clusters without a measured redshift are assumed to be at $z=0.2$, and the circle is dashed in those cases. Clusters without publicly available \emph{XMM}-Newton and \emph{Chandra} observations are shown without an X-ray image. Images of all clusters are made available online. Note that the image-based filtering retains the resolution and brightness units as the original robust $+0.25$ map, and so appears with the same brightness scale as the original robust $+0.25$ image after filtering.}
\end{figure*}

\begin{figure*}[t!]
    \centering
    \begin{subfigure}[b]{0.5\linewidth}
    \includegraphics[width=1\linewidth]{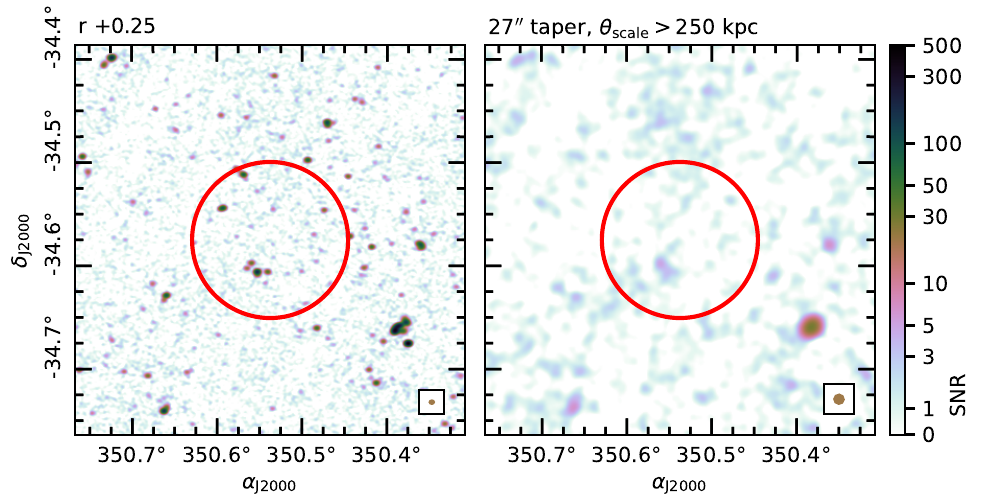}
    \caption{\label{fig:2panel:1} PSZ2~G006.16$-$69.49.}
    \end{subfigure}%
    \begin{subfigure}[b]{0.5\linewidth}
    \includegraphics[width=1\linewidth]{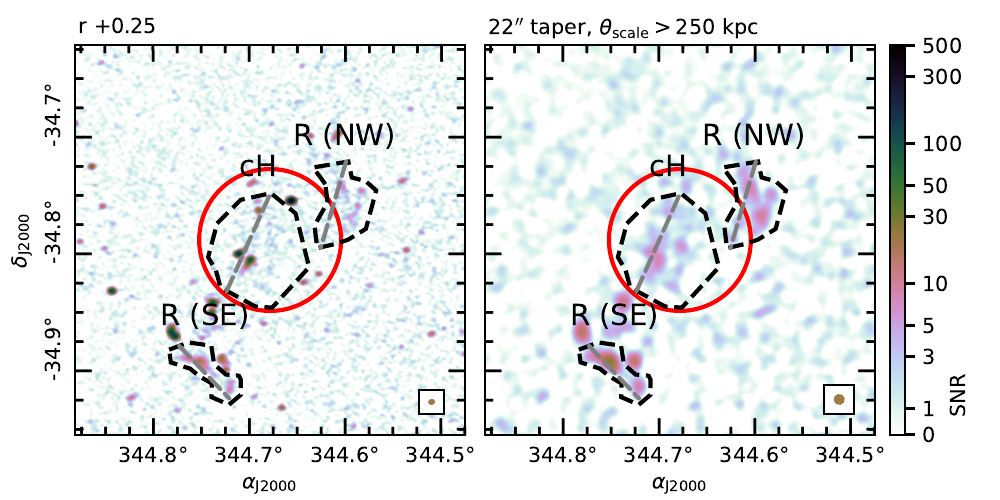}
    \caption{\label{fig:2panel:2} PSZ2~G008.31$-$64.74.}
    \end{subfigure}
    \caption{{\label{fig:2panel} Example images of PSZ2~G006.16$-$69.49 \subref{fig:2panel:1} with no diffuse sources and PSZ2~G008.31$-$64.74 \subref{fig:2panel:2} with two relics and a candidate halo. Similar images for all clusters are included in Appendix~\ref{sec:app:images}. \emph{Left panels.} The robust $+0.25$ reference image. \emph{Right panels.} The robust $+0.25$ image, tapered, after subtraction of sources of scales $<250$~kpc. In all panels, the red circle has a 1~Mpc radius at the redshift of the clusters (in Appendix~\ref{sec:app:images} a dashed circle indicates an assumed redshift of 0.2). For clusters with diffuse emission, the dashed polygon regions indicate the diffuse sources of interest and are the regions used for integrated flux density measurements. The PSF of each image is shown in the bottom right corner.}}
\end{figure*}

The clusters in our sample that are found to host diffuse emission are reported in Table~\ref{tab:source:flux} with measured and derived quantities where possible.  The classification scheme we follow and the source measurements are described in the following sections, along with notes on the individual systems. 



\subsection{Classification scheme}

For classification of sources, we largely follow the scheme outlined by \citetalias{Botteon2022} though we relax the criteria for considering objects `candidate' sources. The classifications generally only consider three types of emission: \begin{enumerate}
    \item \emph{Radio halo (H).} A diffuse, extended radio source located at the centre of a cluster. For the purpose of classifying diffuse radio sources, the cluster centre can be considered the X-ray centroid (if there is X-ray data available), the SZ peak (the location of the PSZ2 coordinates), or the optical centre (as seen in the available optical images), with preference in that order. No distinction is made between different types of radio halos (mega, giant, mini).
    \item \emph{Radio relic (R).}  An extended radio source towards the periphery of a cluster, assuming the same concept of cluster centre as described above. We relax the sharp surface brightness criterion from \citetalias{Botteon2022}, {noting relics viewed at different angles can have a range of morphologies \citep[e.g.][]{Skillman2013,Wittor2023} and that the lower resolution of the ASKAP observations is less able to detect such features in the images.} {We require that any sources we classify as radio relics do not have obvious features of a radio galaxy (lobes, hotspot, AGN core, optical host). We do not require morphological features such as `arc-like shape' as this is largely dependent on the geometry of the merger/projection as mentioned above. While we do find the projected size of the detected relics to be $> 300$\,kpc in line with \citetalias{Botteon2022}, we did not use this criterion.}
    \item \emph{Unclassified/other diffuse emission (U).} An extended radio source that does not fit into the radio halo and relic classifications, but is not obviously a radio galaxy (or similar active radio source) and does not have an obvious optical identification. This classification collects diffuse radio sources such as phoenices/(revived) fossil plasmas \citep{kbc+04}.   We do not distinguish between these types of emission as there is no meaningful way to do so with the single frequency ASKAP data available. 
    \item {\emph{Nothing applicable (NA).} Clusters that do not feature a source that can be classified as above are classified as `NA'. Note this applies to the cluster as a whole rather than individual sources as opposed to the other classifiers.}
\end{enumerate}

A halo or relic might be considered a candidate (cH, cR) if it is not clear whether the emission is from other unsubtracted radio sources in the cluster. While this classification scheme is similar to that used by \citetalias{Botteon2022}, it is not as rigorous. \citet{Hoang2022} use a similar visual classification approach rather than a rigorous decision tree looking at non-PSZ2 clusters in the LoTSS-DR2 data. They find similar results for radio halos and relics when comparing the classification methods using the same decision tree. For other sources (the unclassified diffuse sources and AGN-related emission) the two approaches may be less consistent. 

{Figure~\ref{fig:example} shows a set of images used for `quick-look' classification of the sources (using PSZ2~G008.31$-$64.74 as an example) and highlights the range of images available for each cluster. Generally, images with interesting sources are followed-up with more in-depth inspection of the FITS images. In Figure~\ref{fig:2panel} we show an example of the set of images provided in Appendix~\ref{sec:app:images} for all clusters that highlight the main robust 0.25 image, the source-subtracted, tapered image, and any sources of interest.}


\subsection{Source measurements}\label{sec:flux}

\begin{figure}[t!]
    \centering
    \includegraphics[width=1\linewidth]{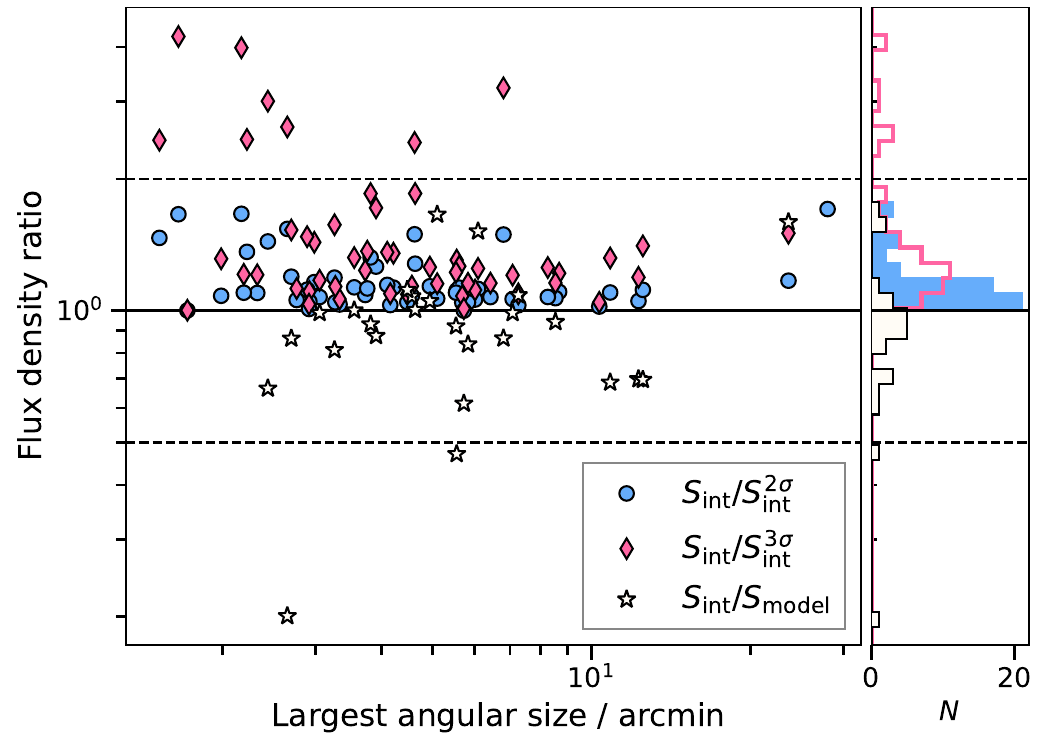}
    \caption{\label{fig:flux_comparison} Flux density ratios as a function of largest angular size/extent. We show the comparison between full measurement (all pixels) and three alternatives: integration over $2\,\sigma_\text{rms}$ (blue, circles), $3\,\sigma_\text{rms}$ (pink, diamonds), and the model flux densities for radio halos (white, stars). A histogram of the distribution of the flux density ratios is also shown. The solid black line indicates a ratio of 1, and the dashed black lines are drawn at flux density ratios of 0.5 and 2.}
\end{figure}


{Measured properties of sources are reported in Table~\ref{tab:source:flux} and relevant measurements are described below. When measuring properties of radio halos we make use of the compact source-subtracted [$(u,v)$-plane subtraction] robust 0.25 map with tapering. As relics and smaller-scale diffuse emission tend to have small-scale features, we generally make use of the non-filtered images for those sources to avoid loss of flux density. For relics/unclassified sources with embedded compact emission, we instead subtract the peak flux density of the intervening sources from the total integrated flux density measurements.}


We define polygon regions that cover the sources of interest and first estimate the largest angular size (LAS) of the source using the largest angular separation between any pair of pixels within the region above. {Figure~\ref{fig:2panel:2} [and Figures~\ref{fig:app:PSZ2G006.16-69.49}--\ref{fig:app:PSZ2G347.58-35.35} in Appendix~\ref{sec:app:images}] shows the line (grey, dashed) between the two selected pixels used to estimate the source sizes.} We also make use of the polygon regions to determine the flux-weighted centroid of each source, and for relics and other unclassified diffuse emission we calculated the angular separation along with the project distance of the diffuse source from the cluster centre as reported in the PSZ2 catalogue. 

For all diffuse cluster sources reported, we provide a measurement of the integrated flux density, $S_\text{int}$, at the frequency of the relevant ASKAP image, following

\begin{equation}
S_\text{int} = \sum_n^N S_{n} \left( \dfrac{|c_1 c_2| 4 \ln 2}{\pi \theta_\text{M}\theta_\text{m}}\right) \quad \text{Jy} \, ,
\end{equation}

where $N$ is the number of image pixels comprising the integration region (optionally for pixels above a brightness threshold, e.g.\ $3\,\sigma_\text{rms}$), $c_1$ and $c_2$ are the pixel dimensions in R.~A. and declination, and $\theta_\text{M}$ and $\theta_\text{m}$ are the major and minor axes of the PSF. Associated uncertainties, $\sigma_S$, are calculated via 

\begin{equation}\label{eq:fluxunc}
{\sigma_S} = \sqrt{N_\text{PSF}{\sigma_\text{rms}}^2 + \left( S_\text{int} \xi_\text{scale} \right)^2}\quad \text{Jy} \, ,
\end{equation}
where $N_\text{PSF}$ is the number of PSFs covering the full integration region, $\xi_\text{scale}$ is the brightness scale uncertainty (Equation~\ref{eq:brightnessunc}). 

{After visual inspection of the compact source-subtracted datasets we find residual emission for some sources and suggest this creates a bias in the measurements for the radio halos. This scales with the number of compact sources subtracted and is generally more significant for fainter compact sources which are harder to image and subtract when imaging with a $(u,v)$ cut. We define this flux density bias as \begin{equation}
S_\text{sub} = 0.2 \left( 1 - \dfrac{S^{\prime}}{S} \right) S_\text{int} \quad \text{Jy} \, ,
\end{equation}
where $S$ is the integrated flux density within the polygon region of the uniformly weighted image, and $S^\prime$ is the same measurement on the uniform map after compact source subtraction. By construction this is always less than 20\% of the original integrated flux density measurement and is subtracted from measurements made using the $(u,v)$-filtered maps.}

 For the flux density measurement, we include all pixels within the polygon regions. Including all pixels within the polygon allows some reduction in bias of low-significance diffuse emission \citep[e.g.][]{blobcat} and we assume the noise is symmetric and account for measurement of low-SNR pixels by including all pixels within the polygon in the estimation of $\sigma_S$ (in Equation~\ref{eq:fluxunc}). In Figure~\ref{fig:flux_comparison} we show the ratio of flux density measurements between this measurement technique and similar integration on pixels above $2\,\sigma_\text{rms}$ and $3\,\sigma_\text{rms}$ as a function of the largest angular size of the source measured within $2\,\sigma_\text{rms}$ contours. We also show the same flux density ratio between the integrated flux density and the model flux density for radio halos described in the following section. The measured flux density is always larger when including all pixels, but does not appreciably change with source size. {Polygon regions used for measurements are shown on the right panel of Figure~\ref{fig:2panel:2} for PSZ2~G008.31$-$64.74 and in Figures~\ref{fig:app:PSZ2G006.16-69.49}--\ref{fig:app:PSZ2G347.58-35.35} in Appendix~\ref{sec:app:images} for the remaining clusters.}

\subsection{Radio halo models}\label{sec:models}

\begin{figure*}
    \centering
    \begin{subfigure}[b]{0.5\linewidth}        
    \includegraphics[width=1\linewidth]{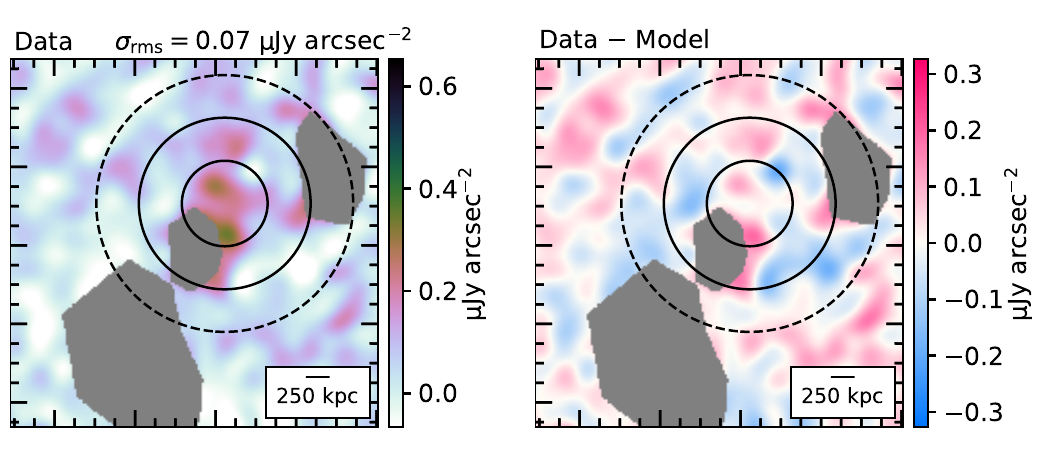}
    \caption{\label{fig:model:example:1} PSZ2 G008.31$-$64.74 (circle model).}
    \end{subfigure}%
    \begin{subfigure}[b]{0.5\linewidth}        
    \includegraphics[width=1\linewidth]{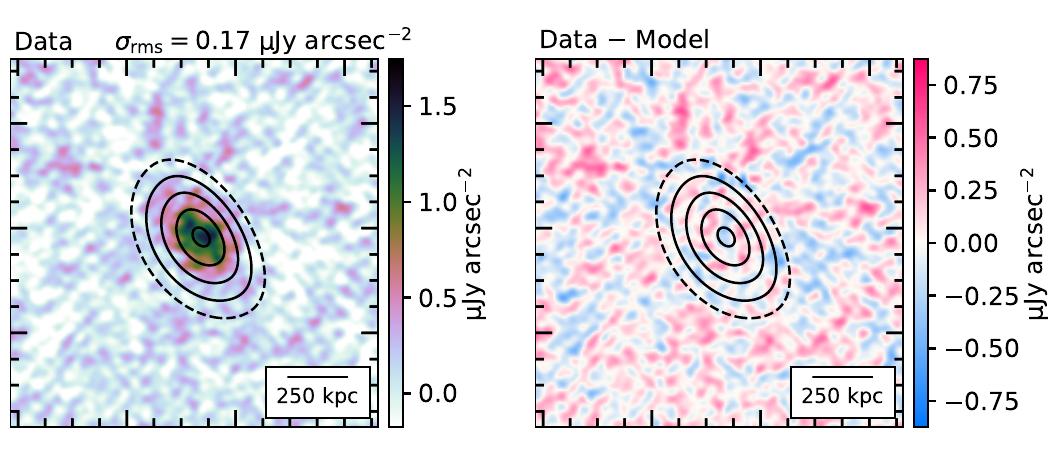}
    \caption{\label{fig:model:example:2} PSZ2 G011.06$-$63.84 (skewed model).}
    \end{subfigure}%
    \caption{\label{fig:model:example} {Example radio halo models fit using \texttt{Halo-FDCA}. \emph{Left panels.} Compact source-subtracted image used for modelling the halo (and flux density measurements). \emph{Right panels.} Residual image after subtraction of the model. The model is shown as black contours in both panels (solid: $[1, 2, 4, 8, 16, 32]\times\sigma_\text{rms}$, dashed: $0.5\,\sigma_\text{rms}$). The left panel colourscales are linear between $[-1, 10]\times\sigma_\text{rms}$ and the right panel colourscales are linear between $[-5, 5]\times\sigma_\text{rms}$. Note the surface brightness units are in \textmu Jy\,arcsec$^{-2}$ for consistency with the literature.}}
\end{figure*}

To help with obtaining flux densities of radio halos, we use \texttt{Halo-FDCA} \citep{Boxelaar2021} to fit a range of 2-D exponential profiles to the surface brightness in the compact source-subtracted images. For each halo (and candidate halo) we use the same image used for integrated flux density measurements described in the previous section, and mask intervening sources/residual emission not associated with the radio halo. We then use \texttt{Halo-FDCA} to fit three profiles to each halo: a standard circular exponential profile \citep[see e.g.][]{Orru2007,Murgia2009,Bonafede2009}, and generalised elliptical and skewed elliptical profiles \citep{Boxelaar2021}. After fitting, we use the reduced $\chi^2$ as a simple model selection parameter. To obtain the model flux density, $S_\text{model}$, the fitted exponential profile is integrated out to three times the $e$-folding radius following \citetalias{Botteon2022}.

In Figure~\ref{fig:model:example} we show the results of fitting the radio halo models with \texttt{Halo-FDCA} in a low-SNR case [a circular model for the halo in PSZ2 G008.31$-$64.74, \ref{fig:model:example:1}] and a high-SNR case [a skewed model for the halo in PSZ2 G011.06$-$63.84, \ref{fig:model:example:2}]. Equivalent images for other halos reported in this work are shown in Appendix~\ref{sec:app:models}. The model flux densities and the selected models are reported in Table~\ref{tab:source:flux} alongside the integrated flux densities described in Section~\ref{sec:flux}. We show the ratio of $S_\text{int}/S_\text{model}$ for all halos in Figure~\ref{fig:flux_comparison}, finding a median $S_\text{int}/S_\text{model} = 0.93_{-0.24}^{+0.16}$. Residual emission from partially subtracted sources or heavily-confused clusters results in difference between the integration within polygon regions and integration of the model profile. For radio halo power calculations, we use the model flux density rather than the integrated flux density measured within the polygon region unless otherwise stated.


\subsection{Notes on individual systems}\label{sec:notes}

{Each cluster hosting a diffuse source of interest is included in Table~\ref{tab:source:flux} along with measured properties and a note indicating if we are reporting the source for the first time. In the following, we include relevant notes about the individual clusters and the diffuse emission we detect.}

\subsubsection{PSZ2 G008.31\texorpdfstring{$-$}{-}64.74 (Abell S1077)}\label{sec:PSZ2G008.31-64.74}
\emph{Figure~\ref{fig:app:PSZ2G008.31-64.74}}. We report the detection of a double radio relic system (SE and NW), with additional residual emission at the cluster center that we consider a candidate radio halo. \citet{DeFilippis2004} report a $\approx 1.5$~arcmin soft X-ray tail of emission in \emph{Chandra} data, in the direction of the newly discovered SE radio relic. \citet{DeFilippis2004} also report two X-ray surface brightness and temperature discontinuities towards the NE of the cluster centre, though these are not coincident with the NW and SE radio relics detected here. 

\subsubsection{PSZ2 G011.06\texorpdfstring{$-$}{-}63.84 (Abell 3934)}\label{sec:PSZ2GG011.06-63.84}
\emph{Figure~\ref{fig:app:PSZ2G011.06-63.84}}. We report a candidate radio halo in this cluster. The cluster has no deep X-ray observation available from neither \emph{Chandra} nor XMM-\emph{Newton}. There is a compact source at the centre of the emission which is often seen with mini-halos, though with a project linear size of 650 kpc (within 2$\sigma_\text{rms}$ contours) the source is considerably larger than a traditional mini-halo. 

\subsubsection{PSZ2 G011.92\texorpdfstring{$-$}{-}63.53 (SPT-CL~J2251\texorpdfstring{$-$}{-}3324)}\label{sec:PSZ2G011.92-63.53}
\emph{Figure~\ref{fig:app:PSZ2G011.92-63.53}}. While no redshift is available in the PSZ2 catalogue, \citet{Bleem2020} report $z=0.24$ for the cluster. We report the detection of 1.5-Mpc diffuse emission near the centre of the cluster present in both the $(u,v)$-plane compact source-subtracted map and the image-filtered map, which we consider a radio halo. There are presently no \emph{Chandra} or \emph{XMM}-Newton observations available. We note that there is a large redshift distribution in the general vicinity of the cluster, in the range $0.06\lesssim z \lesssim 0.25$, indicating there may various clusters along this line of sight.



\subsubsection{PSZ2 G110.28\texorpdfstring{$-$}{-}87.48}\label{sec:PSZ2G110.28-87.48}
\emph{Figure~\ref{fig:app:PSZ2G110.28-87.48}.} We report the detection of a radio halo in PSZ2 G110.28\texorpdfstring{$-$}{-}87.48. The radio halo is co-located with the X-ray emission, and we note there is a $\approx 2.4$\,arcmin offset between the \emph{Planck}-SZ detection and the X-ray centroid. 

\subsubsection{PSZ2 G149.63\texorpdfstring{$-$}{-}84.19 (Abell 133)}\label{sec:PSZ2G149.63-84.19}
\emph{Figure~\ref{fig:app:PSZ2G149.63-84.19}}. A radio phoenix was reported by \citet{sr84} and \citet{Slee2001} in this cluster, and is well-detected in the ASKAP data. While considered a `phoenix', in our classification scheme we do not distinguish between small-scale AGN-related diffuse emission and label the emission as `uncertain/unclassified diffuse emission'. The second component south of the cluster centre is also detected with the ASKAP data. This component has previously been seen with the GMRT \citep{rcn+10}, the MWA \citep{Duchesne2017}, and MeerKAT \citep{Knowles2022} though it is unclear if this component was the lobe of a (possible background) radio galaxy or a diffuse radio source associated with the ICM. A wideband spectral study is required to confirm the nature of southern source, and the cluster and source will be discussed further in upcoming work focused on detection of giant radio galaxies (Koribalski et al., in prep) though we leave the classification of both the northern and southern sources as `U' in this work.


\subsubsection{PSZ2 G172.98\texorpdfstring{$-$}{-}53.55 (Abell 370)}\label{sec:PSZ2G172.98-53.55}
\emph{Figure~\ref{fig:app:PSZ2G172.98-53.55}}. The Frontier Fields cluster Abell 370 was observed with the VLA and GMRT by \citet{Xie2020} and was reported to host a candidate radio halo. Subsequent observations by \citet{Knowles2022} with MeerKAT confirm the detection of the radio halo and the ASKAP data presented here also detect the radio halo at low significance. 



\subsubsection{PSZ2 G175.69\texorpdfstring{$-$}{-}85.98 (Abell 141)}\label{sec:PSZ2G175.69-85.98}
\emph{Figure~\ref{fig:app:PSZ2G175.69-85.98}}. This is a pre-merging system \citep{Caglar2018} with a radio halo reported by \citet{Duchesne2017,Duchesne2021a}. The ASKAP data in this work have a marginally lower noise than in \citet{Duchesne2021a} owing to a new observation with the cluster closer to the PAF beam centre. This combined with better beam models, and a slightly different integration region, provides perhaps a more accurate (if less precise) flux density measurement ($23\pm8$\,mJy cf.\ $13.7\pm1.9$\,mJy reported by \citealt{Duchesne2021a}). 

\subsubsection{PSZ2 G219.88\texorpdfstring{$+$}{+}22.83 (Abell 664)}\label{sec:PSZ2G219.88+22.83}
\emph{Figure~\ref{fig:app:PSZ2G219.88+22.83}}. We report a candidate relic $\approx 2.5$\,Mpc to the NW of the cluster center. The candidate relic has no obvious AGN/host, and there is no high-resolution X-ray data that covers the location of the candidate relic. Artefacts from an off-axis bright source pass through the cluster centre, limiting any detection of a radio halo. 

\subsubsection{PSZ2 G223.47\texorpdfstring{$+$}{+}26.85 (MACS J0845.4\texorpdfstring{$+$}{+}0327)}\label{sec:PSZ2G223.47+26.85}
\emph{Figure~\ref{fig:app:PSZ2G223.47+26.85}}. We report the detection of a radio halo, which is co-located with the X-ray emission detected by \emph{XMM}-Newton. 

\subsubsection{PSZ2 G225.48\texorpdfstring{$+$}{+}29.41 (Abell 732)}\label{sec:PSZ2G225.48+29.41}
\emph{Figure~\ref{fig:app:PSZ2G225.48+29.41}}. We report the detection of a radio halo in this cluster. The emission is partially confused with nearby extended sources. Even after subtraction of emission $<250$\,kpc, the full extent of the halo is difficult to determine. \emph{Chandra} data reveal a disturbed morphology for the cluster. Extended emission is also visible to the north of the cluster (beyond 1\,Mpc), though is likely an unrelated pair of radio sources, both with clear optical hosts. 

\subsubsection{PSZ2 G227.59\texorpdfstring{$+$}{+}22.98 (MaxBCG J129.82432\texorpdfstring{$-$}{-}01.69949)}\label{sec:PSZ2G227.59+22.98}
\emph{Figure~\ref{fig:app:PSZ2G227.59+22.98}}. We report the detection of two candidate relics towards the NW and SE clearly visible in the compact-source subtracted maps and at low resolution. A brighter extended source SW of the cluster centre is prominent in the source-subtracted images and is a radio galaxy but also has diffuse emission extending towards the W. The radio galaxy, which may be a wide-angle tailed radio galaxy (WAT) in projection, has an optical ID, (SDSS~J083917.83$-$014158.1), considered the BCG position in the maxBCG cluster catalogue \citep{Koester2007}. This would suggest an offset of $\approx 2.5$\,arcmin from the SZ coordinates and the BCG. The nature of the diffuse extension of this radio galaxy is unclear. No high-resolution X-ray data is available for the cluster.

\subsubsection{PSZ2 G227.89\texorpdfstring{$+$}{+}36.58}\label{sec:PSZ2G227.89+36.58}
\emph{Figure~\ref{fig:app:PSZ2G227.89+36.58}}. The cluster hosts a complex collection of extended emission, including active radio galaxies. The subtraction of $<250$\,kpc sources leaves significant emission in and around the cluster, and we consider the residual emission to be candidate radio halo. Sensitive and high-resolution follow-up observations by, e.g., MeerKAT would be required to confirm this source as a radio halo.

\subsubsection{PSZ2 G228.38\texorpdfstring{$+$}{+}38.58}\label{sec:PSZ2G228.38+38.58}
\emph{Figure~\ref{fig:app:PSZ2G228.38+38.58}}. There is unclassified $<300$\,kpc emission $\approx 1.6$\,arcmin to the NE of the reported SZ cluster coordinates. The location of the diffuse emission is the centroid of cluster WHL~J093439.0$+$054144 \citep{Wen2009} at $z\approx0.54$ and is likely the same system. Given the location of the small diffuse source at the centre of optical density for this system, this may be a mini-halo. Follow-up high-resolution X-ray observations would be required to confirm this.

\subsubsection{PSZ2 G228.50\texorpdfstring{$+$}{+}34.95}\label{sec:PSZ2G228.50+34.95}
\emph{Figure~\ref{fig:app:PSZ2G228.50+34.95}}. While the cluster has a a significant number of radio sources projected onto it, we are able to detect a residual extended component after subtraction of compact sources. This extended radio component coincides with the X-ray emission centroid, and we consider this a candidate radio halo.  

\subsubsection{PSZ2 G231.79\texorpdfstring{$+$}{+}31.48 (Abell 776)}\label{sec:PSZ2G231.79+31.48}
\emph{Figure~\ref{fig:app:PSZ2G231.79+31.48}}. We report the detection of a radio halo and radio relic. The radio halo is located co-spatial with the X-ray emission and the radio relic lies towards the edge of the X-ray emission region to the W. As the relic is embedded in the western portion of the halo (see Figure~\ref{fig:app:PSZ2G231.79+31.48}), we subtract the relic's integrated flux density from the flux density measurement of the radio halo, though the full extent of the halo towards the west is unclear. 

\subsubsection{PSZ2 G232.84\texorpdfstring{$+$}{+}38.13 (Abell 847)}\label{sec:PSZ2G232.84+38.13}
\emph{Figure~\ref{fig:app:PSZ2G232.84+38.13}}. The cluster hosts an extended radio source with uncertain classification. The source has no obvious optical host and may be a relic or fossil/phoenix. Given the distance from the cluster centre, we consider it a candidate relic. The 16 cluster members with spectroscopic redshifts from the SDSS have a velocity dispersion of $\approx 740$\,km\,s$^{-1}$. There is also diffuse emission near the cluster centre $\approx 15$\,arcsec south of the BCG (i.e. separated by one PSF width). The small angular scale (230 kpc) of the emission and its elongated morphology are not suggestive of a radio halo or mini-halo source. We consider this unclassified diffuse emission, and for both diffuse sources in the cluster X-ray observations would help in further clarification of the their nature.

\subsubsection{PSZ2 G233.68\texorpdfstring{$+$}{+}36.14}\label{sec:PSZ2G233.68+36.14}
\emph{Figure~\ref{fig:app:PSZ2G233.68+36.14}}. We report the detection of a radio halo and two radio relics to the N and SE of the radio halo emission. While there is no X-ray data available to confirm the cluster dynamics, the nature of the radio emission is clear from the morphology alone in this case. 

\subsubsection{PSZ2 G236.92\texorpdfstring{$-$}{-}26.65 (Abell~3364)}\label{sec:PSZ2G236.92-26.65}
\emph{Figure~\ref{fig:app:PSZ2G236.92-26.65}}. This relaxed cluster hosts diffuse emission at its centre that not only coincides with the X-ray peak but also features a compact radio source near its centre, consistent with radio mini-halos. However, a second diffuse source with the same morphology as the first is located directly towards the west of the cluster separated by $\approx 4$\,arcmin. A radio source associated with the optical galaxy DES J054726.18$-$315210.8 (with a photometric redshift of 0.28) is located equidistant between the two diffuse sources and is also extended E-W in the direction of the two diffuse sources. We suggest DES J054726.18$-$315210.8 hosts a background radio galaxy with the diffuse sources the lobes and the E-W extension jets.

\subsubsection{PSZ2 G239.27\texorpdfstring{$-$}{-}26.01 (MACS J0553.4\texorpdfstring{$-$}{-}3342)}\label{sec:PSZ2G239.27-26.01}
\emph{Figure~\ref{fig:app:PSZ2G239.27-26.01}}. The cluster hosts a previously detected radio halo \citep{Bonafede2012} and previous ASKAP data reported by \citet{Wilber2020} show the radio halo as well. The ASKAP data here are the same observations used by \citet{Wilber2020}, though our self-calibration process and compact source subtraction is different. We end up with a marginally better detection with compact sources removed, though our directional self-calibration process in this case has similar results to the  full direction-dependent calibration used by \citet{Wilber2020}. We report a higher flux density, though note \citet{Wilber2020} use a different integration region within $3\sigma_\text{rms}$ contours. No further diffuse emission is found. 


\subsubsection{PSZ2 G241.79\texorpdfstring{$-$}{-}24.01 (Abell 3378)}\label{sec:PSZ2G241.79-24.01}
\emph{Figure~\ref{fig:app:PSZ2G241.79-24.01}}. The cluster hosts a bright compact source at the centre (PKS~0604$-$352, associated with the BCG), though artefacts are reduced during the directional self-calibration process (see Section~\ref{sec:ddselfcal}). The cluster is relaxed and features at least one diffuse component off-centre. A second component coincident with the central bright source is seen after compact source subtraction, though will need further confirmation with higher dynamic range imaging due to the possibility of residual artefacts left over after compact source subtraction. 


\subsubsection{PSZ2 G260.80\texorpdfstring{$+$}{+}06.71}\label{sec:PSZ2G260.80+06.71}

\emph{Figure~\ref{fig:app:PSZ2G260.80+06.71}}. We consider an elongated source near the edge of the cluster (assuming $z=0.2$) a candidate radio relic. There is also residual diffuse emission at the cluster centre after compact source subtraction that resembles a radio halo, though it is unclear if this is residual emission from partial subtraction or is a double radio source at the cluster centre. We note that the cluster is at a low Galactic latitude, has not been confirmed by other surveys/at other wavelengths, and has no reported redshift, and the candidate sources may be therefore Galactic in origin. 

\subsubsection{PSZ2 G262.36\texorpdfstring{$-$}{-}25.15 (Abell 3391)}\label{sec:PSZ2G262.36-25.15}
\emph{Figure~\ref{fig:app:PSZ2G262.36-25.15}}. A small ($\approx 160$\,kpc) diffuse source is located towards the SE of the cluster centre, though it is unclear what the source is. \citet{Bruggen2020} show the same ASKAP data but with full direction-dependent calibration. They do not comment on this source as it lies within the region of large-scale artefacts from the bright radio galaxy at the centre of the cluster. However, after compact source subtraction and image-based filtering the source remains and we suggest it is a real diffuse component.

\subsubsection{PSZ2 G263.14\texorpdfstring{$-$}{-}23.41 (Abell S592)}\label{sec:PSZ2G263.14-23.41}
\emph{Figure~\ref{fig:app:PSZ2G263.14-23.41}}. This cluster hosts a radio halo, originally detected by \citet{Wilber2020} with the same ASKAP observations. As with PSZ2 G239.27$-$26.01, differences in the integration region and thresholds used yield differences in the flux density measurements. We note as well that this dataset has one of the poorer PAF beam models, with $\approx 26$\% uncertainty from the primary beam modelling alone. 

\subsubsection{PSZ2 G263.19\texorpdfstring{$-$}{-}25.19 (Abell 3395)}\label{sec:PSZ2G263.19-25.19}
\emph{Figure~\ref{fig:app:PSZ2G263.19-25.19}}. \citet{Bruggen2020} report the detection of a complex extended radio source with diffuse components (their sources `S2' and `S3'). The source comprises both active radio sources as well as diffuse components which may be revived fossil plasma. 

\subsubsection{PSZ2 G263.68\texorpdfstring{$-$}{-}22.55 (Abell 3404)}\label{sec:PSZ2G263.68-22.55}
\emph{Figure~\ref{fig:app:PSZ2G263.68-22.55}}. This radio halo was detected by \citet{Duchesne2021a} with the same ASKAP data. We report a higher flux density in this work, again due to either integration of the full polygon region or model fitting, though note that due to the density of sources in the cluster that are subtracted, the associated uncertainty in the measurement is $50$\%.  We note that while \citet{planck16} report $z=0.1644$ for this cluster, only four galaxies in the vicinity of the cluster have reported redshifts: two at $z\approx 0.164$ and two at $z\approx 0.338$ \citep{Jones2009,Guzzo2009,Bocquet2019}, suggesting a possible second cluster along the line of sight.


\subsubsection{PSZ2 G272.08\texorpdfstring{$-$}{-}40.16 (Abell 3266)}\label{sec:PSZ2 G272.08-40.16}
\emph{Figure~\ref{fig:app:PSZ2G272.08-40.16}}. A radio relic, fossil source, and other ambiguous diffuse emission were detected at multiple frequencies \citep{murphy99,Duchesne2022,Riseley2022} and a radio halo was also detected in these ASKAP observations by \citet{Riseley2022}. The compact source-subtraction and directional self-calibration procedure have revealed more of the radio halo. Some residual artefacts around a bright WAT source limit the full detection of the halo to the SW. The model flux density of the radio halo is larger than that reported by \citet{Riseley2022}, though this is a combination of general increase in flux density from integrating a model and a much larger region over which we detect the halo. From the upper limit at 216-MHz reported by \citet{Duchesne2022}, we place a limit on the spectral index of $\alpha_{216}^{944} \gtrsim -1.4$.



\subsubsection{PSZ2 G286.28\texorpdfstring{$-$}{-}38.36}\label{sec:PSZ2G286.28-38.36}
\emph{Figure~\ref{fig:app:PSZ2G286.28-38.36}}. We report the detection of a radio halo in this cluster. The halo is almost perpendicular to the elongation of the X-ray emission, though it is unclear how much of the E-W extension in the radio is associated with the halo. Previously the cluster was observed with the ATCA \footnote{Australia Telescope Compact Array.} but no diffuse emission was detected \citep{Martinez2018}. An upper limit to the radio halo luminosity of $P_\text{1.4\,GHz} \lesssim 1.55\times 10^{24}$\,W\,Hz$^{-1}$ (assuming $\alpha=-1.3$) was reported by \citet{Martinez2018}. Extrapolating from our measurement and assuming the same spectral index, we find $P_\text{1.4\,GHz} = \left(4\pm1\right)\times 10^{24}$\,W\,Hz$^{-1}$, inconsistent with the upper limit, requiring $-1.5 \lesssim \alpha_{888}^{1400} \lesssim -2.6$ to be consistent, suggesting this may be an ultra-steep spectrum halo \citep[e.g.][]{bgc+08}. 

\subsubsection{PSZ2 G286.75\texorpdfstring{$-$}{-}37.35}\label{sec:PSZ2G286.75-37.35}
\emph{Figure~\ref{fig:app:PSZ2G286.75-37.35}}. We report the detection of a radio halo in this cluster. While there is no high-resolution X-ray data available, it is is clear from the location and morphology of the radio emission that it represents a radio halo.

\subsubsection{PSZ2 G311.98\texorpdfstring{$+$}{+}30.71 (Abell 3558)}\label{sec:PSZ2G311.98+30.71}
\emph{Figure~\ref{fig:app:PSZ2G311.98+30.71}}. The cluster is part of the Shapley supercluster and \citet{Venturi2022} detected a radio halo with ASKAP. The ASKAP dataset in this work is a different (but similar) observation, and the radio halo is detected again in this work.

\subsubsection{PSZ2 G313.33\texorpdfstring{$+$}{+}30.29 (Abell 3562)}\label{sec:PSZ2G313.33+30.29}
\emph{Figure~\ref{fig:app:PSZ2G313.33+30.29}}. Also in the Shapley supercluster, The ASKAP data detect the well-known radio halo \citep{Venturi2003} along with the recently detected bridge between the cluster and the nearby group SC 1329$-$313 \citep{Venturi2022}. 


\subsubsection{PSZ2 G332.23\texorpdfstring{$-$}{-}46.37 (Abell 3827)}\label{sec:G332.23-46.37}
\emph{Figure~\ref{fig:app:PSZ2G332.23-46.37}}. We report the detection of a radio halo in this cluster. The halo aligns well with X-ray emission detected by \emph{XMM}-Newton, though has a concentration parameter, $c\approx 0.23$, (\citealt{Lovisari2017}, with a similar value reported by \citealt{Yuan2022}: $\approx 0.21$). Giant radio halos are typically found in clusters with $c\lesssim0.2$ \citep{Cassano2010,Cassano2023}. The largest linear size within $2\sigma_\text{rms}$ contours is 620\,kpc and with a point source at the centre of the emission in this comparatively relaxed cluster the halo may be considered a mini-halo. \citet{bvc+16} show observations of the cluster with KAT-7, though with the low angular resolution of the KAT-7 data they could not separate any diffuse emission from the compact sources at the cluster centre.

\subsubsection{PSZ2 G333.89\texorpdfstring{$-$}{-}43.60 (SPT-CL J2138\texorpdfstring{$-$}{-}6007)}\label{sec:PSZ2G333.89-43.60}
\emph{Figure~\ref{fig:app:PSZ2G333.89-43.60}}. We report the detection of a radio halo in this cluster. The residual diffuse emission in the cluster centre after subtraction of compact sources is co-spatial with the X-ray emission detected by \emph{XMM}-Newton.


\subsubsection{PSZ2 G335.58\texorpdfstring{$-$}{-}46.44 (Abell 3822)}\label{sec:PSZ2G335.58-46.44}
\emph{Figure~\ref{fig:app:PSZ2G335.58-46.44}}. We report the detection of a radio halo and radio relic. The dynamical state of the cluster from X-ray observations is considered `Mixed' by \citet{Lovisari2017}, with concentration parameter, of $\approx 0.11$. \citet{Yuan2022} report a slightly higher concentration parameter ($\approx 0.17$), consistent with most halo-hosting clusters.


\subsubsection{PSZ2 G341.19\texorpdfstring{$-$}{-}36.12 (Abell 3685)}\label{sec:PSZ2 G341.19-36.12}
\emph{Figure~\ref{fig:app:PSZ2G341.19-36.12}}. \citet{Duchesne2020b} report the detection of double relics with these ASKAP observations. The re-imaged data here do not reveal any new diffuse emission in this cluster, and flux density measurements are reasonably consistent with those reported by \citet{Duchesne2020b} for the two relics. 



\subsubsection{PSZ2 G342.33\texorpdfstring{$-$}{-}34.93 (SPT-CL J2023\texorpdfstring{$-$}{-}5535)}\label{sec:PSZ2G342.33-34.93}
\emph{Figure~\ref{fig:app:PSZ2G342.33-34.93}}. \citet{HyeongHan2020} reported the detection of a radio halo and relic. The same observations are used here, and with our tapered and compact-source subtracted images we detect a marginally larger extent of the radio halo. Flux density measurements are consistent with \citet{HyeongHan2020}, though we note the integrated model flux density for the radio halo is $\approx 1.6$ times the direct integration from the map. A secondary relic is also detected. This second relic is not reported by \citet{HyeongHan2020}, and though it is visible in the deeper MeerKAT image from the MGCLS it is also not reported by \citet{Knowles2022}.

\subsubsection{PSZ2 G342.62\texorpdfstring{$-$}{-}39.60 (Abell 3718)}\label{sec:PSZ2G342.62-39.60}
\emph{Figure~\ref{fig:app:PSZ2G342.62-39.60}}. \citet{Loi2023} reported the detection of an unclassified extended source at the cluster centre using the same ASKAP observations. The source has small-scale features that are subtracted during the $(u,v)$-filtering and is also blended with unassociated point sources that make the integrated flux density measurement unreliable, therefore this measurement (and radio power) is not included in Table~\ref{tab:source:flux}. The source is elongated, and its nature remains unclear.  


\subsubsection{PSZ2 G346.86\texorpdfstring{$-$}{-}45.38 (Abell 3771)}\label{sec:PSZ2G346.86-45.38}
\emph{Figure~\ref{fig:app:PSZ2G346.86-45.38}}. We report the detection of a radio halo. The cluster was part of the ATCA \footnote{Australia Telescope Compact Array.} \textsf{REXCESS} \footnote{Representative \emph{XMM}-Newton Cluster Structure Survey; \citet{rexcess}.} Diffuse Emission Survey \citep[ARDES;][]{Shakouri2016}, though no halo was detected in the ATCA data. The ATCA data detected the brighter head-tail radio galaxy in the cluster, offset from the radio halo by $\approx 5$\,arcmin ($\approx 450$\,kpc).

\subsubsection{PSZ2 G347.58\texorpdfstring{$-$}{-}35.35 (Abell S871)}\label{sec:PSZ2G347.58-35.35}
\emph{Figure~\ref{fig:app:PSZ2G347.58-35.35}}. We report a candidate halo. The candidate halo is located between two extended radio sources (and related to AGN) which are not fully subtracted during the $(u,v)$-based compact source subtraction, but this is mitigated during the model fitting where the residuals from those sources are masked. No X-ray data are available from \emph{Chandra} or \emph{XMM}-Newton. The candidate halo is offset from the PSZ2-reported position by $\approx 2$\,arcmin ($\approx 430$\,kpc), but is centered on the position reported in the Abell catalogue \citep{aco89}. \citet{aco89} also note that the distribution of optical galaxies is bimodal (following the aforementioned radio galaxies) and the the diffuse radio source sits somewhat between these optical concentrations as in the case of PSZ2~G175.69$0-$85.98 (Abell~141, Section~\ref{sec:PSZ2G175.69-85.98}; \citealt{Duchesne2021a}). 



\section{Discussion}

\subsection{The number of diffuse cluster sources}

\begin{figure}[t!]
    \includegraphics[width=1\linewidth]{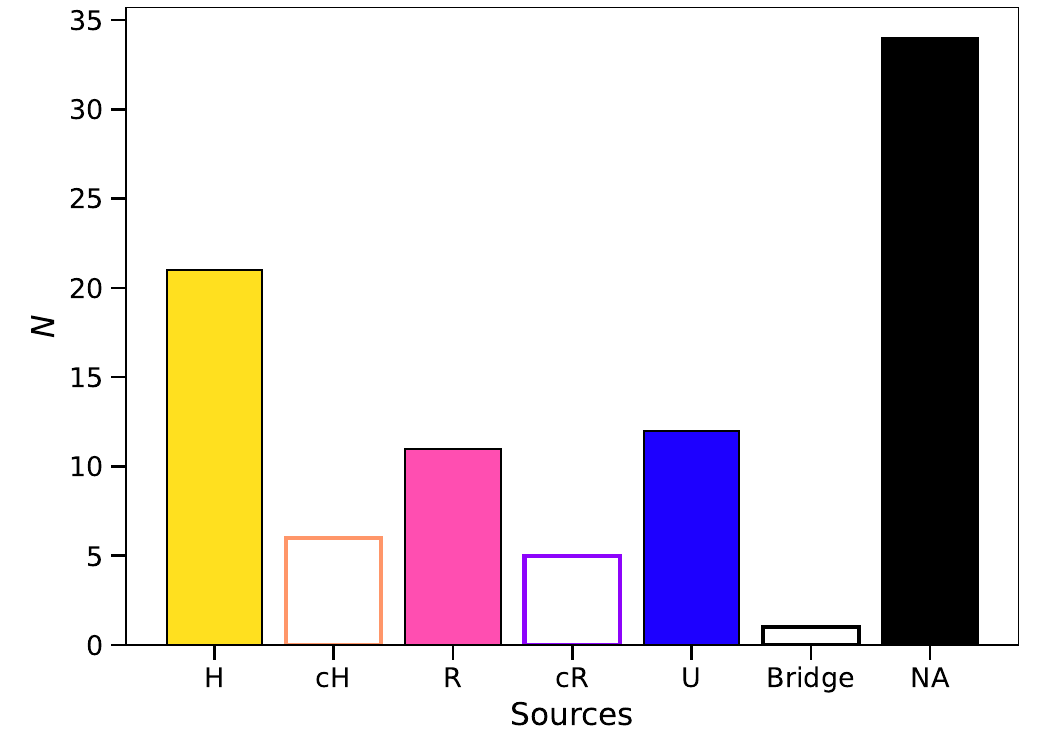}
    \caption{\label{fig:bar} Counts of the halos (H), candidate halos (cH), relics (R), candidate relics (cR), unclassified sources (U), bridge, and clusters without diffuse emission (NA) in the ASKAP data for all 71 clusters in the sample.}
\end{figure}

\begin{figure}[t!]
\centering
\includegraphics[width=1\linewidth]{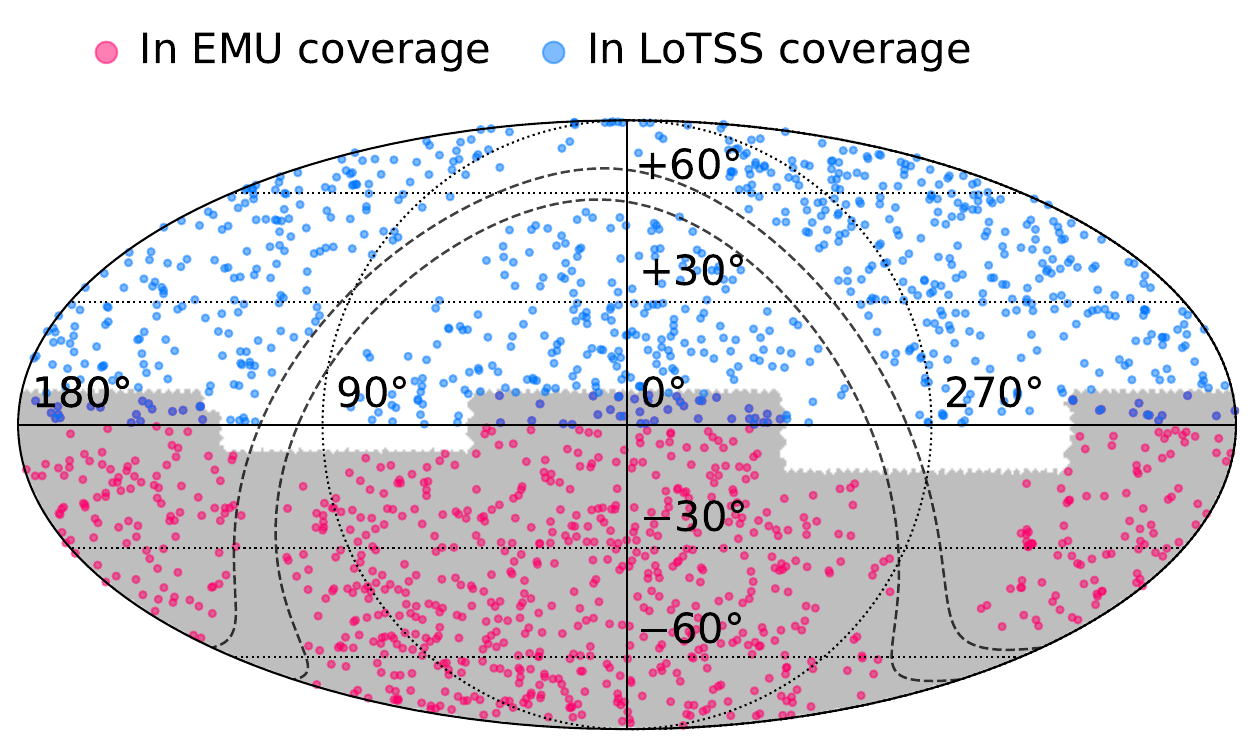}
\caption{\label{fig:emu_full} The distribution of PSZ2 clusters across the sky, coloured by their eventual presence in the EMU survey (pink) and LoTSS (blue). Clusters appearing in overlap regions are coloured purple. The expected full EMU survey coverage is coloured grey.}
\end{figure}

Some type of diffuse radio emission, i.e., not associated with active radio galaxies, is found in 37 (52\%) of the 71 clusters in our sample (including candidate sources). Figure~\ref{fig:bar} shows the numbers of each source type---halo (H), candidate halo (cH), relic (R), candidate relic (cR), unclassified diffuse emission (U), bridge, and clusters with no diffuse sources (NA). We include the class `bridge' for a single case of an unambiguous and previously-detected bridge of emission between a cluster and group (PSZ2~G313.33$+$30.29 and SC~1329$-$313; \citealt{Venturi2022}). For halos and relics, the total numbers are \haloPilot\ halos (\haloP\ of clusters, with \chaloPilot\ candidates) and \relicPilot\ relics (\relicP\ of clusters, with an additional \crelicPilot\ candidates). These sources are found in \uPilot\ clusters; \percHR\ of the PSZ2 subsample in this work are found to host halos and/or relics. 



In the current observation plan for the ASKAP main surveys, EMU will cover the full sky south of declination $-7^\circ$ with additional coverage between  $-7^\circ$ and $+7^\circ$ in declination for certain RA ranges. In Figure~\ref{fig:emu_full} we show the planned full EMU sky coverage and highlight the 858 PSZ2 clusters that lie within 0.75 degrees of a PAF beam in this region. Extrapolating the detection fractions for our PSZ2 subsample, we can expect up to \haloN\ and \relicN\ PSZ2 clusters hosting halos and relics, respectively, in the EMU survey, assuming the number of candidates provides an estimate to the upper limit in the number. 

{\citet{cbn+12} use the original EMU survey description (at 1.4\,GHz) to predict up to $\approx 250$ radio halos detected in clusters with redshifts between $0 < z \leq 0.6$ within the planned EMU coverage shown in Figure~\ref{fig:coverage}. This includes the assumption that a significant fraction of giant radio halos are formed through non-turbulent processes (i.e.\,emission via secondary electrons generated by collisions in the ICM). We note that while this number agrees with our prediction for the full EMU survey, we do not distinguish between giant radio halos and mini-halos, the latter of which is predominantly found in non-merging clusters. Conversely,} the total number of halos expected in PSZ2 clusters is lower than the $\approx 1\,000$ radio halos expected below $z<0.5$ reported by \citet{Nishiwaki2022} from a comparison of theoretical models, though we note one halo in this work is detected in PSZ2~G110.28$-$87.48 ($z=0.52$). Moreover, \citet{Nishiwaki2022} consider EMU to cover the full sky up to declination $+30^\circ$, and the PSZ2 catalogue itself does not contain all clusters. With these caveats in mind, we can consider the expected number of halos extrapolated from this work in reasonable agreement with theoretical predictions.

The full LoTSS-DR2 and EMU surveys are expected to have similar total sky coverage---we show the distribution of PSZ2 cluster across the full EMU and LoTSS regions in Figure~\ref{fig:emu_full}. Assuming the median LoTSS-DR2 noise of 83\,\textmu Jy\,PSF$^{-1}$ at 6\,arcsec \citep{lotss:dr2} and a median 30\,\textmu Jy\,PSF$^{-1}$ at 15\,arcsec for the EMU survey, the higher-frequency observations at 943\,MHz are less sensitive to the steep-spectrum diffuse cluster emission and we would not expect to detect as many sources as LoTSS\footnote{Unless sources have $\alpha \gtrsim -0.6$, which is not the case for diffuse cluster sources.}. Despite this, the percentage of clusters found to host a radio halo and/or relic is the same as the result from the LoTSS-DR2 PSZ2 survey. \citetalias{Botteon2022} find $(30\pm11)$\% and $(10\pm6)$\% of PSZ2 clusters to host halos and relics, respectively, and suggest the full LoTSS will uncover $251\pm92$ and $83\pm50$ PSZ2 clusters hosting halos and relics, respectively. We caution that the fraction of clusters found to host diffuse emission in this work is likely higher at low redshift than what can be expected for the full survey. Three of the archival ASKAP observations targeted nearby clusters: PSZ2 G272.08$-$40.16 (Abell~3266 in SB10636; see \citealt{Riseley2022}), the cluster pair PSZ2~G262.36$-$25.15 and PSZ2~G263.19$-$25.19 (Abell 3391 and Abell~3395 in SB8275; see \citealt{Bruggen2020}), and PSZ2~G311.98$+$30.71 and PSZ2~G313.33$+$30.29 (Abell~3558 and Abell~3562 in SB34120; \citealt{Venturi2022}). These low-redshift clusters were targeted specifically for known or expected diffuse cluster emission.


\subsection{{The discovery space of the EMU survey}}\label{sec:scaling}

\begin{figure*}[t!]
    \centering
    \begin{subfigure}[b]{0.5\linewidth}
    \includegraphics[width=1\linewidth]{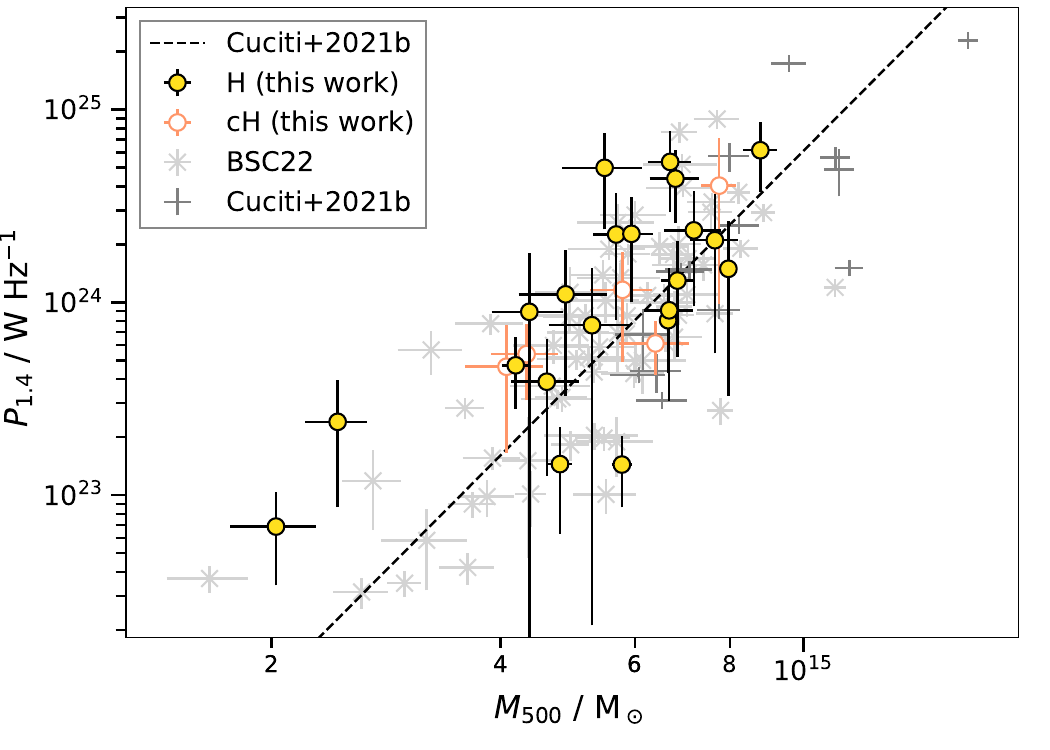}
    \caption{\label{fig:pm:halos}Radio halos.}
    \end{subfigure}%
    \begin{subfigure}[b]{0.5\linewidth}
    \includegraphics[width=1\linewidth]{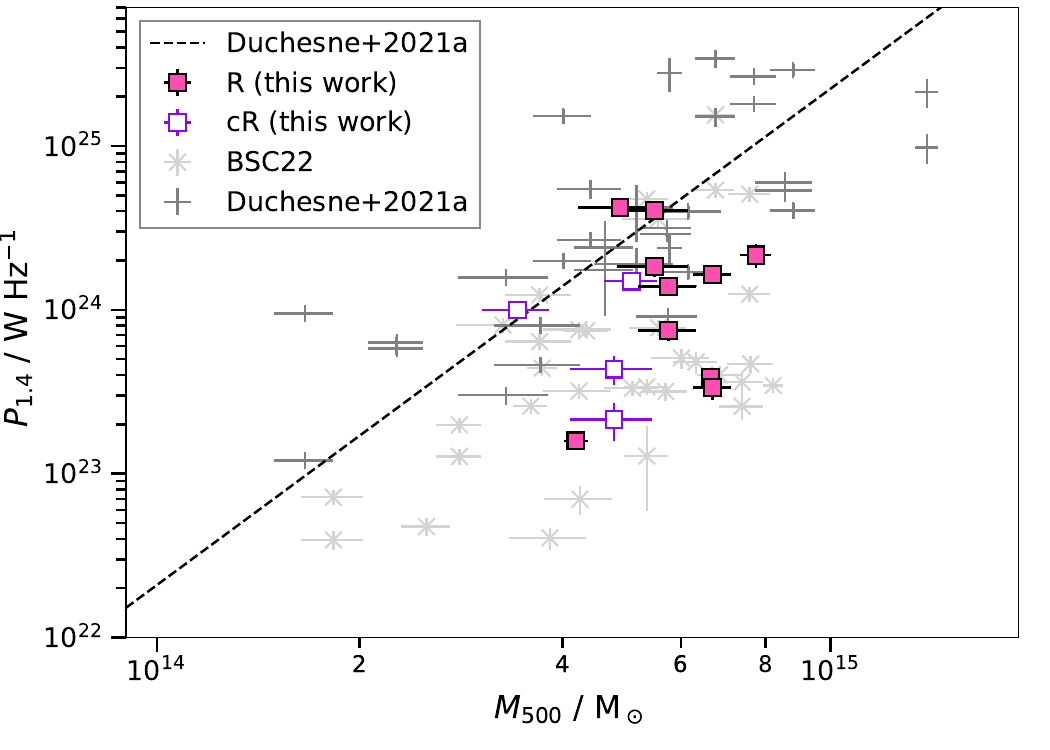}
    \caption{\label{fig:pm:relics}Radio relics.}
    \end{subfigure}\\%
    \caption{\label{fig:pm} $P_\text{1.4\,GHz}$--$M_{500}$ scaling relation for radio halos (\subref{fig:pm:halos}) and radio relics (\subref{fig:pm:relics}). We show radio halos and relics (and candidates) detected in this work, along with the halo and relic samples discovered in the LoTSS-DR2 data \citep[at 144 MHz;][]{Botteon2022} and the samples curated by \citet[][and see references therein]{Cuciti2021a,Cuciti2021b} for radio halos and \citet[][and see references therein]{Duchesne2020b} for radio relics, largely detected at frequencies above $\approx 1$\,GHz. {Flux densities and luminosities have been scaled to 1.4\,GHz assuming $\alpha = -1.3$ for radio halos and $\alpha = -1.2$ for relics as described in the text. Best-fit $P_{1.4}$--$M_{500}$ correlations from \citet{Cuciti2021b} and \citet{Duchesne2020b} are shown for halos and relics, respectively.}} 
\end{figure*}


\begin{figure}[t!]
    \includegraphics[width=1\linewidth]{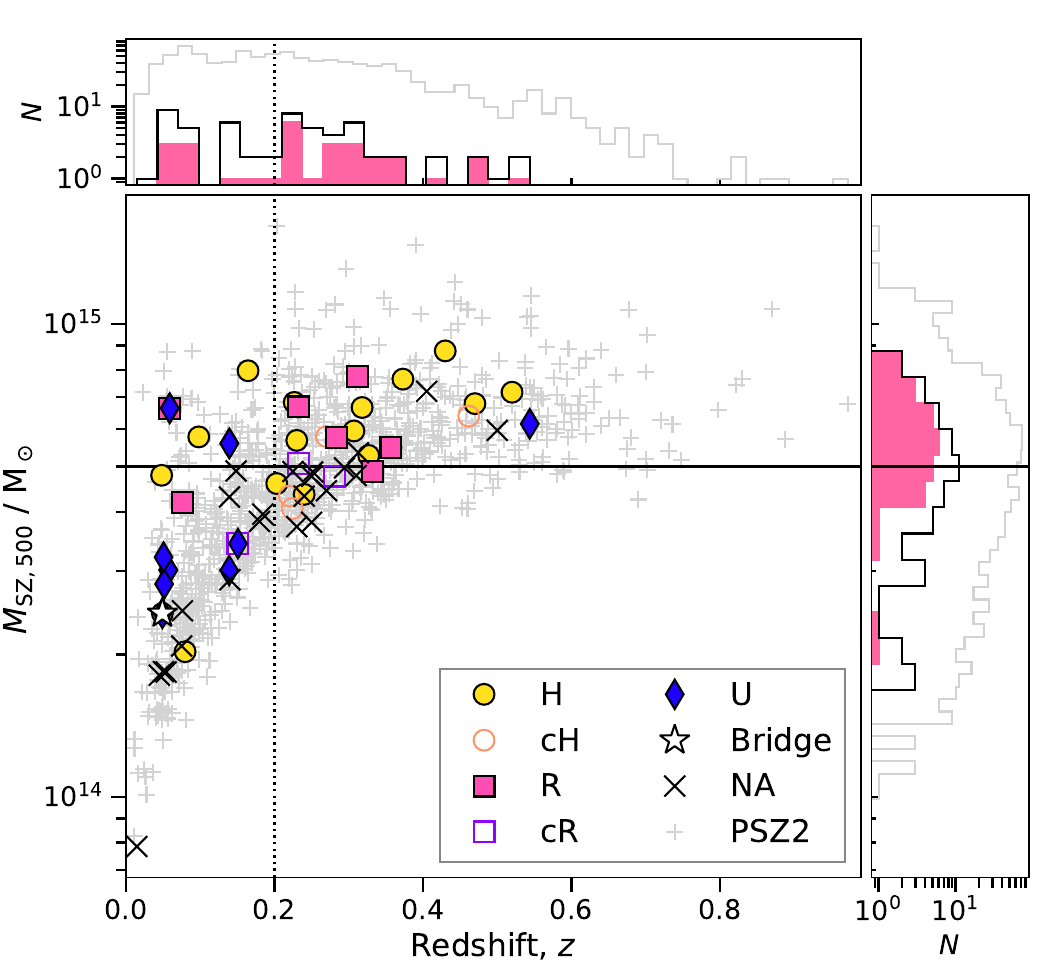}
    \caption{\label{fig:mass_sources} The mass-redshift distribution of the PSZ2 catalogue (with redshifts) as in Figure~\ref{fig:psz2} but with clusters and their sources from this survey marked as appropriate. {Clusters in our sample without detected diffuse emission are labelled `NA'.} The dashed and dotted black lines indicate $z=0.09$ and $z=0.2$, respectively. The histograms show the distributions of the combined halo and relic counts (pink, including candidates) across the redshift and mass range along with the full PSZ2 sample (grey) and cluster searched in this work (black, with redshifts).}
\end{figure}

The radio power of both halos and relics have been observed to scale with host cluster mass \citep[e.g.][]{ceb+13,dvb+14,Cuciti2021b,Duchesne2020b,Duchesne2021a} and other related cluster and source morphological properties (e.g.\ X-ray luminosity and temperature \citealt{lhba00}; source size \citealt{Bonafede2009}). Traditionally, scaling relations have been explored at 1.4\,GHz, though with the advent of LOFAR and the MWA they are now being explored at 150\,MHz as well \citep{vanWeeren2020,Duchesne2021a}. In addition to the low-frequency exploration, recent results from the LoTSS-DR2 data release and the MWA have revealed radio halos and relics in clusters of lower mass than previously seen \citep{Dwarakanath2018,Botteon2019,vanWeeren2020,Botteon2021b,Botteon2022,Duchesne2017,Duchesne2021b}. Detection of diffuse emission in clusters with masses less than $\approx 5\times10^{14}$\,M$_\odot$ has been required to investigate the scaling relations in the low mass regime (and presumably with low turbulent energy for the generation of the radio halos). Recent statistical works with the LoTSS-DR2 cluster sample \citep{Cassano2023,Cuciti2023,Jones2023} have begun to probe this low-mass regime, and find that the scaling relations continue into this previously unexplored space. The lower luminosity halos ($<10^{24}$\,W\,Hz$^{-1}$) are comparable to the stacked results reported in \cite{brown11}, although they could not filter out any diffuse emission associated with radio galaxies that was below the individual cluster detection limits.

The cluster sample in this work is not statistically complete and we do not provide upper limits to non-detections. We do not update the scaling relations but we do show our detections of halos and relics with the existing power-mass ($P_{1.4}$--$M_{500}$) relations in Figure~\ref{fig:pm} for halos [\ref{fig:pm:halos}] and relics [\ref{fig:pm:relics}]. For comparison, the power calculations for sources detected in this work are scaled to 1.4\,GHz assuming $\alpha = -1.3 \pm 0.2$ for radio halos and $\alpha = -1.2 \pm 0.2$ for relics \citep{Duchesne2021b}. Similarly, the LoTSS-DR2 sources are shown and scaled to 1.4\,GHz assuming the same mean spectral indices. We also show the best-fit $P_{1.4}$--$M_{500}$ relations from \citet{Cuciti2021b} and \citet{Duchesne2020b} for radio halos and relics, respectively. Due to the sensitivity of the ASKAP sample in this work, we are pushing into the low-mass regime for both radio halos and radio relics, and in the case of halos, we find six halos (and two candidate halos) in clusters with masses of $<5\times10^{14}$\,M$_\odot$. Four of these low-mass clusters are nearby ($z<0.09$) with the lowest mass cluster Abell~3771 (see Section~\ref{sec:PSZ2G346.86-45.38}) with a redshift of $0.0796$. The candidate radio halos in low-mass clusters, conversely, are all hosted by higher-redshift clusters. For clusters with relics, only two have masses $<5\times10^{14}$\,M$_\odot$ (with an additional two candidate relics in low-mass clusters). One system has a low redshift, and no double relic systems are detected in the low-mass clusters in these ASKAP data.

Figure~\ref{fig:mass_sources} shows the distribution of PSZ2 clusters as a function of redshift as in Figure~\ref{fig:psz2} with clusters and sources in this survey overlaid. We also show histograms of the redshift and mass distributions, highlighting the full PSZ2 catalogue (grey), the 71 clusters used in this work (black), and PSZ2 clusters with a radio halo and/or relic (pink). We are detecting diffuse emission over the majority of the mass and redshift range, though with a steady decrease in detections as mass is decreased. Considering the empirical scaling relations, we expect to see only low-power halos and relics in low-mass clusters. Low-power radio sources are naturally harder to detect. For halos in particular, this remains true even at low redshifts where their large size and low surface brightness limits detectability despite a large total flux density. Despite these limitations, the results from this survey with archival ASKAP data products yields promising prospects for the full EMU survey.


\subsection{Detection and measurement of low-brightness, extended emission}\label{sec:comparison_filtering}

\begin{figure}[t!]
    \centering
    \includegraphics[width=1\linewidth]{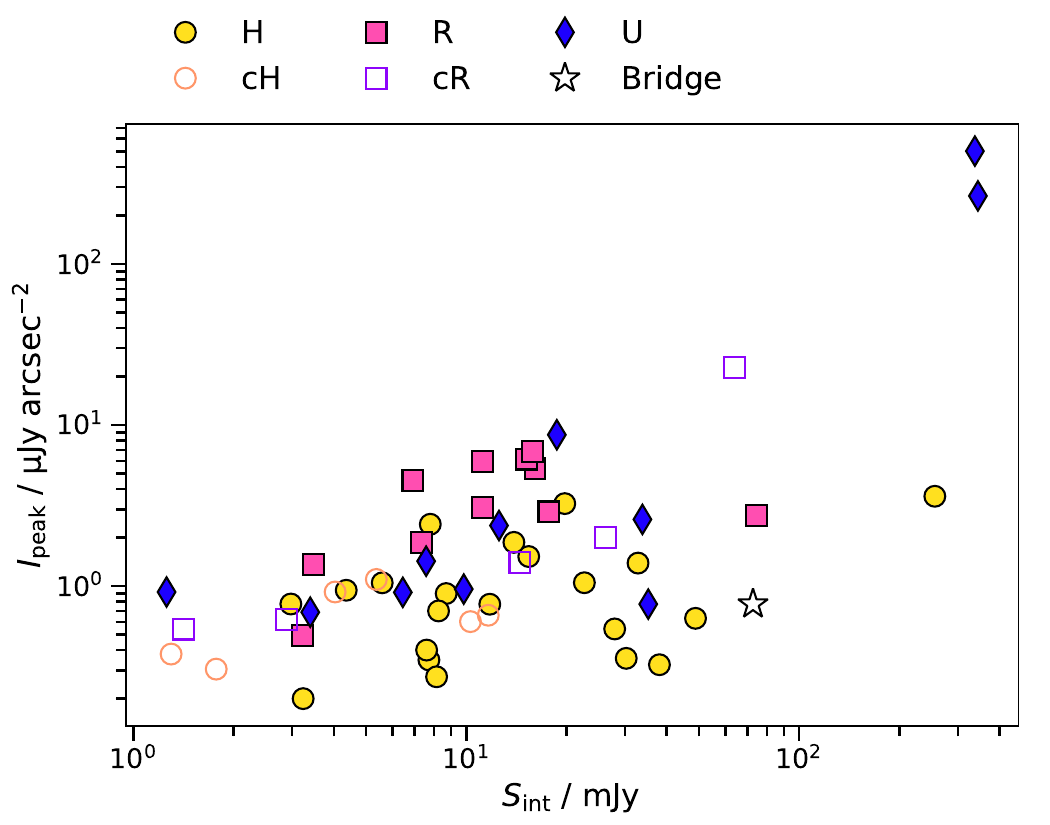}
    \caption{{\label{fig:sb_limit} The peak surface brightness as a function of integrated flux density for the diffuse sources.}}
\end{figure}

\begin{figure}[t!]
    \centering
    \includegraphics[width=1\linewidth]{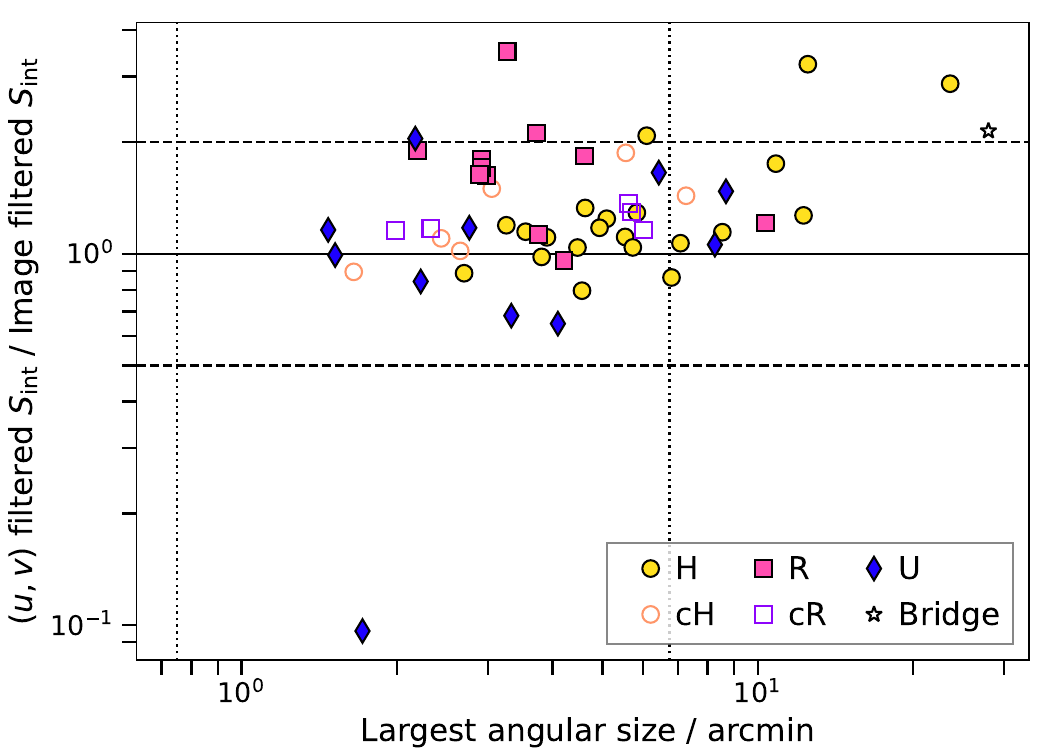}
    \caption{\label{fig:filtering} The ratio of integrated flux density measurements from the $(u,v)$-filtered and image-filtered maps as a function of largest angular scale of the source. The solid black line is drawn at 1, with the dashed lines indicating ratios of 0.5 and 2. The vertical dotted lines indicate 45\,arcsec and 405\,arcsec---approximately the image filtering scale, $3\,\theta_\text{M}$ and $27\,\theta_\text{M}$, respectively.}
\end{figure}

{Many of the lowest brightness extended features are only detectable at lower resolutions, after the confusing emission from more compact features are removed. Figure~\ref{fig:sb_limit} shows the peak surface brightness as a function of integrated flux density for the diffuse radio sources reported in this work. We find radio halos down to a peak surface brightness of $\approx 0.2$\,\textmu Jy\,arcsec$^{-2}$ and relics down to $\approx 0.5$\,\textmu Jy\,arcsec$^{-2}$. Radio halos are found with both lower peak surface brightness and lower average surface brightness (not shown here) than other diffuse sources.

}

{To help both detect and measure this low surface brightness emission, it is common to model and subtract embedded, unassociated (typically compact) sources in the $(u,v)$ plane as done in this work (see Section~\ref{sec:ddselfcal}). In the current era of large sky surveys and ever-increasing data volumes, this additional computationally expensive imaging step becomes prohibitive. This is exacerbated when filtering multiple angular scales.
An alternative is removing compact sources or small-scale features in the image plane. Imaged-based angular scale filtering methods are being used on radio data to look at both large-scale features \citep[e.g.][]{Rudnick2002a,Knowles2022,Riseley2022,deJong2022,Velovic2023}, and small-scale features \citep[e.g.\ edges;][]{Murgia2001,Ramatsoku2020,Botteon2023}. EMU is making use of image-based filtering and for this work, as described in Section~\ref{sec:filtering}, we have produced equivalent filtered images for comparison with traditional $(u,v)$-plane filtering.}


Upon inspection of the output maps, we find that the image-based filtering of both small and very large angular scales is useful in identifying diffuse sources while discarding artefacts from poor short baseline calibration and interference from the off-axis sources of large angular scales (including the Sun). For the full EMU survey, the image-based filtering is being done on all images, retaining scales between $3\,\theta_\text{M} \lesssim \theta_\text{scale} \lesssim 27\,\theta_\text{M}$. However, it is important to understand the limitations of this image-based approach, so we take advantage of the current analysis to make a comparison with image-based filtering.

First, the EMU default range of filter angular scales is not suitable for all clusters: for clusters with $z \lesssim 0.1$ the large-scale filter begins to remove sources of interest. An example of this is PSZ2~G272.08$-$40.16 (Abell~3266), which was found to host a $\approx 10$-arcmin ($\approx 700$\,kpc) radio halo via similarly filtered images \citep{Riseley2022}. The radio halo is clearly detected in the pre-filtered data and the $(u,v)$-filtered data (after removing angular scales $>88$~arcsec). In our re-imaged data, the halo is detected within $2\,\sigma_\text{rms}$ contours out to $25$\,arcmin ($\approx 1.7$\,Mpc) which is $>27\,\theta_\text{M}$ ($\approx 7.5$\,arcmin) and is completely removed during image-based filtering with the default parameters. This is highlighted in the right panel of Figure~\ref{fig:filter:example1}. 

In some situations, neither $(u,v)$- or image-based filtering can be used effectively, such as in the presence of bright point sources. Abell~3378, for example (Figure~\ref{fig:filter:example2}) hosts a bright point source at its centre. After application of image-based filtering, the artefacts accompanying bright sources result in regions of negative emission around the source.  After $(u,v)$ filtering and removal of the point source, some residual emission remains at the cluster centre but it is unclear whether this is also related to residual artefacts from the bright source. In this case, neither the image-based or $(u,v)$-based subtraction provide a clear image of the centre of the cluster.   

We also investigate the utility of the image-filtered maps for measuring flux density. \citet{Rudnick2002a} showed that the flux remaining in the filtered maps is a function of the relative size of the feature and the filter scale. For the EMU survey, the filter is chosen to ensure that sources at $\approx \theta_\text{M}$ (i.e. compact sources) are removed completely from the diffuse emission maps. Figure~\ref{fig:filtering} shows the ratio of measured flux densities in the $(u,v)$-filtered and image-filtered maps as a function of source size. We find that the image-filtered measurements are on average $\approx 70$\% of the $(u,v)$-filtered flux density, though this is more extreme for relics (median $\approx 60$\%) than for halos (median $\approx 90$\%). This results from the removal of some small-scale fluctuations in the relic by the image-based filter. With the EMU survey filter, it is thus important to recognize that relics can be significantly underestimated, and sources with angular sizes $>27\,\theta_\text{M}$ will be difficult to detect and measure reliably.

\section{Future work}

The full EMU survey commenced in November 2022 and is expected to take about five years, covering most of the southern sky (see Figure~\ref{fig:emu_full}). So far $\approx 100$ fields (out of $\approx 850$) have been observed, validated, and made public in CASDA, and we expect observing efficiency and data flow to increase over the next year. Equatorial fields are observed in two 5-h sessions; after combination and subsequent mosaicking of all tiles this will result in a homogeneous radio continuum sky survey where no or very little re-processing should be required before applying the analysis methods developed here to search for diffuse, non-thermal radio emission in PSZ2 clusters. Of particular note is the utility of the image-based filtering for searching for this emission in lieu of $(u,v)$-plane subtraction methods.

There are a number of avenues for future work involving diffuse emission in clusters with EMU. This includes the natural extension of the present work to the whole EMU survey. With the predicted numbers of sources, we should approach a statistically significant sample for which secondary correlations, e.g., with cluster morphology, can be examined. While the statistical results obtained at 150\,MHz and at 1.4\,GHz so far have not been substantially different \citep[e.g.][]{vanWeeren2020}, having two independent samples of cluster sources at 144\,MHz (from the full LoTSS) and 943\,MHz (from the full EMU survey) will allow a deep exploration of the scaling relations for halos and relics as a function of frequency. We will also be able to expand the cluster sample, including non-PZS2 cluster catalogues like the Abell catalogues \citep[][with 3\,511 clusters in the planned EMU coverage]{aco89}, X-ray cluster samples like the Meta-Catalogue of X-ray detected Clusters of galaxies \citep[MCXC;][796 in EMU]{pap+11} or recent (and upcoming) cluster catalogues from the eROSITA\footnote{extended ROentgen Survey with an Imaging Telescope Array; \citet{erosita2}.}, including the Final Equatorial Depths Survey \citep[eFEDS;][with all 542 clusters in EMU]{Liu2022} and the All-Sky Survey (eRASS:1; Bulbul et al., in prep) and other SZ-selected cluster catalogues from the South Pole Telescope \citep[e.g.][with 677 in EMU]{Bleem2020} and the Atacama Cosmology Telescope \citep[e.g.][with 2\,826 in EMU]{Hilton2021}. While many of these catalogues have significant overlap, the number of non-PSZ2 clusters will also provide an additional sample to search, with similar LoTSS searches showing promising results \citep{Hoang2022}.


EMU is well-suited for exploration of radio galaxies in clusters. Recently, \citet{Boeckmann2023} combined the EMU Pilot Survey \citep{Norris2021} and eRASS:1 cluster catalogue, analysing the interaction between the X-ray--emitting ICM and the central radio galaxy/AGN. This work, too, is expected to be expanded to EMU images that overlap with the eRASS:1 coverage. Within the context of searching for low-surface brightness radio sources, work is being done to employ methods that do not rely on individual visual inspection alone. \citet{Gupta2022} used unsupervised machine learning to highlight sources with perculiar morphologies, finding a handful of diffuse cluster sources in archival ASKAP data. Other methods include the combination of complexity metrics followed by crowd-sourced inspection of classified sources \citep{Segal2023}. Such techniques may be employed to search for diffuse cluster emission, which will be particularly useful when considering expanded cluster samples. EMU will provide an important first step towards samples of diffuse radio sources in Southern Sky, providing the groundwork for future high-sensitivity, and high-resolution surveys with the SKA.


\section{Summary}

In this work we have performed a survey of diffuse radio emission in galaxy clusters using archival ASKAP data. The purpose of this work is to assess the prospects of the full Evolutionary Map of the Universe (EMU) survey with respect to the numbers of radio halos and relics that we should expect to find once the survey is completed, and to inform on the type of statistical work that will be possible.

We follow \citetalias{Botteon2022} and investigate clusters from the PSZ2 catalogue, identifying 71 PSZ2 clusters within deep archival datasets around $\approx 1$\,GHz. We re-calibrate and re-image the archival data to both improve consistency in the imaging quality and to generate images suitable for both finding diffuse cluster emission and characterising it. This includes producing low-resolution, tapered images and images with compact sources filtered out via $(u,v)$-plane subtraction and image-based angular scale filtering. We make a brief comparison of these filtering methods, highlighting the utility of the (less resource-intensive) image-based filtering which is being used to generate additional data products for the full EMU survey. The filtering methods all perform well at detecting diffuse sources and removing the compact source contributions, though there is some discrepancy between measured flux densities that arises due to differences in angular scales being filtered.   

In this survey we see a number of previously detected diffuse cluster sources, including those detected in ASKAP data for the first time. In total, we report the detection of \haloPilot\ radio halos (with an additional \chaloPilot\ candidates), of which {12} (and all candidates) are reported here for the first time. {We note that five of the remaining halos have been previously reported using the same or similar ASKAP observations.} We also detect \relicPilot\ relics across \relicPilotClusters\ clusters (with \crelicPilot\ additional candidates), {six (and all candidates) reported here for the first time, and four previously detected in the same ASKAP data}. In addition to the radio halo and relic detections, we identify \uPilot\ unclassified diffuse sources, including the previously detected phoenix in Abell~133. We also confirm the radio bridge connecting to Abell~3562 and a nearby group. Based on this survey of archival data, we estimate we may detect halos in \haloN\ PSZ2 clusters and relics in \relicN\ PSZ2 clusters in the full EMU survey. We find radio halos and relics down to peak surface brightnesses of $\approx 0.2$\,\textmu Jy\,arcsec$^{-2}$ and $\approx 0.5$\,\textmu Jy\,arcsec$^{-2}$, respectively. We also find that we are detecting diffuse sources in low mass ($<5\times10^{14}$\,M$_\odot$) clusters, highlighting overall exciting prospects for the full EMU survey in complementing Northern Hemisphere clusters surveys with LOFAR.

\begin{availability}
{Observatory-processed ASKAP data products are available through CASDA: \url{https://data.csiro.au/domain/casdaObservation}, listed under the SBIDs in Table~\ref{tab:observations}. X-ray observations can be accessed through \url{https://cda.harvard.edu/chaser} and \url{https://nxsa.esac.esa.int/nxsa-web} for \emph{Chandra} and \emph{XMM}-Newton data products, respectively. Optical data products can be accessed through \url{https://des.ncsa.illinois.edu/desaccess}, \url{https://ps1images.stsci.edu/cgi-bin/ps1cutouts}, and \url{http://archive.stsci.edu/cgi-bin/dss_form} for DES, PS1, and DSS2 respectively. The specific images produced for this work are available through the PASA Datastore (\url{https://data-portal.hpc.swin.edu.au/institute/pasa}).}
\end{availability}

\begin{acknowledgement}
We thank the anonymous referee for their useful comments.
AB acknowledges financial support from the European Union - Next Generation EU.
This scientific work uses data obtained from Inyarrimanha Ilgari Bundara / the Murchison Radio-astronomy Observatory. We acknowledge the Wajarri Yamaji People as the Traditional Owners and native title holders of the Observatory site. CSIRO’s ASKAP radio telescope is part of the Australia Telescope National Facility (\url{https://ror.org/05qajvd42}). Operation of ASKAP is funded by the Australian Government with support from the National Collaborative Research Infrastructure Strategy. ASKAP uses the resources of the Pawsey Supercomputing Research Centre. Establishment of ASKAP, Inyarrimanha Ilgari Bundara, the CSIRO Murchison Radio-astronomy Observatory and the Pawsey Supercomputing Research Centre are initiatives of the Australian Government, with support from the Government of Western Australia and the Science and Industry Endowment Fund.
%

Numerous \texttt{python} and other software packages have been used during to production of this manuscript, including \texttt{aplpy} \citep{Robitaille2012}, \texttt{astropy} \citep{astropy:2018}, \texttt{matplotlib} \citep{Hunter2007}, \texttt{numpy} \citep{numpy}, and \texttt{scipy} \citep{scipy}. We make use of \texttt{ds9} \citep{ds9} and \texttt{topcat} \citep{topcat} for visualisation, as well as the ``Aladin sky atlas'' developed at CDS, Strasbourg Observatory, France \citep{aladin1,aladin2} for obtaining catalogue data. We make use of \texttt{CASA} \citep{casa} including its modular \texttt{python} implementation \citep{Raba2019} and \texttt{casacore} (\url{https://github.com/casacore/casacore}) including \texttt{python-casacore} (\url{https://github.com/casacore/python-casacore}). We make use of the \texttt{cubehelix} colour scheme \citep{cubehelix} to represent radio maps. 

This project used public archival data from the Dark Energy Survey (DES). Funding for the DES Projects has been provided by the U.S. Department of Energy, the U.S. National Science Foundation, the Ministry of Science and Education of Spain, the Science and Technology Facilities Council of the United Kingdom, the Higher Education Funding Council for England, the National Center for Supercomputing Applications at the University of Illinois at Urbana-Champaign, the Kavli Institute of Cosmological Physics at the University of Chicago, the Center for Cosmology and Astro-Particle Physics at the Ohio State University, the Mitchell Institute for Fundamental Physics and Astronomy at Texas A\&M University, Financiadora de Estudos e Projetos, Funda{\c c}{\~a}o Carlos Chagas Filho de Amparo {\`a} Pesquisa do Estado do Rio de Janeiro, Conselho Nacional de Desenvolvimento Cient{\'i}fico e Tecnol{\'o}gico and the Minist{\'e}rio da Ci{\^e}ncia, Tecnologia e Inova{\c c}{\~a}o, the Deutsche Forschungsgemeinschaft, and the Collaborating Institutions in the Dark Energy Survey.
The Collaborating Institutions are Argonne National Laboratory, the University of California at Santa Cruz, the University of Cambridge, Centro de Investigaciones Energ{\'e}ticas, Medioambientales y Tecnol{\'o}gicas-Madrid, the University of Chicago, University College London, the DES-Brazil Consortium, the University of Edinburgh, the Eidgen{\"o}ssische Technische Hochschule (ETH) Z{\"u}rich,  Fermi National Accelerator Laboratory, the University of Illinois at Urbana-Champaign, the Institut de Ci{\`e}ncies de l'Espai (IEEC/CSIC), the Institut de F{\'i}sica d'Altes Energies, Lawrence Berkeley National Laboratory, the Ludwig-Maximilians Universit{\"a}t M{\"u}nchen and the associated Excellence Cluster Universe, the University of Michigan, the National Optical Astronomy Observatory, the University of Nottingham, The Ohio State University, the OzDES Membership Consortium, the University of Pennsylvania, the University of Portsmouth, SLAC National Accelerator Laboratory, Stanford University, the University of Sussex, and Texas A\&M University.
Based in part on observations at Cerro Tololo Inter-American Observatory, National Optical Astronomy Observatory, which is operated by the Association of Universities for Research in Astronomy (AURA) under a cooperative agreement with the National Science Foundation.

The Digitized Sky Surveys were produced at the Space Telescope Science Institute under U.S. Government grant NAG W-2166. The images of these surveys are based on photographic data obtained using the Oschin Schmidt Telescope on Palomar Mountain and the UK Schmidt Telescope. The plates were processed into the present compressed digital form with the permission of these institutions.

This research has made use of data obtained from the Chandra Data Archive and also makes use of observations obtained with \emph{XMM}-Newton, an ESA science mission with instruments and contributions directly funded by ESA Member States and NASA.
\end{acknowledgement}


\printendnotes

\bibliography{references}

\appendix

\renewcommand{\thesubfigure}{(\roman{subfigure})}
\renewcommand\thefigure{\thesection\arabic{figure}}
\renewcommand\thetable{\thesection\arabic{table}}

\section{Cluster images}\label{sec:app:images}
\setcounter{figure}{0}
\setcounter{table}{0}


\begin{table*}[t!]
\centering
\begin{threeparttable}
\caption{\label{tab:rms_properties} Image noise properties within 2\,Mpc of the cluster in \textmu Jy\,PSF$^{-1}$.}
\begin{tabular}{l c c c c c c}\toprule
    Cluster &Archive& robust $0.0$ & robust $+0.25$ & uniform & 100 kpc taper & 250 kpc taper  \\\midrule
PSZ2 G006.16$-$69.49 & 48.0 & 39.4 & 38.9 & 68.5 & 126 & 561 \\
PSZ2 G008.31$-$64.74 & 66.6 & 75.5 & 75.1 & 140.7 & 202 & 943 \\
PSZ2 G011.06$-$63.84 & 43.8 & 38.9 & 38.8 & 66.7 & 108 & 359 \\
PSZ2 G011.92$-$63.53 & 43.9 & 36.9 & 38.3 & 53.9 & 121 & 513 \\
PSZ2 G014.72$-$62.49 & 40.5 & 36.6 & 37.8 & 52.0 & 68 & 255 \\
PSZ2 G017.25$-$70.71 & 57.9 & 49.0 & 46.3 & 91.1 & 84 & 288 \\
PSZ2 G018.18$-$60.00 & 52.3 & 41.8 & 40.2 & 79.8 & 98 & 347 \\
PSZ2 G018.76$-$61.65 & 43.1 & 35.1 & 35.4 & 55.1 & 126 & 588 \\
PSZ2 G110.28$-$87.48 & 33.0 & 27.8 & 30.4 & 44.1 & 63 & 166 \\
PSZ2 G149.63$-$84.19 & 40.5 & 32.3 & 32.6 & 45.5 & 323 & 2149 \\
PSZ2 G167.43$-$53.67 & 67.5 & 44.7 & 44.6 & 65.9 & 233 & 977 \\
PSZ2 G167.66$-$65.59 & 75.6 & 42.0 & 41.7 & 63.1 & 112 & 516 \\
PSZ2 G167.98$-$59.95 & 60.0 & 32.9 & 32.4 & 49.5 & 181 & 887 \\
PSZ2 G172.98$-$53.55 & 96.0 & 52.6 & 53.2 & 74.1 & 120 & 462 \\
PSZ2 G174.40$-$57.33 & 67.3 & 37.5 & 38.2 & 51.2 & 174 & 1017 \\
PSZ2 G175.69$-$85.98 & 40.7 & 42.5 & 46.0 & 64.7 & 239 & 1003 \\
PSZ2 G180.74$-$85.21 & 43.3 & 40.1 & 39.2 & 69.2 & 131 & 473 \\
PSZ2 G219.88$+$22.83 & 83.6 & 75.9 & 76.2 & 105.7 & 218 & 825 \\
PSZ2 G220.11$+$22.91 & 67.5 & 63.3 & 61.5 & 88.4 & 202 & 960 \\
PSZ2 G223.47$+$26.85 & 56.7 & 42.1 & 42.6 & 71.6 & 99 & 307 \\
PSZ2 G225.48$+$29.41 & 67.8 & 43.4 & 49.1 & 52.4 & 381 & 1916 \\
PSZ2 G227.59$+$22.98 & 77.4 & 52.7 & 54.8 & 77.4 & 314 & 1482 \\
PSZ2 G227.89$+$36.58 & 64.8 & 56.7 & 66.2 & 65.6 & 211 & 967 \\
PSZ2 G228.38$+$38.58 & 46.3 & 40.6 & 41.9 & 66.0 & 82 & 286 \\
PSZ2 G228.50$+$34.95 & 45.4 & 39.1 & 40.0 & 61.1 & 142 & 779 \\
PSZ2 G230.73$+$27.70 & 53.9 & 42.6 & 44.8 & 64.9 & 159 & 800 \\
PSZ2 G231.79$+$31.48 & 66.0 & 50.9 & 50.3 & 90.1 & 122 & 579 \\
PSZ2 G232.84$+$38.13 & 61.1 & 54.2 & 55.9 & 85.8 & 415 & 1809 \\
PSZ2 G233.68$+$36.14 & 67.2 & 56.8 & 62.2 & 79.7 & 209 & 982 \\
PSZ2 G236.92$-$26.65 & 37.2 & 34.3 & 35.3 & 53.6 & 206 & 1044 \\
PSZ2 G239.27$-$26.01 & 40.4 & 37.4 & 42.8 & 56.6 & 132 & 617 \\
PSZ2 G241.79$-$24.01 & 79.9 & 58.4 & 63.6 & 76.5 & 779 & 5433 \\
PSZ2 G241.98$+$19.56 & 33.7 & 29.4 & 29.2 & 47.8 & 78 & 379 \\
PSZ2 G254.52$+$08.27 & 36.6 & 31.8 & 32.1 & 57.4 & 150 & 898 \\
PSZ2 G260.80$+$06.71 & 39.6 & 37.7 & 42.4 & 49.8 & 278 & 1885 \\
PSZ2 G262.36$-$25.15 & 47.8 & 44.1 & 50.2 & 55.1 & 958 & 3281 \\
PSZ2 G263.14$-$23.41 & 33.1 & 32.1 & 35.2 & 51.6 & 185 & 1005 \\
PSZ2 G263.19$-$25.19 & 37.6 & 43.8 & 51.1 & 45.3 & 422 & 2252 \\
PSZ2 G263.68$-$22.55 & 40.7 & 45.0 & 43.7 & 84.3 & 222 & 1309 \\
PSZ2 G265.21$-$24.83 & 31.2 & 31.3 & 30.4 & 56.6 & 640 & 4309 \\
PSZ2 G270.63$-$35.67 & 38.6 & 36.5 & 38.5 & 79.7 & 83 & 253 \\
PSZ2 G271.28$-$36.11 & 37.4 & 34.4 & 37.8 & 62.5 & 118 & 412 \\
PSZ2 G272.08$-$40.16 & 69.2 & 61.3 & 70.2 & 59.1 & 629 & 4049 \\

    \bottomrule
\end{tabular}
\end{threeparttable}
\end{table*}

\setcounter{table}{0}

\begin{table*}[t!]
\centering
\begin{threeparttable}
\caption{(Continued.) Image noise properties within 2\,Mpc of the cluster in \textmu Jy\,PSF$^{-1}$.}
\begin{tabular}{l c c c c c c}\toprule
    Cluster &Archive&robust $0.0$ & robust $+0.25$ & uniform & 100 kpc taper & 250 kpc taper  \\\midrule
PSZ2 G275.24$-$40.42 & 47.8 & 46.5 & 48.0 & 75.4 & 137 & 506 \\
PSZ2 G275.73$-$06.12 & 30.9 & 31.0 & 33.3 & 41.0 & 164 & 895 \\
PSZ2 G276.09$-$41.53 & 59.8 & 58.3 & 57.3 & 104.1 & 297 & 1913 \\
PSZ2 G276.14$-$07.68 & 32.7 & 34.5 & 39.1 & 52.7 & 255 & 1421 \\
PSZ2 G282.32$-$40.15 & 32.4 & 33.4 & 34.0 & 73.5 & 91 & 309 \\
PSZ2 G286.28$-$38.36 & 44.2 & 40.7 & 46.2 & 69.5 & 123 & 618 \\
PSZ2 G286.75$-$37.35 & 50.1 & 55.0 & 59.6 & 116.9 & 116 & 427 \\
PSZ2 G311.98$+$30.71 & 34.6 & 32.9 & 34.9 & 51.6 & 1608 & 9935 \\
PSZ2 G313.33$+$30.29 & 33.2 & 33.2 & 35.1 & 53.2 & 1274 & 7113 \\
PSZ2 G328.58$-$25.25 & 49.5 & 44.0 & 42.5 & 89.5 & 138 & 517 \\
PSZ2 G331.96$-$45.74 & 31.7 & 29.9 & 31.0 & 49.7 & 510 & 4021 \\
PSZ2 G332.11$-$23.63 & 33.8 & 34.3 & 34.7 & 67.8 & 128 & 639 \\
PSZ2 G332.23$-$46.37 & 32.5 & 28.6 & 29.9 & 48.5 & 326 & 1701 \\
PSZ2 G332.29$-$23.57 & 55.2 & 110.6 & 100.6 & 212.5 & 8406 & 7869 \\
PSZ2 G333.89$-$43.60 & 36.8 & 31.4 & 32.7 & 58.1 & 89 & 310 \\
PSZ2 G335.58$-$46.44 & 35.5 & 30.9 & 33.6 & 48.7 & 565 & 3852 \\
PSZ2 G336.95$-$45.75 & 33.7 & 31.7 & 33.1 & 53.5 & 621 & 2801 \\
PSZ2 G337.99$-$33.61 & 36.7 & 35.8 & 34.1 & 72.7 & 87 & 290 \\
PSZ2 G339.74$-$51.08 & 58.3 & 60.0 & 54.9 & 125.0 & 189 & 1209 \\
PSZ2 G340.35$-$42.80 & 31.3 & 30.3 & 31.8 & 49.7 & 100 & 405 \\
PSZ2 G341.19$-$36.12 & 41.8 & 40.2 & 43.1 & 66.5 & 138 & 688 \\
PSZ2 G341.44$-$40.19 & 39.0 & 34.4 & 38.3 & 51.6 & 182 & 1043 \\
PSZ2 G342.33$-$34.93 & 59.2 & 65.1 & 74.2 & 86.4 & 358 & 1966 \\
PSZ2 G342.62$-$39.60 & 52.9 & 47.8 & 53.5 & 64.8 & 437 & 3321 \\
PSZ2 G345.38$-$39.32 & 37.1 & 36.1 & 38.4 & 57.0 & 124 & 583 \\
PSZ2 G345.82$-$34.29 & 38.4 & 24.1 & 24.8 & 37.8 & 858 & 6319 \\
PSZ2 G346.86$-$45.38 & 31.1 & 33.7 & 36.7 & 47.2 & 462 & 2591 \\
PSZ2 G347.58$-$35.35 & 31.9 & 30.1 & 30.2 & 56.3 & 126 & 525 \\\bottomrule
\end{tabular}
\end{threeparttable}
\end{table*}

\begin{table*}[t!]
\centering
\begin{threeparttable}
\caption{\label{tab:psf_properties} Image PSFs (arcsec $\times$ arcsec, deg).}
\begin{tabular}{l c c c c c c}\toprule
    Cluster &Archive& robust 0.0 & robust $+0.25$ & uniform & 100 kpc taper & 250 kpc taper  \\\midrule
PSZ2 G006.16$-$69.49 & $10.1 \times 8.5$, $82.6$ & $13.3 \times 10.5$, $87.8$ & $15.2 \times 12.0$, $92.1$ & $9.9 \times 7.6$, $79.2$ & $33 \times 32$, $87.2$ & $70 \times 68$, $88.6$ \\
PSZ2 G008.31$-$64.74 & $10.0 \times 8.5$, $83.7$ & $13.0 \times 10.6$, $87.0$ & $14.7 \times 11.8$, $95.5$ & $9.5 \times 7.3$, $76.6$ & $28 \times 27$, $104.1$ & $58 \times 56$, $97.3$ \\
PSZ2 G011.06$-$63.84 & $10.0 \times 8.5$, $83.7$ & $13.0 \times 10.6$, $88.2$ & $14.8 \times 11.8$, $96.8$ & $9.6 \times 7.3$, $77.4$ & $34 \times 33$, $83.9$ & $70 \times 68$, $95.6$ \\
PSZ2 G011.92$-$63.53 & $10.0 \times 8.5$, $83.7$ & $13.2 \times 10.7$, $88.8$ & $15.0 \times 11.9$, $97.3$ & $9.9 \times 7.7$, $79.1$ & $32 \times 32$, $89.8$ & $67 \times 66$, $84.0$ \\
PSZ2 G014.72$-$62.49 & $10.0 \times 8.5$, $83.7$ & $13.2 \times 10.7$, $89.9$ & $15.0 \times 11.9$, $98.5$ & $10.0 \times 7.7$, $79.7$ & $23 \times 21$, $106.3$ & $46 \times 44$, $80.4$ \\
PSZ2 G017.25$-$70.71 & $10.1 \times 8.5$, $82.6$ & $13.3 \times 10.5$, $89.4$ & $15.2 \times 12.0$, $93.3$ & $10.1 \times 7.7$, $80.2$ & $29 \times 27$, $100.1$ & $58 \times 56$, $83.2$ \\
PSZ2 G018.18$-$60.00 & $10.0 \times 8.5$, $83.7$ & $13.1 \times 10.6$, $91.5$ & $15.0 \times 11.8$, $100.3$ & $9.7 \times 7.3$, $79.2$ & $31 \times 30$, $105.0$ & $66 \times 64$, $94.7$ \\
PSZ2 G018.76$-$61.65 & $10.0 \times 8.5$, $83.7$ & $13.2 \times 10.7$, $91.1$ & $15.1 \times 11.9$, $99.8$ & $10.1 \times 7.7$, $80.7$ & $32 \times 32$, $98.4$ & $68 \times 66$, $84.5$ \\
PSZ2 G110.28$-$87.48 & $12.4 \times 9.9$, $86.5$ & $12.9 \times 9.8$, $96.1$ & $15.0 \times 10.8$, $102.6$ & $9.6 \times 7.0$, $89.6$ & $25 \times 21$, $115.0$ & $45 \times 43$, $90.6$ \\
PSZ2 G149.63$-$84.19 & $12.6 \times 9.7$, $-87.5$ & $13.7 \times 9.7$, $100.4$ & $15.6 \times 11.1$, $102.6$ & $11.1 \times 7.8$, $89.9$ & $51 \times 48$, $79.4$ & $112 \times 110$, $80.2$ \\
PSZ2 G167.43$-$53.67 & $18.0 \times 18.0$, $0.0$ & $13.6 \times 10.9$, $85.0$ & $14.8 \times 12.3$, $79.4$ & $12.1 \times 8.6$, $86.2$ & $40 \times 36$, $100.8$ & $81 \times 74$, $110.8$ \\
PSZ2 G167.66$-$65.59 & $18.0 \times 18.0$, $0.0$ & $12.7 \times 10.7$, $84.7$ & $14.0 \times 11.8$, $92.2$ & $10.7 \times 8.1$, $94.3$ & $26 \times 24$, $122.9$ & $51 \times 48$, $105.9$ \\
PSZ2 G167.98$-$59.95 & $18.0 \times 18.0$, $0.0$ & $13.8 \times 10.8$, $84.4$ & $15.4 \times 12.3$, $79.7$ & $11.9 \times 8.3$, $87.3$ & $49 \times 44$, $88.1$ & $103 \times 93$, $112.2$ \\
PSZ2 G172.98$-$53.55 & $18.0 \times 18.0$, $0.0$ & $13.5 \times 10.7$, $86.0$ & $14.7 \times 12.1$, $79.7$ & $12.0 \times 8.5$, $88.4$ & $28 \times 25$, $108.3$ & $56 \times 52$, $94.7$ \\
PSZ2 G174.40$-$57.33 & $18.0 \times 18.0$, $0.0$ & $12.9 \times 11.0$, $86.0$ & $14.1 \times 12.2$, $96.0$ & $11.1 \times 8.2$, $91.6$ & $39 \times 37$, $100.3$ & $81 \times 74$, $109.1$ \\
PSZ2 G175.69$-$85.98 & $12.6 \times 9.7$, $-87.5$ & $13.1 \times 10.0$, $95.1$ & $15.1 \times 11.0$, $101.3$ & $9.7 \times 7.0$, $89.1$ & $35 \times 33$, $98.7$ & $70 \times 68$, $92.2$ \\
PSZ2 G180.74$-$85.21 & $12.6 \times 9.7$, $-87.5$ & $13.2 \times 10.0$, $94.9$ & $15.2 \times 11.1$, $101.0$ & $9.7 \times 7.0$, $89.0$ & $37 \times 35$, $93.1$ & $77 \times 75$, $93.0$ \\
PSZ2 G219.88$+$22.83 & $13.4 \times 12.4$, $-89.5$ & $14.9 \times 11.8$, $113.3$ & $16.4 \times 13.2$, $117.5$ & $12.8 \times 9.4$, $103.7$ & $37 \times 34$, $124.8$ & $72 \times 68$, $100.0$ \\
PSZ2 G220.11$+$22.91 & $13.4 \times 12.4$, $-89.5$ & $14.9 \times 11.8$, $112.8$ & $16.4 \times 13.2$, $117.1$ & $12.8 \times 9.4$, $103.3$ & $39 \times 36$, $124.6$ & $74 \times 69$, $99.8$ \\
PSZ2 G223.47$+$26.85 & $13.4 \times 12.4$, $-89.5$ & $14.1 \times 12.2$, $129.3$ & $16.1 \times 13.7$, $136.6$ & $11.3 \times 9.7$, $90.9$ & $31 \times 28$, $143.0$ & $58 \times 56$, $113.4$ \\
PSZ2 G225.48$+$29.41 & $13.5 \times 12.5$, $-80.5$ & $13.4 \times 11.4$, $103.2$ & $15.2 \times 12.4$, $114.4$ & $11.2 \times 8.2$, $90.7$ & $38 \times 36$, $123.7$ & $77 \times 74$, $104.8$ \\
PSZ2 G227.59$+$22.98 & $13.3 \times 11.9$, $83.9$ & $14.4 \times 10.8$, $93.6$ & $15.8 \times 12.3$, $97.0$ & $12.5 \times 7.6$, $88.1$ & $36 \times 32$, $69.0$ & $64 \times 61$, $77.9$ \\
PSZ2 G227.89$+$36.58 & $13.4 \times 12.5$, $83.0$ & $13.3 \times 12.3$, $109.3$ & $15.1 \times 13.1$, $123.4$ & $11.2 \times 9.2$, $86.3$ & $25 \times 23$, $133.9$ & $50 \times 48$, $165.6$ \\
PSZ2 G228.38$+$38.58 & $13.4 \times 12.5$, $83.0$ & $13.8 \times 11.7$, $95.8$ & $15.4 \times 13.2$, $101.6$ & $12.0 \times 8.8$, $76.4$ & $25 \times 22$, $100.2$ & $46 \times 45$, $131.7$ \\
PSZ2 G228.50$+$34.95 & $13.4 \times 12.5$, $83.0$ & $13.8 \times 11.8$, $111.1$ & $15.4 \times 13.2$, $116.9$ & $11.7 \times 9.0$, $88.8$ & $32 \times 30$, $134.7$ & $64 \times 62$, $104.1$ \\
PSZ2 G230.73$+$27.70 & $13.2 \times 11.6$, $89.5$ & $14.0 \times 11.5$, $118.2$ & $15.8 \times 13.0$, $126.0$ & $11.9 \times 9.3$, $90.2$ & $33 \times 30$, $129.8$ & $62 \times 58$, $104.3$ \\
PSZ2 G231.79$+$31.48 & $13.3 \times 12.1$, $86.9$ & $14.4 \times 10.6$, $107.2$ & $15.8 \times 12.0$, $111.0$ & $12.5 \times 7.8$, $95.6$ & $30 \times 27$, $119.0$ & $59 \times 54$, $106.3$ \\
PSZ2 G232.84$+$38.13 & $13.4 \times 12.5$, $83.0$ & $14.0 \times 11.8$, $106.3$ & $15.5 \times 13.4$, $112.3$ & $12.0 \times 8.8$, $86.6$ & $46 \times 45$, $163.8$ & $95 \times 91$, $99.3$ \\
PSZ2 G233.68$+$36.14 & $13.3 \times 12.1$, $86.9$ & $14.4 \times 11.2$, $107.2$ & $15.8 \times 12.7$, $111.3$ & $12.5 \times 8.4$, $90.6$ & $29 \times 27$, $126.1$ & $56 \times 54$, $117.3$ \\
PSZ2 G236.92$-$26.65 & $12.1 \times 9.7$, $85.4$ & $12.5 \times 10.0$, $85.9$ & $14.1 \times 11.1$, $91.8$ & $9.3 \times 7.0$, $80.2$ & $44 \times 42$, $73.1$ & $97 \times 95$, $89.8$ \\
PSZ2 G239.27$-$26.01 & $12.1 \times 9.7$, $85.4$ & $12.4 \times 10.1$, $84.0$ & $13.9 \times 11.0$, $91.0$ & $9.1 \times 6.9$, $78.3$ & $24 \times 22$, $97.9$ & $49 \times 47$, $100.8$ \\
PSZ2 G241.79$-$24.01 & $12.1 \times 9.7$, $85.4$ & $12.5 \times 10.2$, $82.5$ & $14.0 \times 11.2$, $89.8$ & $9.1 \times 7.0$, $77.4$ & $45 \times 44$, $79.7$ & $100 \times 99$, $111.3$ \\
PSZ2 G241.98$+$19.56 & $12.4 \times 10.5$, $81.4$ & $13.1 \times 10.5$, $92.0$ & $14.8 \times 11.6$, $99.9$ & $10.3 \times 7.7$, $83.2$ & $31 \times 30$, $105.5$ & $63 \times 61$, $94.2$ \\
PSZ2 G254.52$+$08.27 & $12.3 \times 10.1$, $80.8$ & $12.6 \times 10.1$, $84.3$ & $14.3 \times 11.3$, $90.1$ & $9.1 \times 6.9$, $77.7$ & $36 \times 35$, $86.0$ & $77 \times 75$, $101.9$ \\
PSZ2 G260.80$+$06.71 & $12.5 \times 10.3$, $80.2$ & $12.6 \times 10.2$, $80.9$ & $14.3 \times 11.4$, $85.7$ & $9.0 \times 6.9$, $74.4$ & $36 \times 35$, $84.1$ & $77 \times 75$, $102.5$ \\
PSZ2 G262.36$-$25.15 & $11.2 \times 9.5$, $86.7$ & $11.5 \times 9.7$, $83.8$ & $12.9 \times 10.9$, $88.1$ & $8.4 \times 6.8$, $73.5$ & $37 \times 36$, $86.0$ & $81 \times 80$, $87.7$ \\
PSZ2 G263.14$-$23.41 & $11.2 \times 9.5$, $86.7$ & $11.3 \times 10.0$, $79.4$ & $12.6 \times 11.5$, $85.4$ & $8.2 \times 6.7$, $69.9$ & $32 \times 32$, $81.0$ & $69 \times 68$, $108.9$ \\
PSZ2 G263.19$-$25.19 & $11.2 \times 9.5$, $86.7$ & $11.7 \times 9.9$, $83.4$ & $13.2 \times 11.1$, $87.9$ & $9.2 \times 7.4$, $75.6$ & $38 \times 37$, $87.7$ & $82 \times 81$, $88.6$ \\
PSZ2 G263.68$-$22.55 & $11.2 \times 9.5$, $86.7$ & $11.3 \times 10.0$, $79.1$ & $12.7 \times 11.6$, $85.2$ & $8.2 \times 6.8$, $69.8$ & $39 \times 38$, $80.6$ & $87 \times 87$, $99.1$ \\
PSZ2 G265.21$-$24.83 & $11.2 \times 9.5$, $86.7$ & $11.8 \times 10.1$, $83.2$ & $13.4 \times 11.5$, $88.1$ & $8.5 \times 6.9$, $71.8$ & $92 \times 89$, $44.5$ & $221 \times 202$, $14.3$ \\
PSZ2 G270.63$-$35.67 & $13.9 \times 11.4$, $-88.8$ & $13.9 \times 11.1$, $87.3$ & $16.0 \times 12.9$, $91.9$ & $9.1 \times 7.1$, $77.8$ & $30 \times 26$, $104.8$ & $61 \times 58$, $102.3$ \\
PSZ2 G271.28$-$36.11 & $13.9 \times 11.4$, $-88.8$ & $14.1 \times 11.3$, $88.2$ & $16.3 \times 13.1$, $92.8$ & $9.1 \times 7.2$, $78.2$ & $33 \times 30$, $102.5$ & $68 \times 65$, $104.1$ \\
PSZ2 G272.08$-$40.16 & $15.5 \times 13.7$, $-6.2$ & $15.2 \times 13.9$, $35.9$ & $16.8 \times 15.8$, $59.3$ & $12.8 \times 10.5$, $9.6$ & $51 \times 46$, $65.7$ & $107 \times 103$, $61.2$ \\

\bottomrule
\end{tabular}
\end{threeparttable}
\end{table*}

\setcounter{table}{1}

\begin{table*}[t!]
\centering
\begin{threeparttable}
\caption{(Continued.) Image PSFs (arcsec $\times$ arcsec, deg).}
\begin{tabular}{l c c c c c c}\toprule
    Cluster &Archive& robust $0.0$ & robust $+0.25$ & uniform & 100 kpc taper & 250 kpc taper  \\\midrule

PSZ2 G275.24$-$40.42 & $15.5 \times 13.7$, $-6.2$ & $15.2 \times 13.6$, $33.2$ & $16.9 \times 15.6$, $53.5$ & $12.0 \times 9.5$, $177.1$ & $37 \times 34$, $75.8$ & $77 \times 75$, $89.9$ \\
PSZ2 G275.73$-$06.12 & $12.3 \times 10.7$, $81.3$ & $12.8 \times 11.0$, $78.4$ & $14.5 \times 12.5$, $83.9$ & $9.7 \times 8.1$, $72.2$ & $36 \times 35$, $106.7$ & $77 \times 76$, $98.6$ \\
PSZ2 G276.09$-$41.53 & $15.5 \times 13.7$, $-6.2$ & $15.5 \times 13.8$, $40.1$ & $17.3 \times 15.7$, $59.3$ & $12.2 \times 9.6$, $1.6$ & $47 \times 44$, $74.6$ & $100 \times 97$, $83.7$ \\
PSZ2 G276.14$-$07.68 & $12.3 \times 10.7$, $81.3$ & $13.0 \times 10.8$, $74.2$ & $14.9 \times 12.4$, $78.7$ & $9.4 \times 7.0$, $61.9$ & $36 \times 35$, $98.2$ & $77 \times 75$, $102.2$ \\
PSZ2 G282.32$-$40.15 & $13.9 \times 12.0$, $-84.8$ & $14.0 \times 11.8$, $93.0$ & $16.3 \times 13.6$, $98.8$ & $9.0 \times 7.4$, $78.7$ & $37 \times 34$, $108.8$ & $79 \times 76$, $111.0$ \\
PSZ2 G286.28$-$38.36 & $13.9 \times 12.0$, $-84.8$ & $14.0 \times 12.0$, $92.8$ & $16.2 \times 13.8$, $99.4$ & $8.9 \times 7.8$, $71.7$ & $29 \times 26$, $105.7$ & $59 \times 57$, $109.2$ \\
PSZ2 G286.75$-$37.35 & $13.9 \times 12.0$, $-84.8$ & $14.0 \times 12.0$, $90.8$ & $16.2 \times 13.9$, $97.7$ & $8.9 \times 7.9$, $67.1$ & $24 \times 22$, $102.6$ & $48 \times 45$, $111.3$ \\
PSZ2 G311.98$+$30.71 & $12.3 \times 9.9$, $85.5$ & $12.9 \times 10.4$, $86.7$ & $14.7 \times 11.6$, $92.7$ & $9.3 \times 7.3$, $80.7$ & $102 \times 98$, $60.3$ & $245 \times 230$, $177.6$ \\
PSZ2 G313.33$+$30.29 & $12.3 \times 9.9$, $85.5$ & $12.9 \times 10.4$, $86.1$ & $14.6 \times 11.6$, $91.9$ & $9.3 \times 7.3$, $80.5$ & $100 \times 96$, $61.0$ & $240 \times 225$, $172.6$ \\
PSZ2 G328.58$-$25.25 & $12.3 \times 11.2$, $83.0$ & $12.6 \times 11.4$, $77.1$ & $14.3 \times 13.1$, $87.5$ & $8.6 \times 7.3$, $64.2$ & $38 \times 37$, $110.2$ & $83 \times 81$, $114.4$ \\
PSZ2 G331.96$-$45.74 & $12.1 \times 10.7$, $85.3$ & $12.4 \times 10.8$, $82.2$ & $13.8 \times 12.0$, $82.8$ & $9.0 \times 7.3$, $71.5$ & $72 \times 70$, $72.9$ & $162 \times 156$, $30.1$ \\
PSZ2 G332.11$-$23.63 & $12.3 \times 11.2$, $83.0$ & $12.6 \times 11.1$, $77.7$ & $14.2 \times 12.8$, $85.3$ & $8.6 \times 7.1$, $66.7$ & $36 \times 34$, $111.4$ & $77 \times 76$, $112.0$ \\
PSZ2 G332.23$-$46.37 & $12.1 \times 10.7$, $85.3$ & $12.6 \times 10.9$, $82.6$ & $14.2 \times 12.4$, $87.4$ & $8.8 \times 7.4$, $75.4$ & $58 \times 57$, $74.0$ & $133 \times 129$, $43.3$ \\
PSZ2 G332.29$-$23.57 & $12.3 \times 11.2$, $83.0$ & $12.9 \times 11.5$, $78.2$ & $14.6 \times 13.2$, $86.4$ & $9.3 \times 8.0$, $70.6$ & $106 \times 103$, $40.6$ & $250 \times 232$, $166.7$ \\
PSZ2 G333.89$-$43.60 & $12.1 \times 10.7$, $85.3$ & $12.1 \times 11.0$, $80.3$ & $13.6 \times 12.8$, $87.9$ & $8.7 \times 7.1$, $65.9$ & $28 \times 27$, $128.9$ & $57 \times 55$, $108.3$ \\
PSZ2 G335.58$-$46.44 & $12.1 \times 9.9$, $80.5$ & $12.3 \times 10.6$, $81.1$ & $13.7 \times 12.0$, $84.9$ & $8.8 \times 7.0$, $67.9$ & $69 \times 68$, $50.4$ & $156 \times 149$, $47.3$ \\
PSZ2 G336.95$-$45.75 & $12.1 \times 9.9$, $80.5$ & $12.2 \times 10.5$, $80.7$ & $13.7 \times 11.9$, $83.9$ & $8.8 \times 6.9$, $68.3$ & $68 \times 66$, $52.3$ & $155 \times 145$, $39.8$ \\
PSZ2 G337.99$-$33.61 & $12.5 \times 10.9$, $78.8$ & $12.8 \times 10.9$, $78.8$ & $14.7 \times 12.9$, $86.7$ & $9.4 \times 6.9$, $70.8$ & $37 \times 35$, $94.1$ & $77 \times 76$, $105.3$ \\
PSZ2 G339.74$-$51.08 & $12.1 \times 9.9$, $80.5$ & $12.2 \times 10.3$, $80.7$ & $13.6 \times 11.6$, $84.2$ & $8.7 \times 7.0$, $70.4$ & $35 \times 34$, $158.2$ & $76 \times 75$, $100.6$ \\
PSZ2 G340.35$-$42.80 & $14.0 \times 10.9$, $-58.2$ & $14.0 \times 11.4$, $113.5$ & $15.8 \times 13.1$, $109.4$ & $10.9 \times 7.7$, $120.4$ & $35 \times 34$, $106.2$ & $77 \times 75$, $103.0$ \\
PSZ2 G341.19$-$36.12 & $12.0 \times 10.7$, $84.8$ & $12.3 \times 10.9$, $83.5$ & $13.8 \times 12.6$, $91.8$ & $8.7 \times 7.0$, $72.8$ & $29 \times 28$, $83.6$ & $60 \times 59$, $100.4$ \\
PSZ2 G341.44$-$40.19 & $14.0 \times 10.9$, $-58.2$ & $12.2 \times 10.8$, $81.8$ & $13.7 \times 12.6$, $89.5$ & $8.7 \times 7.1$, $70.6$ & $31 \times 31$, $88.9$ & $66 \times 64$, $101.9$ \\
PSZ2 G342.33$-$34.93 & $12.0 \times 10.7$, $84.8$ & $15.8 \times 10.6$, $65.4$ & $17.0 \times 13.2$, $72.0$ & $12.7 \times 6.3$, $65.8$ & $33 \times 32$, $90.2$ & $70 \times 68$, $101.1$ \\
PSZ2 G342.62$-$39.60 & $12.0 \times 10.7$, $84.8$ & $12.4 \times 11.0$, $83.4$ & $14.0 \times 12.8$, $91.5$ & $8.8 \times 7.1$, $71.6$ & $45 \times 44$, $92.5$ & $101 \times 99$, $93.0$ \\
PSZ2 G345.38$-$39.32 & $14.8 \times 11.1$, $-59.0$ & $14.9 \times 11.6$, $117.4$ & $16.7 \times 13.3$, $111.4$ & $12.1 \times 8.3$, $126.5$ & $34 \times 33$, $96.8$ & $73 \times 71$, $89.6$ \\
PSZ2 G345.82$-$34.29 & $12.0 \times 10.1$, $86.1$ & $12.6 \times 11.1$, $74.9$ & $14.1 \times 12.9$, $84.4$ & $9.6 \times 7.9$, $58.6$ & $97 \times 92$, $54.5$ & $235 \times 213$, $161.6$ \\
PSZ2 G346.86$-$45.38 & $12.1 \times 10.1$, $89.1$ & $14.8 \times 11.5$, $115.4$ & $16.7 \times 13.2$, $109.6$ & $11.5 \times 7.8$, $122.6$ & $68 \times 67$, $74.1$ & $156 \times 153$, $19.3$ \\
PSZ2 G347.58$-$35.35 & $12.0 \times 10.1$, $86.1$ & $12.3 \times 10.3$, $82.4$ & $13.9 \times 11.5$, $87.1$ & $8.8 \times 7.1$, $73.9$ & $33 \times 32$, $100.7$ & $71 \times 70$, $101.6$ \\\bottomrule
\end{tabular}
\end{threeparttable}
\end{table*}

We summarise the image properties in Table~\ref{tab:rms_properties} (rms noise) and Table~\ref{tab:psf_properties} (PSF) for the individual cluster images, including the archival image. For Table~\ref{tab:rms_properties}, the rms noise is calculated within a 2\,Mpc circle around the cluster centre. Figures~\ref{fig:app:PSZ2G006.16-69.49}--\ref{fig:app:PSZ2G347.58-35.35} shows ASKAP images for each cluster in our sample. For each cluster, we show the robust $+0.25$ reference image and the tapered, compact source-subtracted image. In each panel we label the diffuse sources of interest, and show dashed, black polygon regions which are used for integrated flux density measurements. More comprehensive images of all clusters (e.g. Figure~\ref{fig:example}), along with the FITS files for the reprocessed radio data, will be provided in the PASA Datastore. In the PDF document, a link is included in the caption for clusters with a corresponding entry in Section~\ref{sec:notes}.

\begin{figure*}
\centering
\begin{subfigure}[b]{0.5\linewidth}
\includegraphics[width=\linewidth]{2panel_figures/PSZ2G006.16-69.49_2panel.pdf}
\caption{\label{fig:app:PSZ2G006.16-69.49} PSZ2 G006.16$-$69.49.}
\end{subfigure}%
\begin{subfigure}[b]{0.5\linewidth}
\includegraphics[width=\linewidth]{2panel_figures/PSZ2G008.31-64.74_2panel.pdf}
\caption{\label{fig:app:PSZ2G008.31-64.74} \hyperref[sec:PSZ2G008.31-64.74]{PSZ2 G008.31$-$64.74}.}
\end{subfigure}\\%
\begin{subfigure}[b]{0.5\linewidth}
\includegraphics[width=\linewidth]{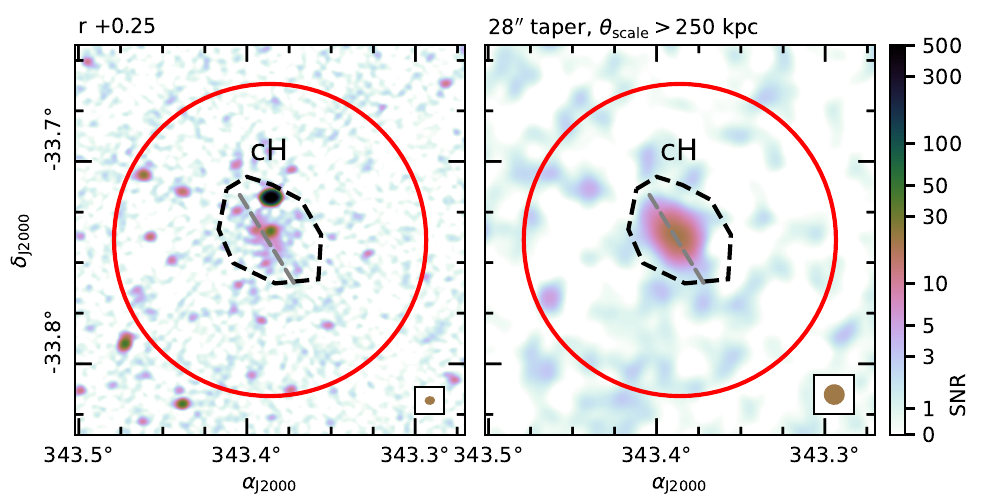}
\caption{\label{fig:app:PSZ2G011.06-63.84} \hyperref[sec:PSZ2GG011.06-63.84]{PSZ2 G011.06$-$63.84}.}
\end{subfigure}%
\begin{subfigure}[b]{0.5\linewidth}
\includegraphics[width=\linewidth]{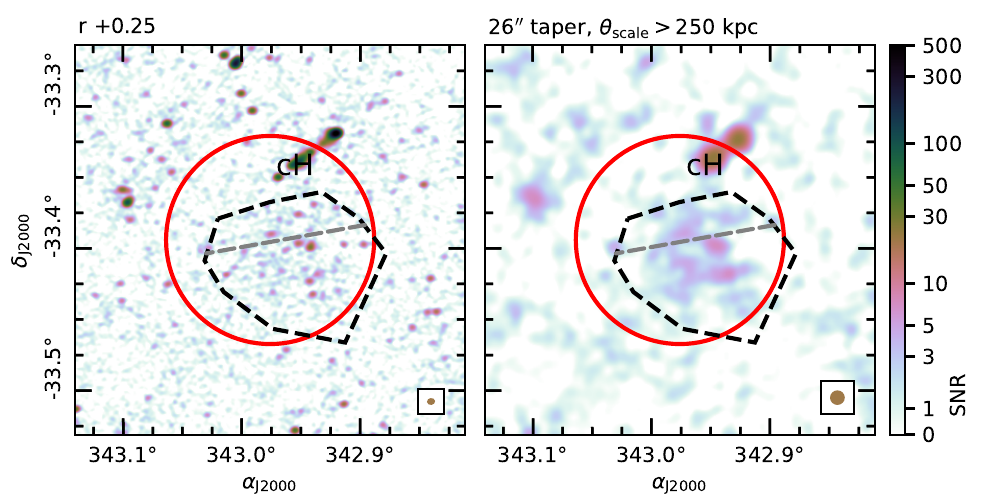}
\caption{\label{fig:app:PSZ2G011.92-63.53} \hyperref[sec:PSZ2G011.92-63.53]{PSZ2 G011.92$-$63.53}.}
\end{subfigure}\\%
\begin{subfigure}[b]{0.5\linewidth}
\includegraphics[width=\linewidth]{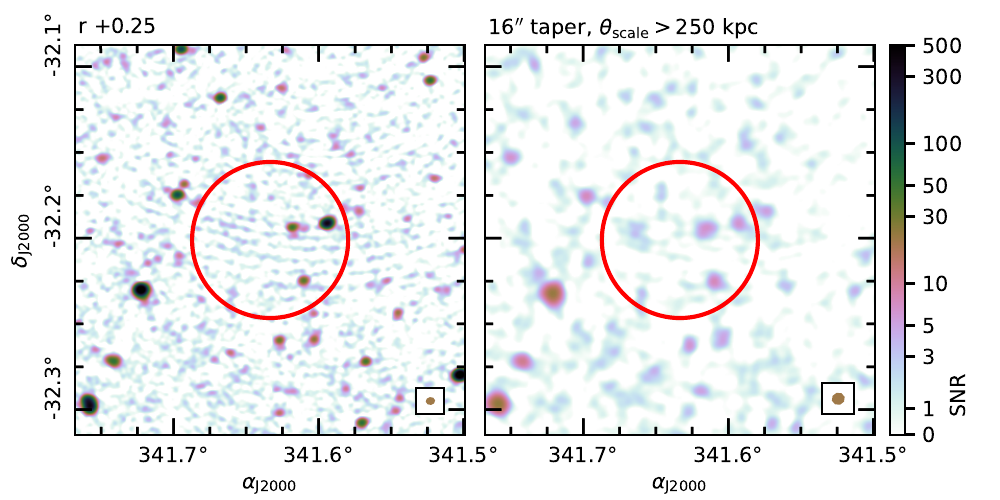}
\caption{\label{fig:app:PSZ2G014.72-62.49} PSZ2 G014.72$-$62.49.}
\end{subfigure}%
\begin{subfigure}[b]{0.5\linewidth}
\includegraphics[width=\linewidth]{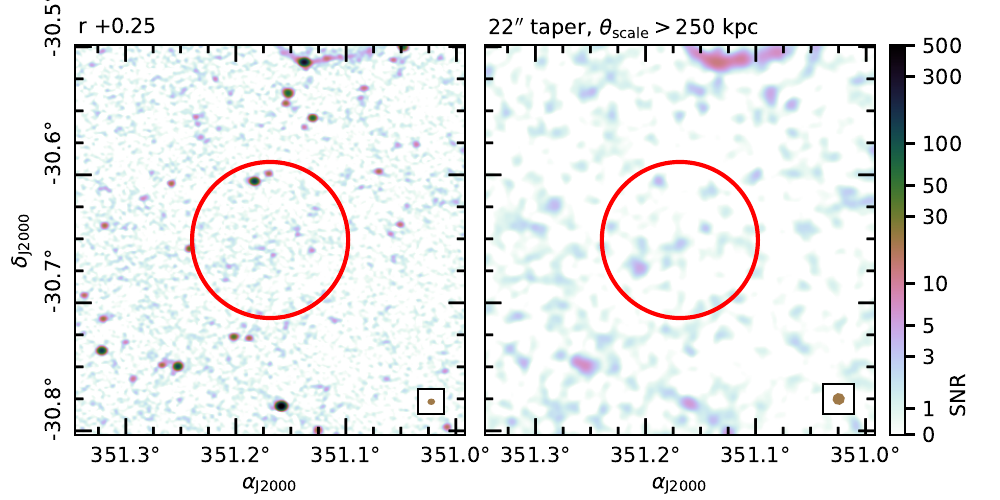}
\caption{\label{fig:app:PSZ2G017.25-70.71} PSZ2 G017.25$-$70.71.}
\end{subfigure}\\%
\begin{subfigure}[b]{0.5\linewidth}
\includegraphics[width=\linewidth]{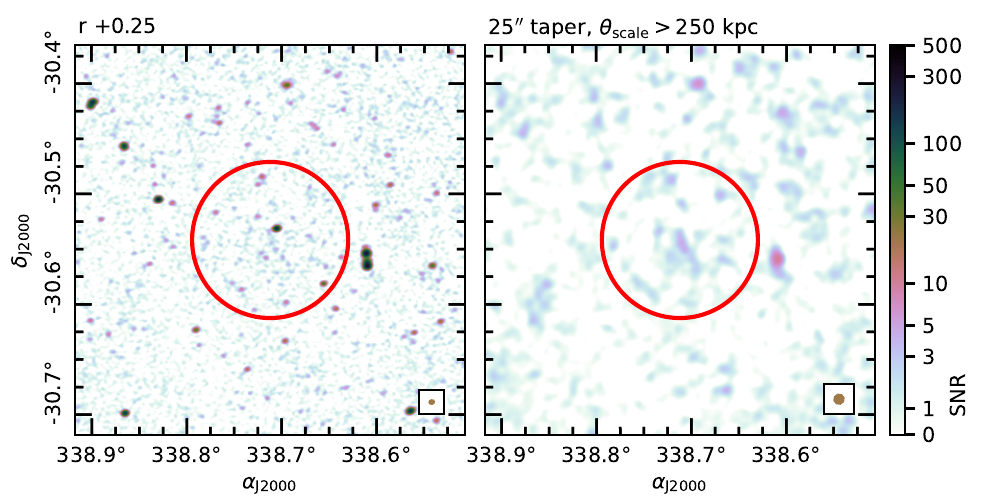}
\caption{\label{fig:app:PSZ2G018.18-60.00} PSZ2 G018.18$-$60.00.}
\end{subfigure}%
\begin{subfigure}[b]{0.5\linewidth}
\includegraphics[width=\linewidth]{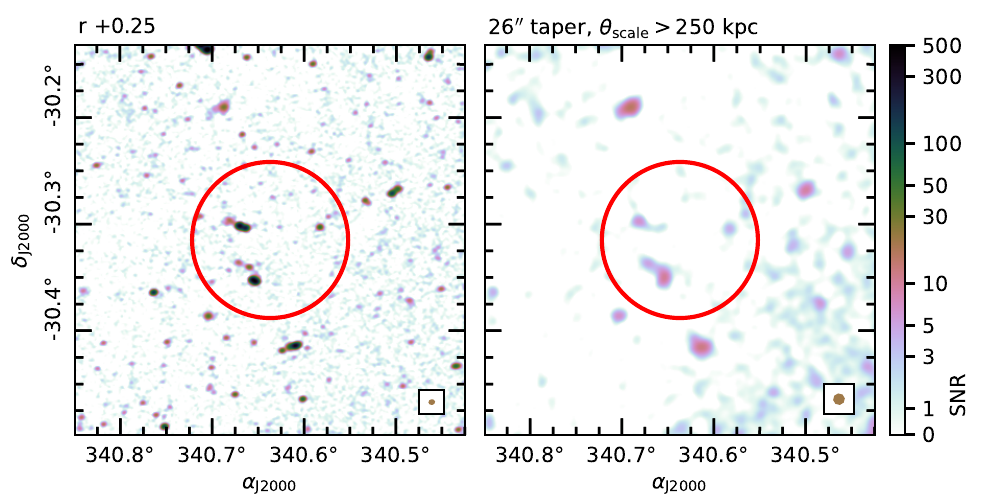}
\caption{\label{fig:app:PSZ2G018.76-61.65} PSZ2 G018.76$-$61.65.}
\end{subfigure}\\%
\caption{\label{fig:app:clusters} Radio images of the clusters. \emph{Left.} The robust $+0.25$ reference image. \emph{Right.} The robust $+0.25$ image, tapered, after subtraction of sources of scales $<250$~kpc. In both panels, the red circle has a 1~Mpc radius at the redshift of the clusters (dashed indicates an assumed redshift of 0.2). Dashed polygon regions indicate the diffuse sources of interest and are the region used for integrated flux density measurements. The PSF of each image is shown in the bottom right corner.}
\end{figure*}

\begin{figure*}
\centering
\ContinuedFloat

\begin{subfigure}[b]{0.5\linewidth}
\includegraphics[width=\linewidth]{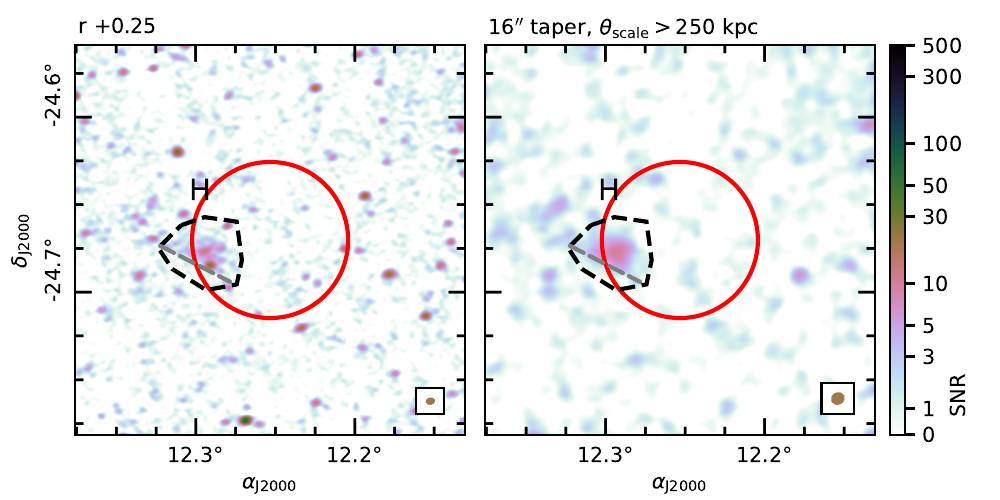}
\subcaption{\label{fig:app:PSZ2G110.28-87.48} \hyperref[sec:PSZ2G110.28-87.48]{PSZ2 G110.28$-$87.48}.}
\end{subfigure}%
\begin{subfigure}[b]{0.5\linewidth}
\includegraphics[width=\linewidth]{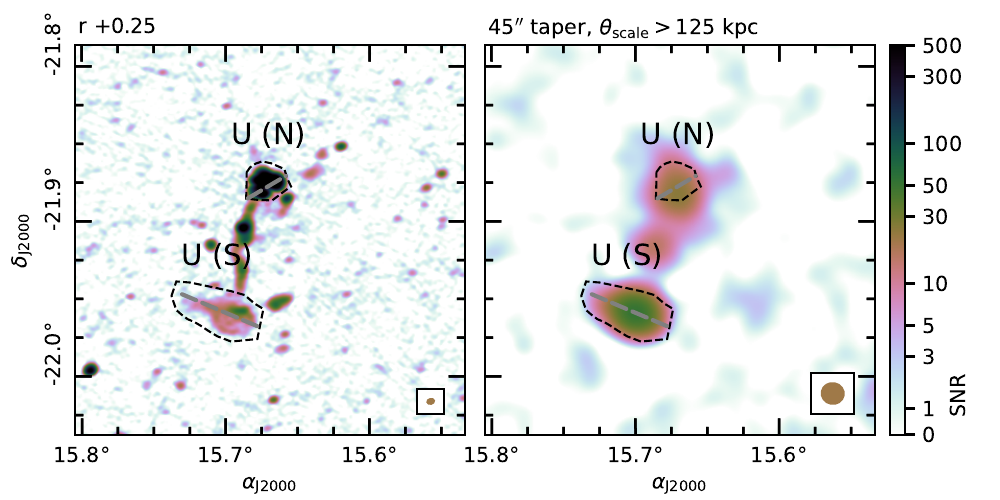}
\caption{\label{fig:app:PSZ2G149.63-84.19} \hyperref[sec:PSZ2G149.63-84.19]{PSZ2 G149.63$-$84.19}.}
\end{subfigure}\\%
\begin{subfigure}[b]{0.5\linewidth}
\includegraphics[width=\linewidth]{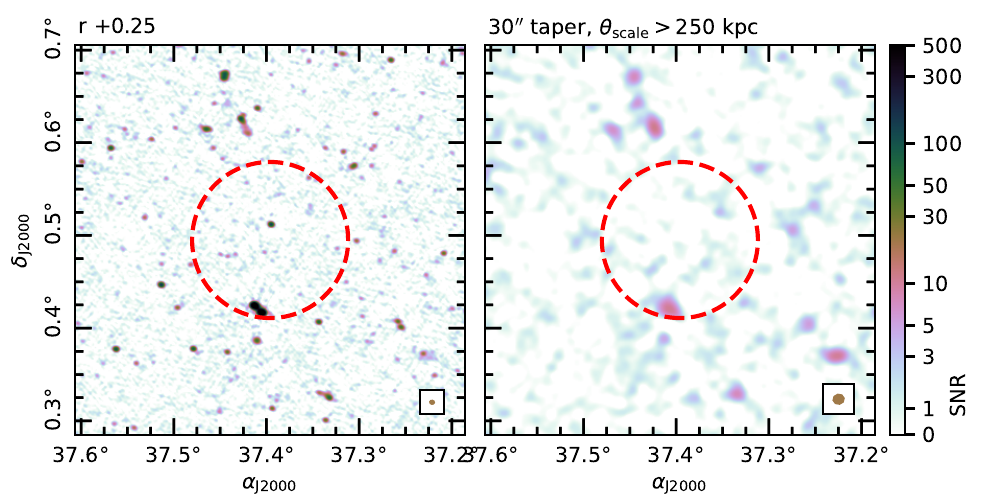}
\caption{\label{fig:app:PSZ2G167.43-53.67} PSZ2 G167.43$-$53.67.}
\end{subfigure}%
\begin{subfigure}[b]{0.5\linewidth}
\includegraphics[width=\linewidth]{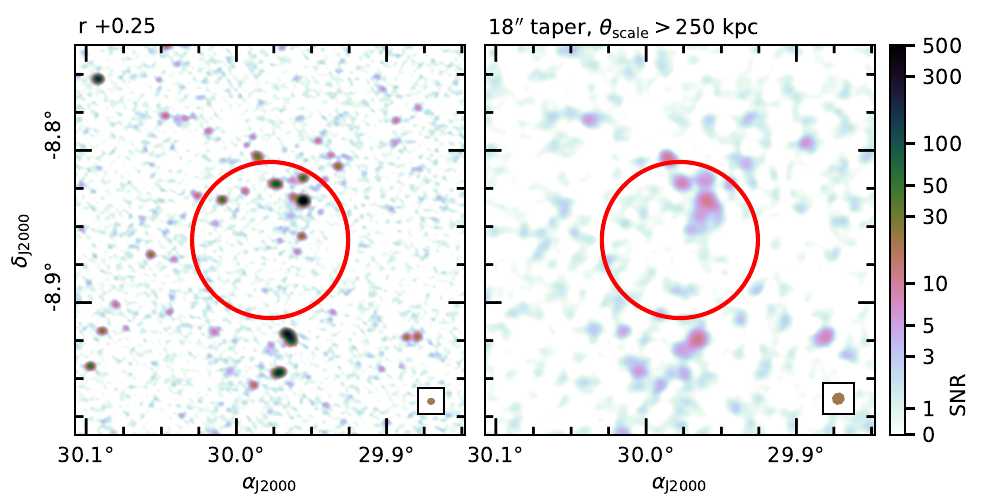}
\caption{\label{fig:app:PSZ2G167.66-65.59} PSZ2 G167.66$-$65.59.}
\end{subfigure}\\%
\begin{subfigure}[b]{0.5\linewidth}
\includegraphics[width=\linewidth]{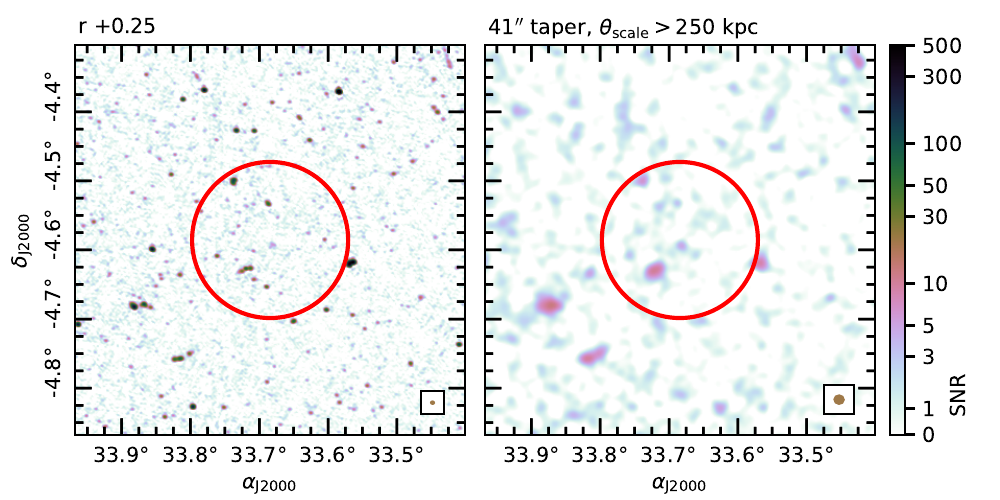}
\caption{\label{fig:app:PSZ2G167.98-59.95} PSZ2 G167.98$-$59.95.}
\end{subfigure}%
\begin{subfigure}[b]{0.5\linewidth}
\includegraphics[width=\linewidth]{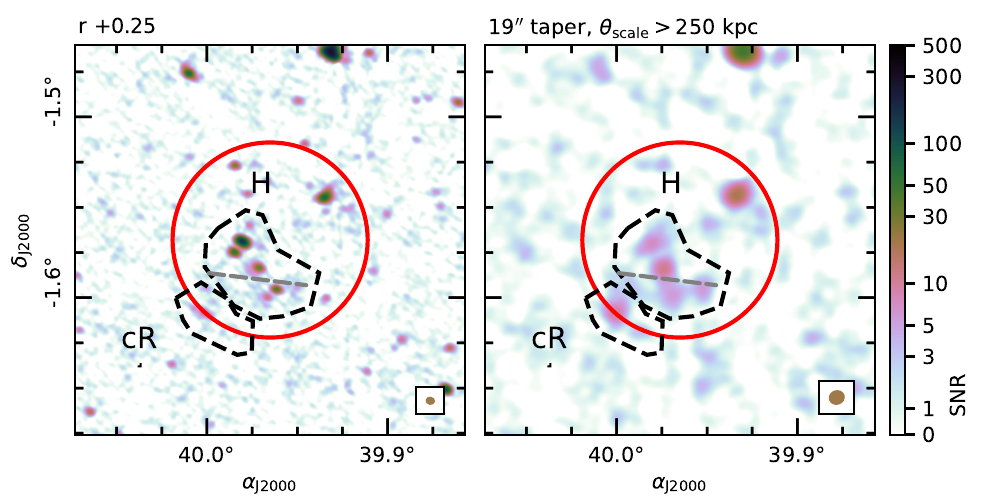}
\caption{\label{fig:app:PSZ2G172.98-53.55} \hyperref[sec:PSZ2G172.98-53.55]{PSZ2 G172.98$-$53.55}.}
\end{subfigure}\\%
\begin{subfigure}[b]{0.5\linewidth}
\includegraphics[width=\linewidth]{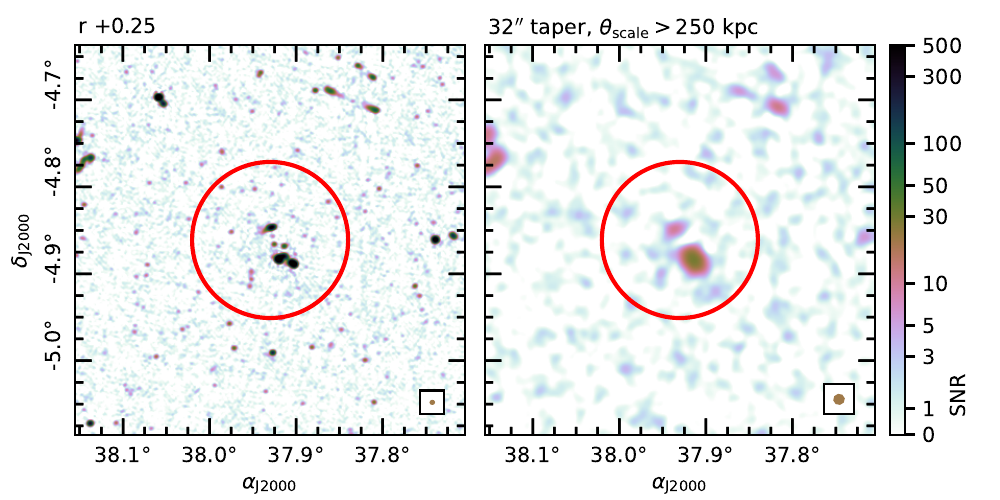}
\caption{\label{fig:app:PSZ2G174.40-57.33} PSZ2 G174.40$-$57.33.}
\end{subfigure}%
\begin{subfigure}[b]{0.5\linewidth}
\includegraphics[width=\linewidth]{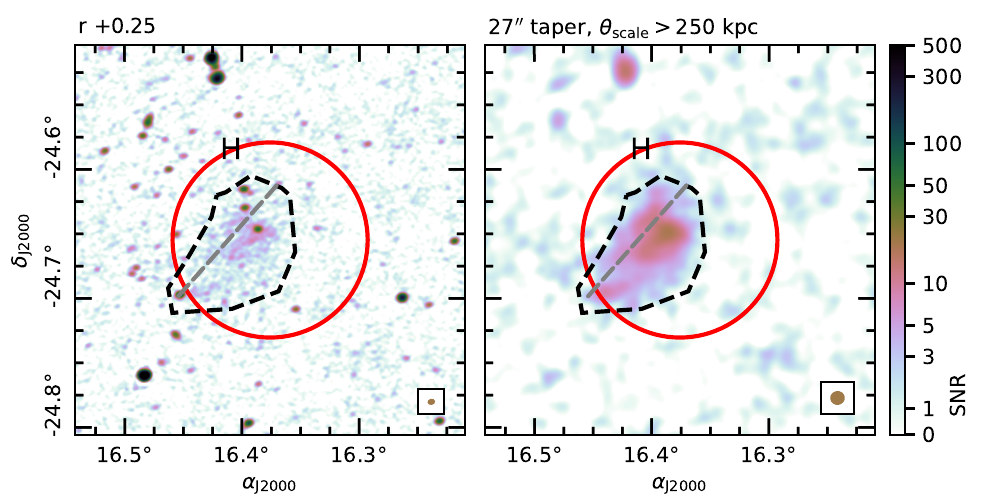}
\caption{\label{fig:app:PSZ2G175.69-85.98} \hyperref[sec:PSZ2G175.69-85.98]{PSZ2 G175.69$-$85.98}.}
\end{subfigure}\\%
\caption{(Continued).}
\end{figure*}

\begin{figure*}
\centering
\ContinuedFloat
\begin{subfigure}[b]{0.5\linewidth}
\includegraphics[width=\linewidth]{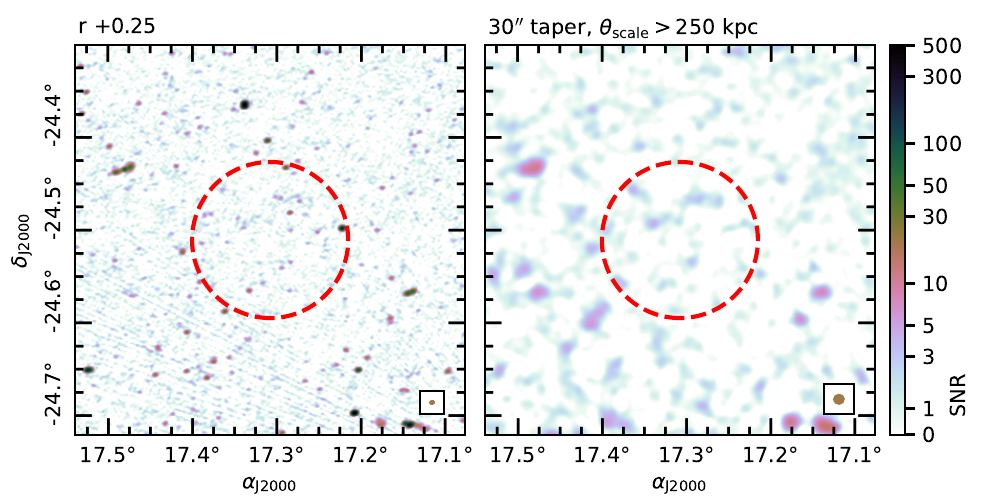}
\caption{\label{fig:app:PSZ2G180.74-85.21} PSZ2 G180.74$-$85.21.}
\end{subfigure}%
\begin{subfigure}[b]{0.5\linewidth}
\includegraphics[width=\linewidth]{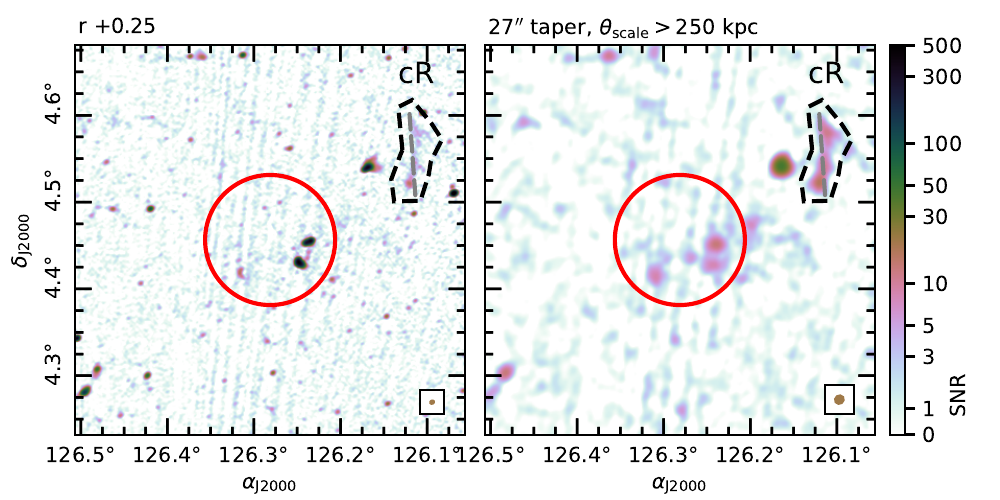}
\caption{\label{fig:app:PSZ2G219.88+22.83} \hyperref[sec:PSZ2G219.88+22.83]{PSZ2 G219.88$+$22.83}.}
\end{subfigure}\\%
\begin{subfigure}[b]{0.5\linewidth}
\includegraphics[width=\linewidth]{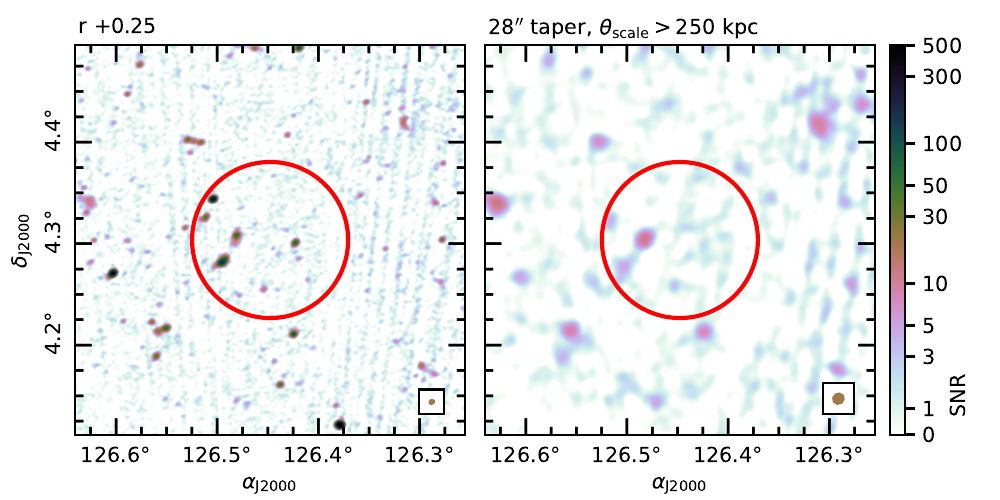}
\caption{\label{fig:app:PSZ2G220.11+22.91} PSZ2 G220.11$+$22.91.}
\end{subfigure}%
\begin{subfigure}[b]{0.5\linewidth}
\includegraphics[width=\linewidth]{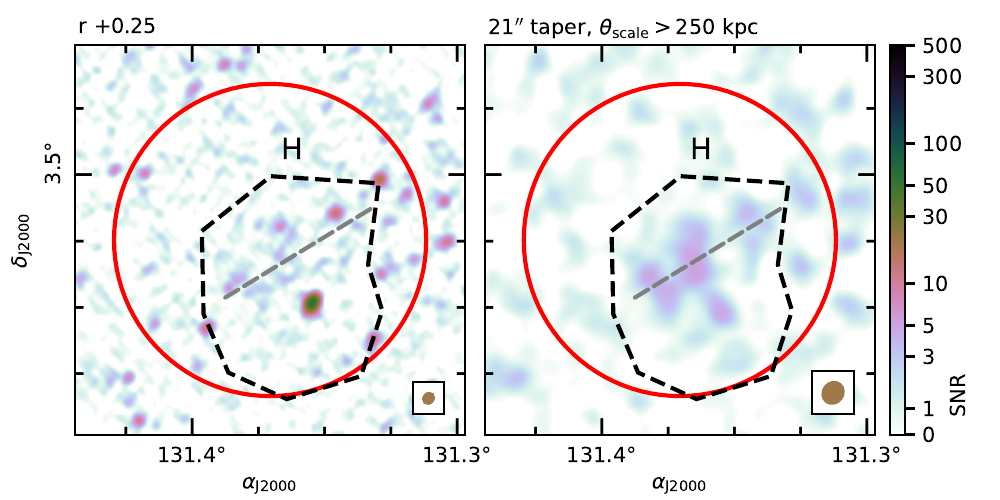}
\caption{\label{fig:app:PSZ2G223.47+26.85} \hyperref[sec:PSZ2G223.47+26.85]{PSZ2 G223.47$+$26.85}.}
\end{subfigure}\\%
\begin{subfigure}[b]{0.5\linewidth}
\includegraphics[width=\linewidth]{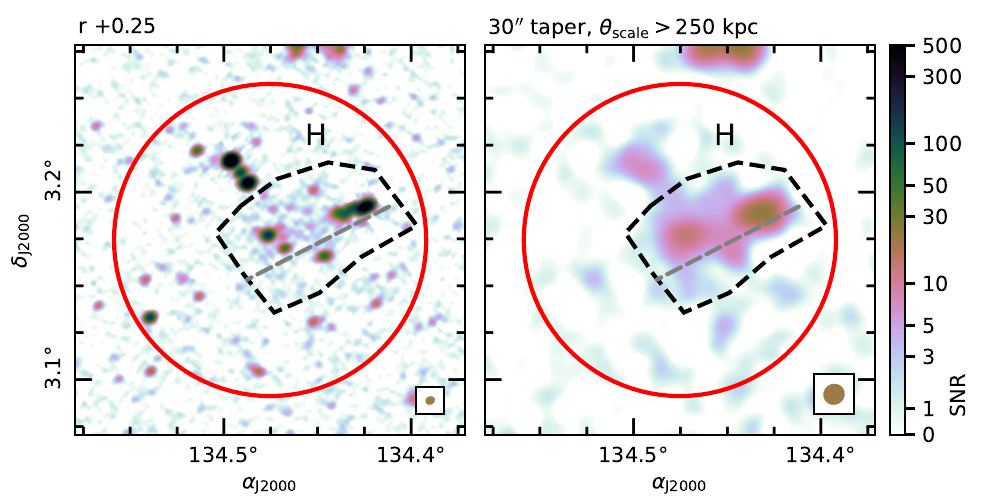}
\caption{\label{fig:app:PSZ2G225.48+29.41} \hyperref[sec:PSZ2G225.48+29.41]{PSZ2 G225.48$+$29.41}.}
\end{subfigure}%
\begin{subfigure}[b]{0.5\linewidth}
\includegraphics[width=\linewidth]{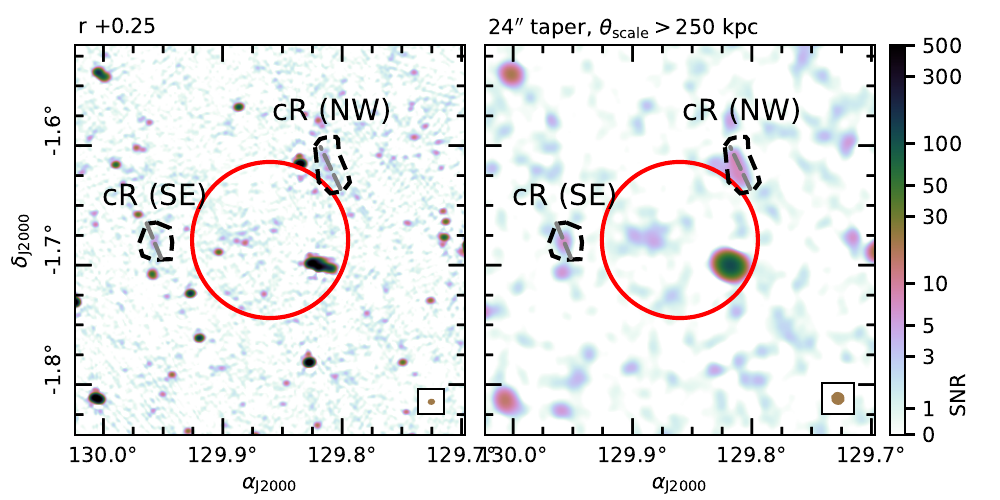}
\caption{\label{fig:app:PSZ2G227.59+22.98} \hyperref[sec:PSZ2G227.59+22.98]{PSZ2 G227.59$+$22.98}.}
\end{subfigure}\\%
\begin{subfigure}[b]{0.5\linewidth}
\includegraphics[width=\linewidth]{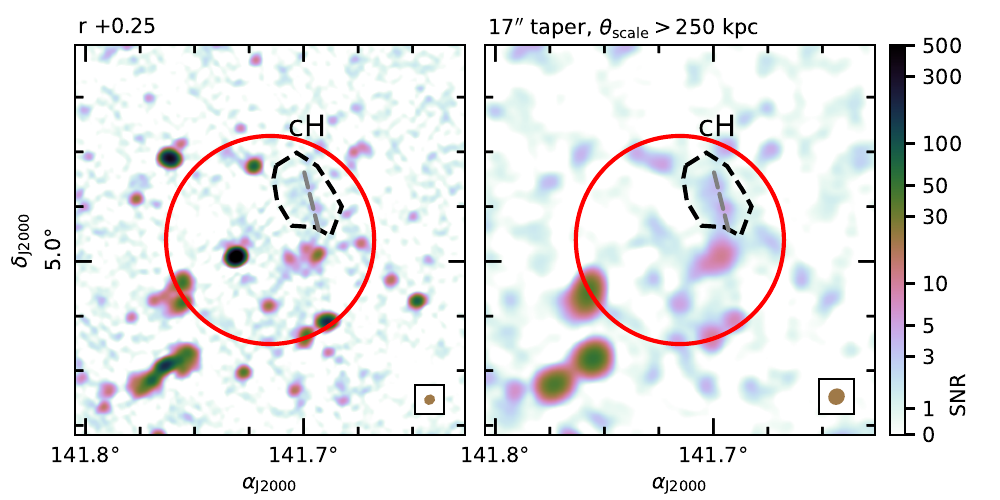}
\caption{\label{fig:app:PSZ2G227.89+36.58} \hyperref[sec:PSZ2G227.89+36.58]{PSZ2 G227.89$+$36.58}.}
\end{subfigure}%
\begin{subfigure}[b]{0.5\linewidth}
\includegraphics[width=\linewidth]{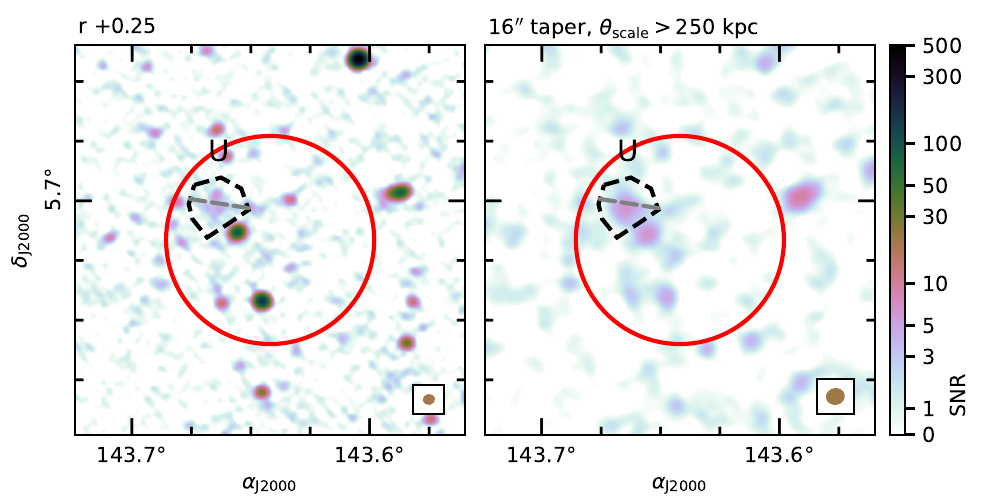}
\caption{\label{fig:app:PSZ2G228.38+38.58} \hyperref[sec:PSZ2G228.38+38.58]{PSZ2 G228.38$+$38.58}.}
\end{subfigure}\\%
\caption{(Continued).}
\end{figure*}

\begin{figure*}
\centering
\ContinuedFloat
\begin{subfigure}[b]{0.5\linewidth}
\includegraphics[width=\linewidth]{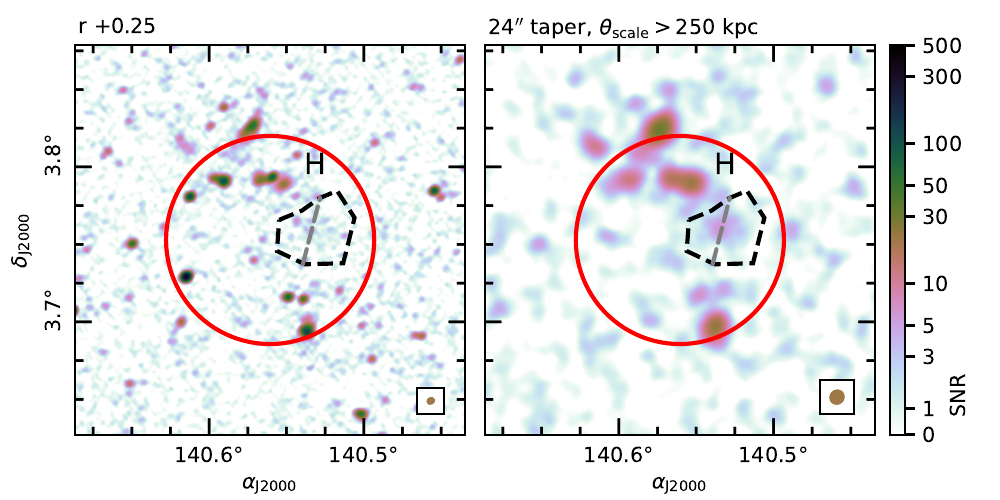}
\caption{\label{fig:app:PSZ2G228.50+34.95} \hyperref[sec:PSZ2G228.50+34.95]{PSZ2 G228.50$+$34.95}.}
\end{subfigure}%
\begin{subfigure}[b]{0.5\linewidth}
\includegraphics[width=\linewidth]{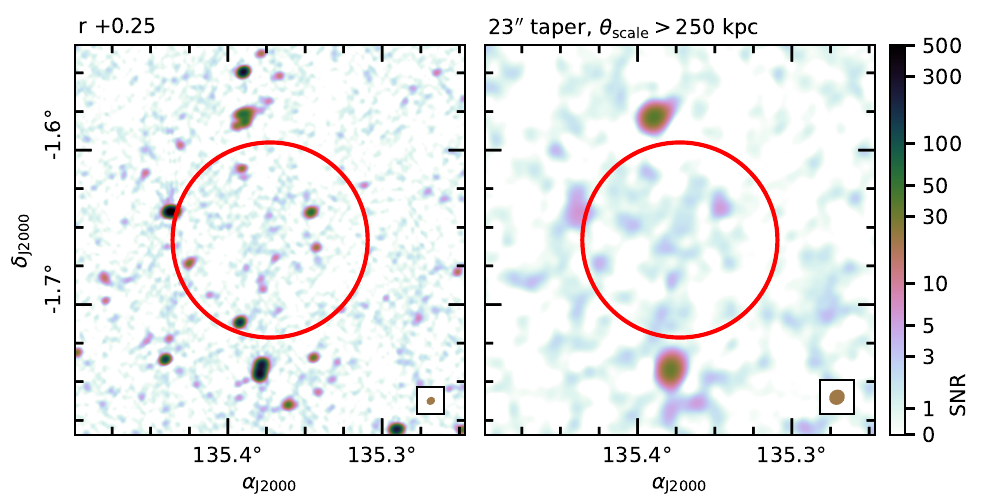}
\caption{\label{fig:app:PSZ2G230.73+27.70} PSZ2 G230.73$+$27.70.}
\end{subfigure}\\%
\begin{subfigure}[b]{0.5\linewidth}
\includegraphics[width=\linewidth]{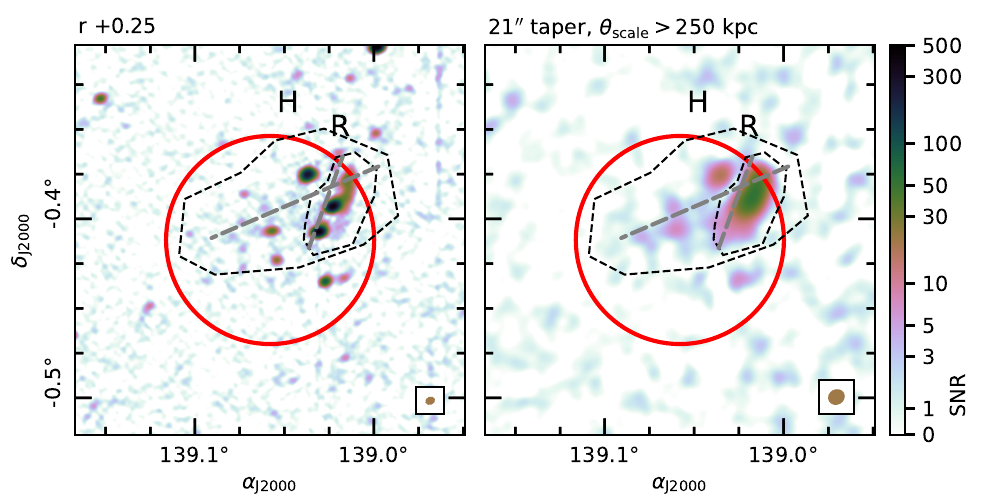}
\caption{\label{fig:app:PSZ2G231.79+31.48} \hyperref[sec:PSZ2G231.79+31.48]{PSZ2 G231.79$+$31.48}.}
\end{subfigure}%
\begin{subfigure}[b]{0.5\linewidth}
\includegraphics[width=\linewidth]{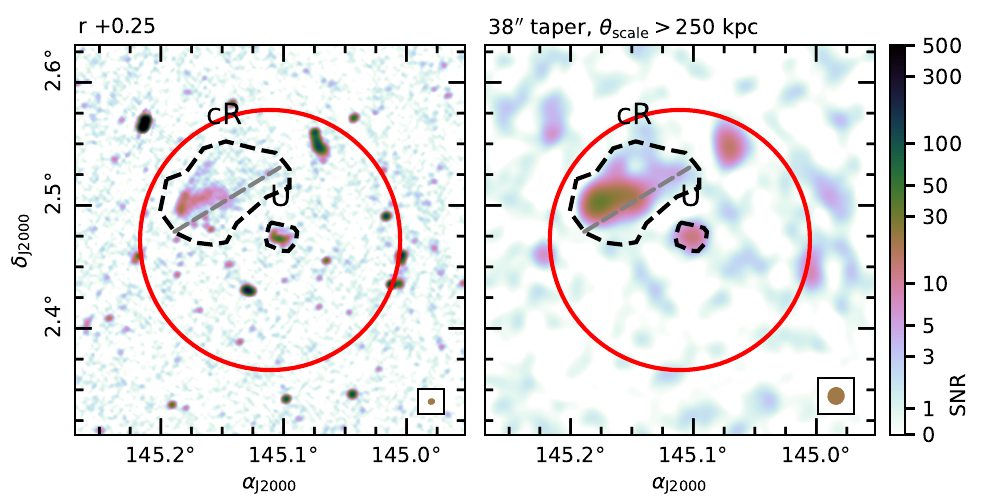}
\caption{\label{fig:app:PSZ2G232.84+38.13} \hyperref[sec:PSZ2G232.84+38.13]{PSZ2 G232.84$+$38.13}.}
\end{subfigure}\\%
\begin{subfigure}[b]{0.5\linewidth}
\includegraphics[width=\linewidth]{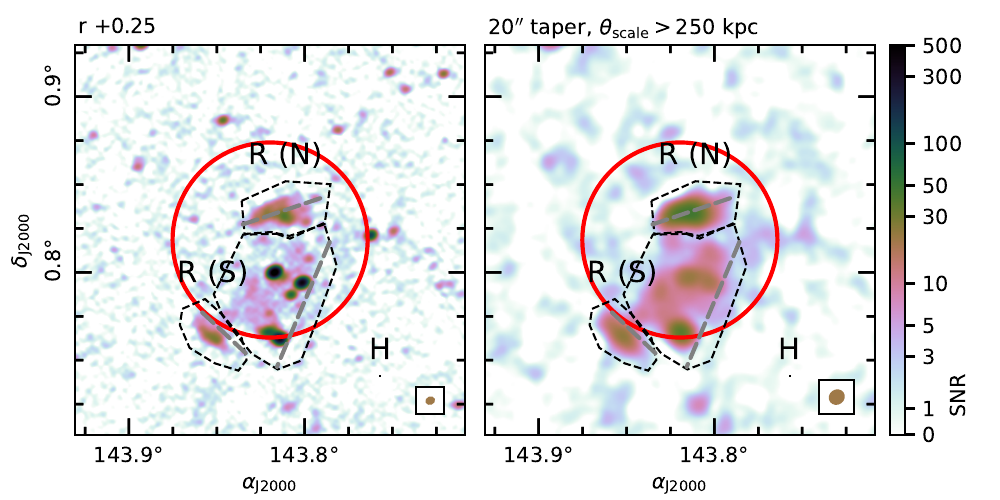}
\caption{\label{fig:app:PSZ2G233.68+36.14} \hyperref[sec:PSZ2G233.68+36.14]{PSZ2 G233.68$+$36.14}.}
\end{subfigure}%
\begin{subfigure}[b]{0.5\linewidth}
\includegraphics[width=\linewidth]{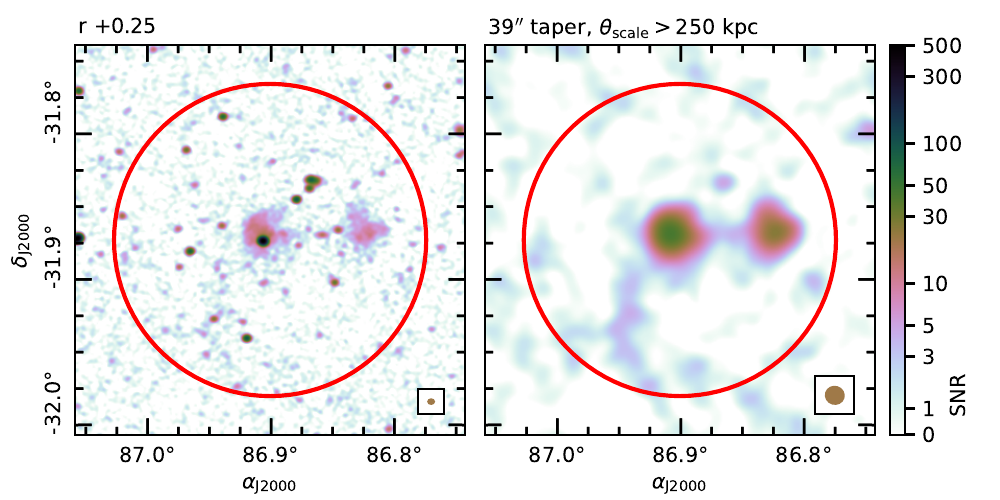}
\caption{\label{fig:app:PSZ2G236.92-26.65} \hyperref[sec:PSZ2G236.92-26.65]{PSZ2 G236.92$-$26.65}.}
\end{subfigure}\\%
\begin{subfigure}[b]{0.5\linewidth}
\includegraphics[width=\linewidth]{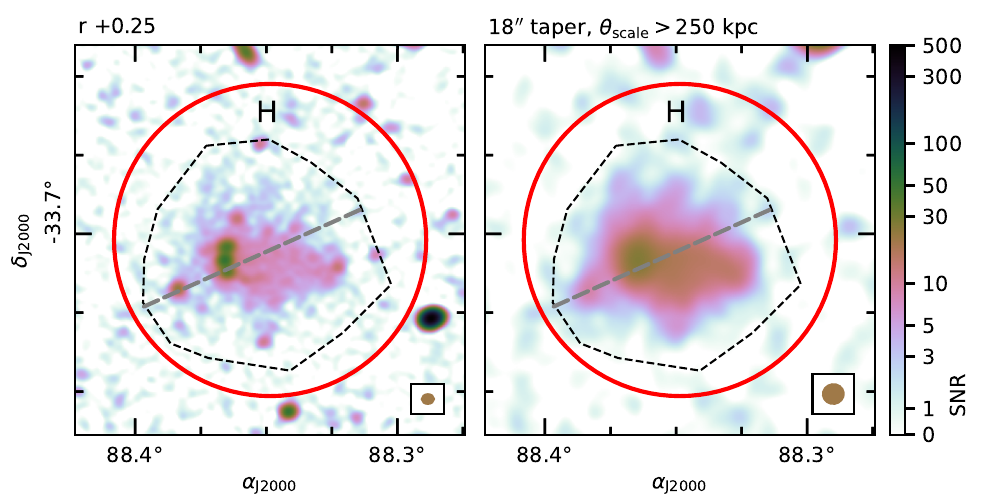}
\caption{\label{fig:app:PSZ2G239.27-26.01} \hyperref[sec:PSZ2G239.27-26.01]{PSZ2 G239.27$-$26.01}.}
\end{subfigure}%
\begin{subfigure}[b]{0.5\linewidth}
\includegraphics[width=\linewidth]{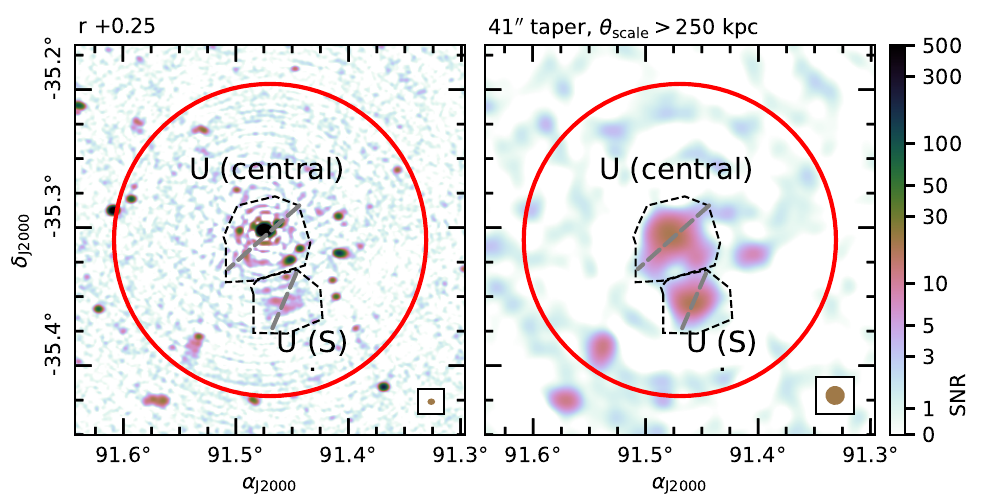}
\caption{\label{fig:app:PSZ2G241.79-24.01} \hyperref[sec:PSZ2G241.79-24.01]{PSZ2 G241.79$-$24.01}.}
\end{subfigure}\\%
\caption{(Continued).}
\end{figure*}

\begin{figure*}
\centering
\ContinuedFloat
\begin{subfigure}[b]{0.5\linewidth}
\includegraphics[width=\linewidth]{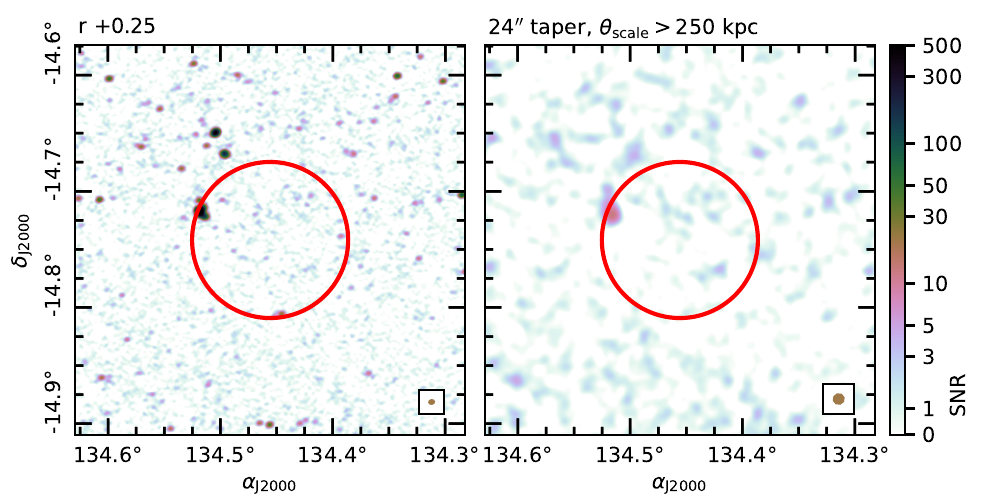}
\caption{\label{fig:app:PSZ2G241.98+19.56} PSZ2 G241.98$+$19.56.}
\end{subfigure}%
\begin{subfigure}[b]{0.5\linewidth}
\includegraphics[width=\linewidth]{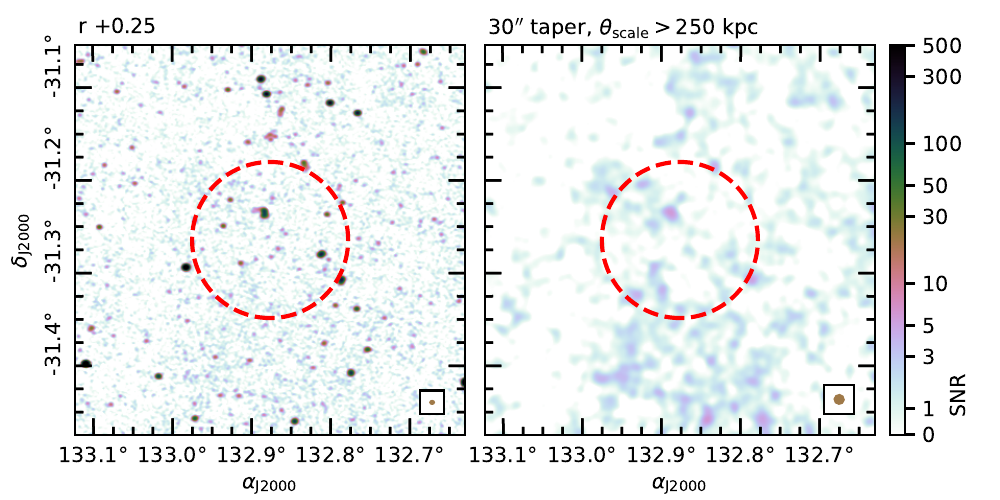}
\caption{\label{fig:app:PSZ2G254.52+08.27} PSZ2 G254.52$+$08.27.}
\end{subfigure}\\%
\begin{subfigure}[b]{0.5\linewidth}
\includegraphics[width=\linewidth]{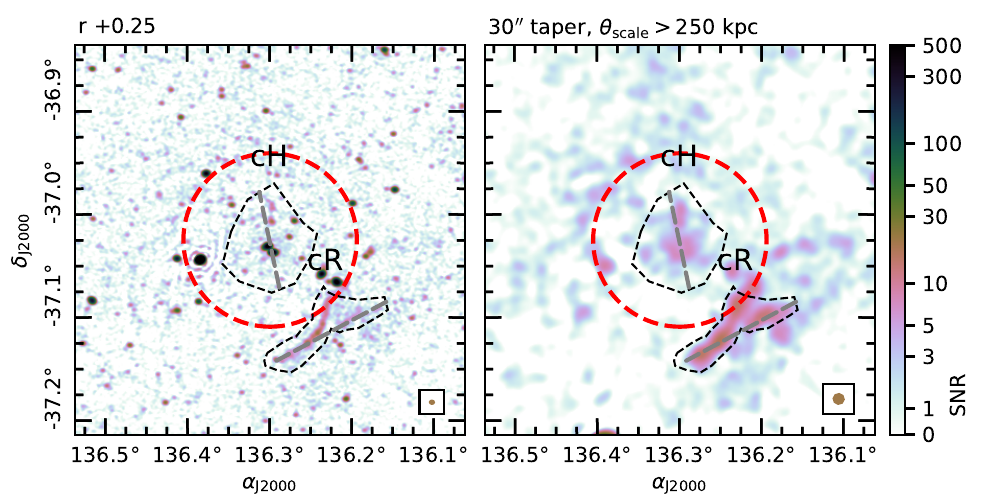}
\caption{\label{fig:app:PSZ2G260.80+06.71} \hyperref[sec:PSZ2G260.80+06.71]{PSZ2 G260.80$+$06.71}.}
\end{subfigure}%
\begin{subfigure}[b]{0.5\linewidth}
\includegraphics[width=\linewidth]{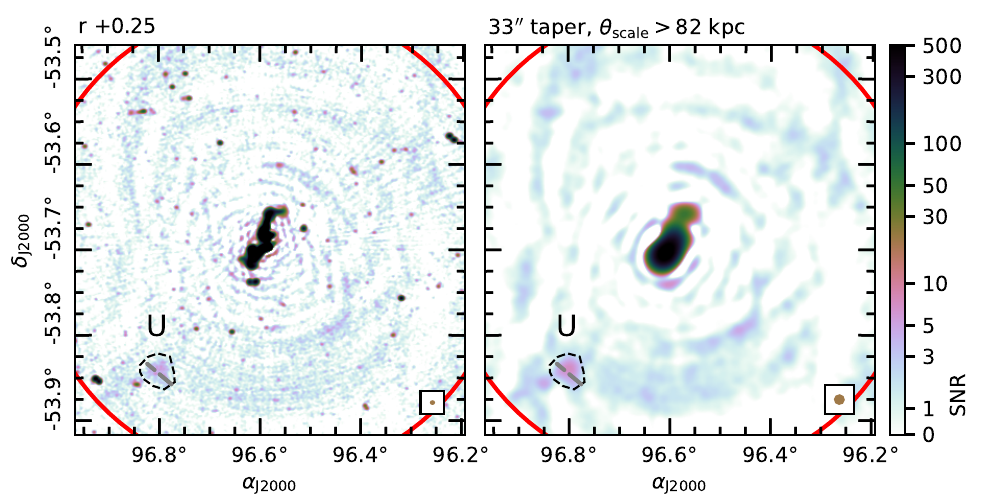}
\caption{\label{fig:app:PSZ2G262.36-25.15} \hyperref[sec:PSZ2G262.36-25.15]{PSZ2 G262.36$-$25.15}.}
\end{subfigure}\\%
\begin{subfigure}[b]{0.5\linewidth}
\includegraphics[width=\linewidth]{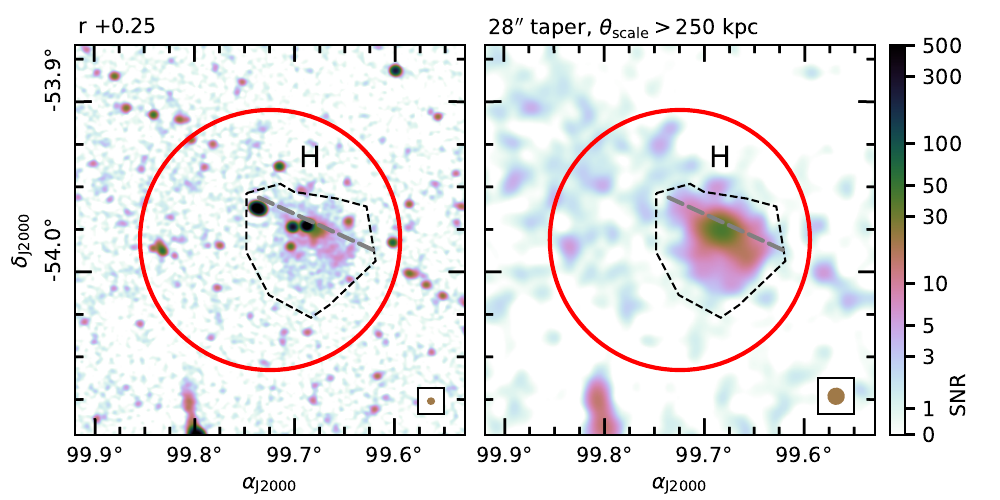}
\caption{\label{fig:app:PSZ2G263.14-23.41} \hyperref[sec:PSZ2G263.14-23.41]{PSZ2 G263.14$-$23.41}.}
\end{subfigure}%
\begin{subfigure}[b]{0.5\linewidth}
\includegraphics[width=\linewidth]{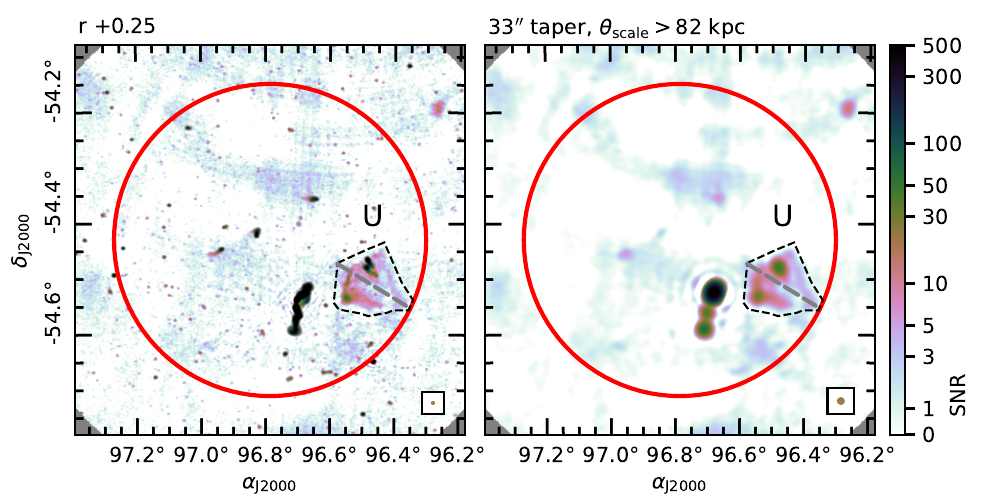}
\caption{\label{fig:app:PSZ2G263.19-25.19} \hyperref[sec:PSZ2G263.19-25.19]{PSZ2 G263.19$-$25.19}.}
\end{subfigure}\\%
\begin{subfigure}[b]{0.5\linewidth}
\includegraphics[width=\linewidth]{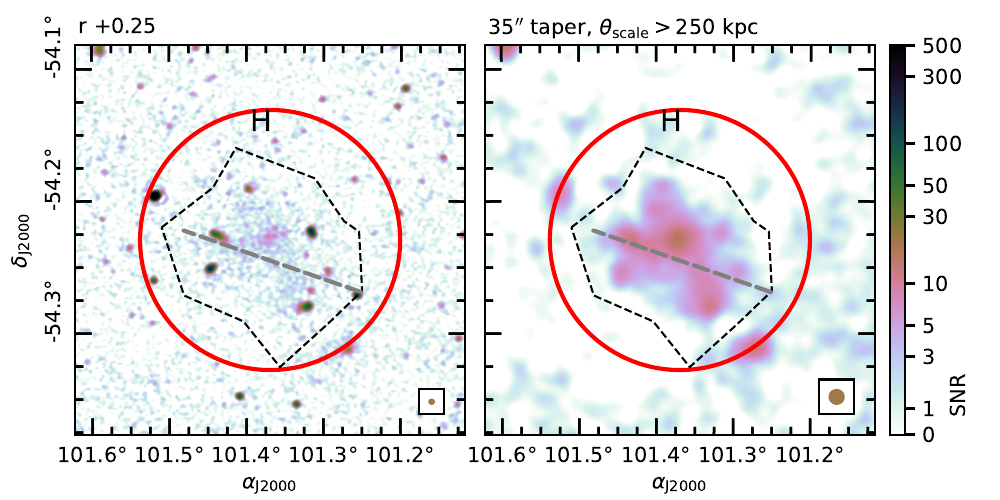}
\caption{\label{fig:app:PSZ2G263.68-22.55} \hyperref[sec:PSZ2G263.68-22.55]{PSZ2 G263.68$-$22.55}.}
\end{subfigure}%
\begin{subfigure}[b]{0.5\linewidth}
\includegraphics[width=\linewidth]{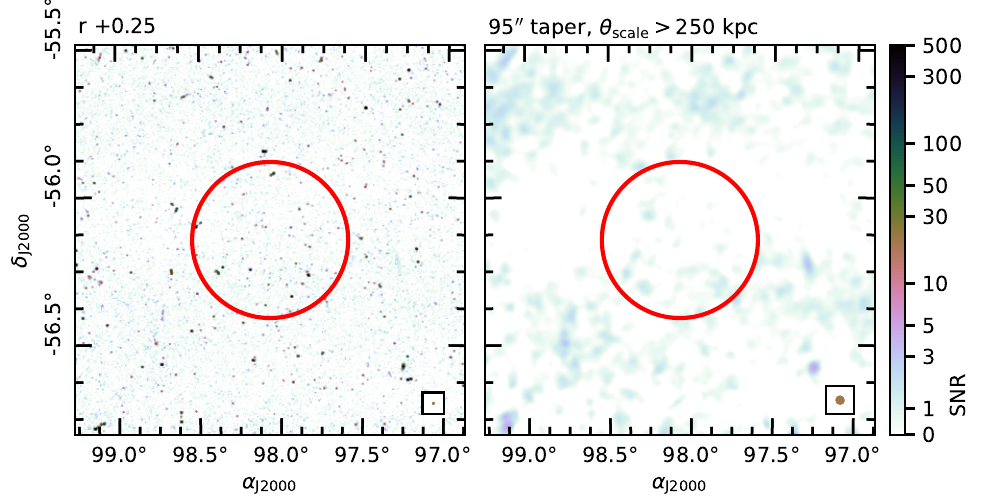}
\caption{\label{fig:app:PSZ2G265.21-24.83} PSZ2 G265.21$-$24.83.}
\end{subfigure}\\%
\caption{(Continued).}
\end{figure*}

\begin{figure*}
\centering
\ContinuedFloat
\begin{subfigure}[b]{0.5\linewidth}
\includegraphics[width=\linewidth]{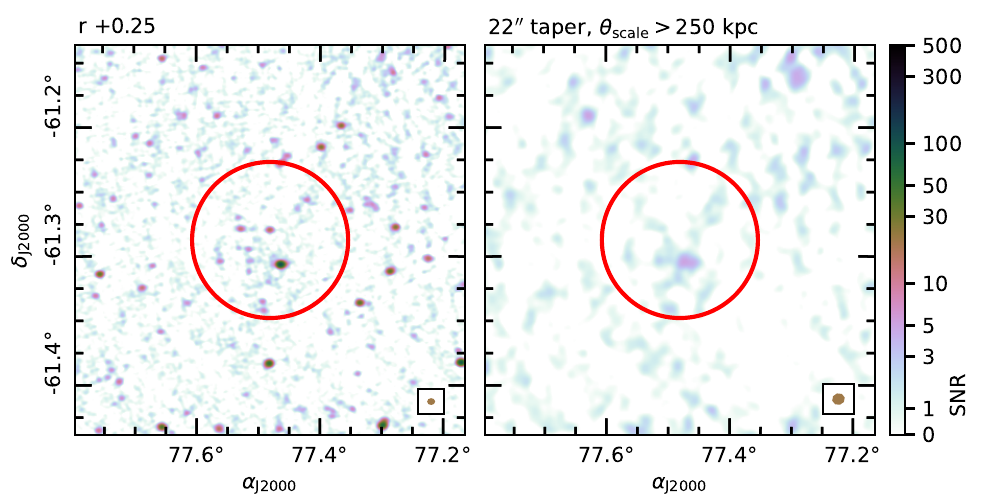}
\caption{\label{fig:app:PSZ2G270.63-35.67} PSZ2 G270.63$-$35.67.}
\end{subfigure}%
\begin{subfigure}[b]{0.5\linewidth}
\includegraphics[width=\linewidth]{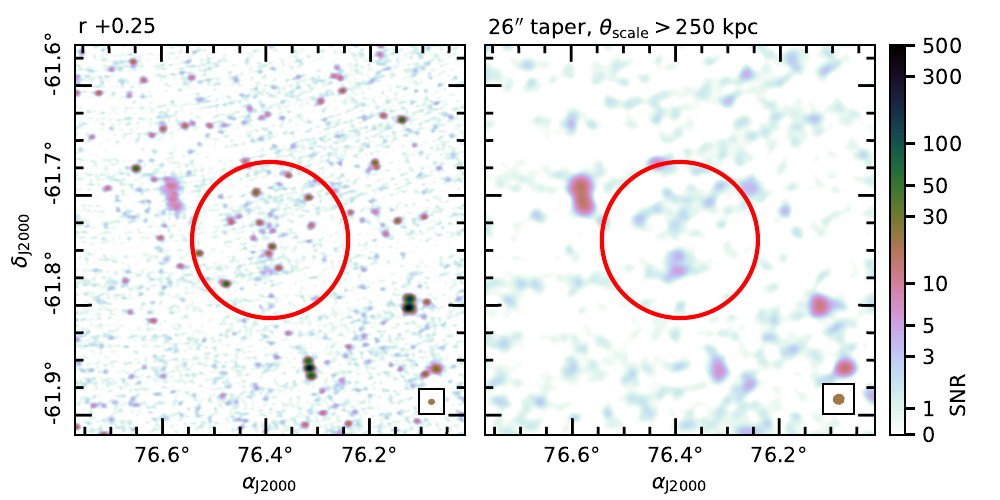}
\caption{\label{fig:app:PSZ2G271.28-36.11} PSZ2 G271.28$-$36.11.}
\end{subfigure}\\%
\begin{subfigure}[b]{0.5\linewidth}
\includegraphics[width=\linewidth]{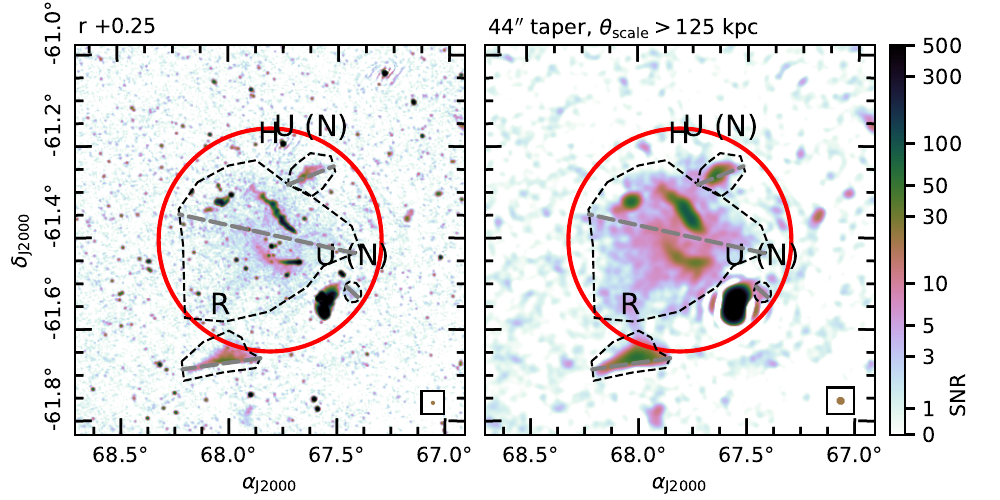}
\caption{\label{fig:app:PSZ2G272.08-40.16} \hyperref[sec:PSZ2 G272.08-40.16]{PSZ2 G272.08$-$40.16}.}
\end{subfigure}%
\begin{subfigure}[b]{0.5\linewidth}
\includegraphics[width=\linewidth]{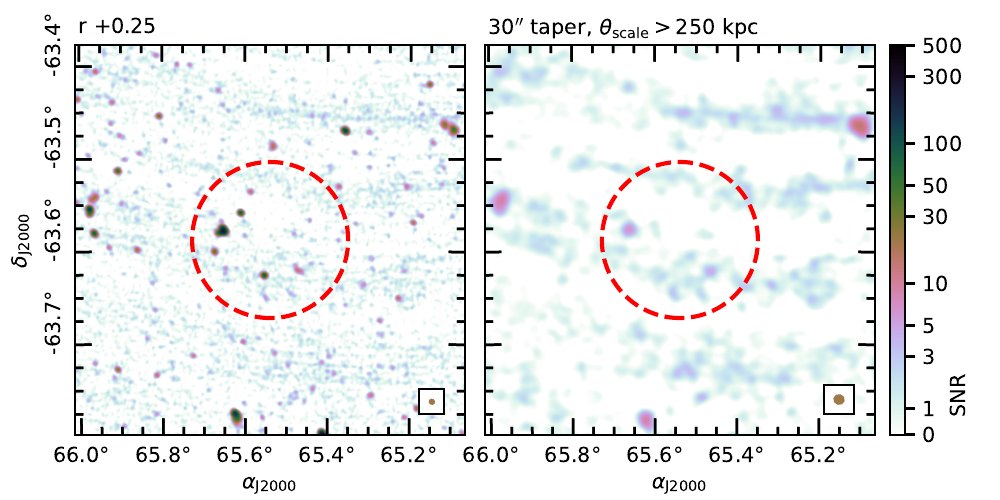}
\caption{\label{fig:app:PSZ2G275.24-40.42} PSZ2 G275.24$-$40.42.}
\end{subfigure}\\%
\begin{subfigure}[b]{0.5\linewidth}
\includegraphics[width=\linewidth]{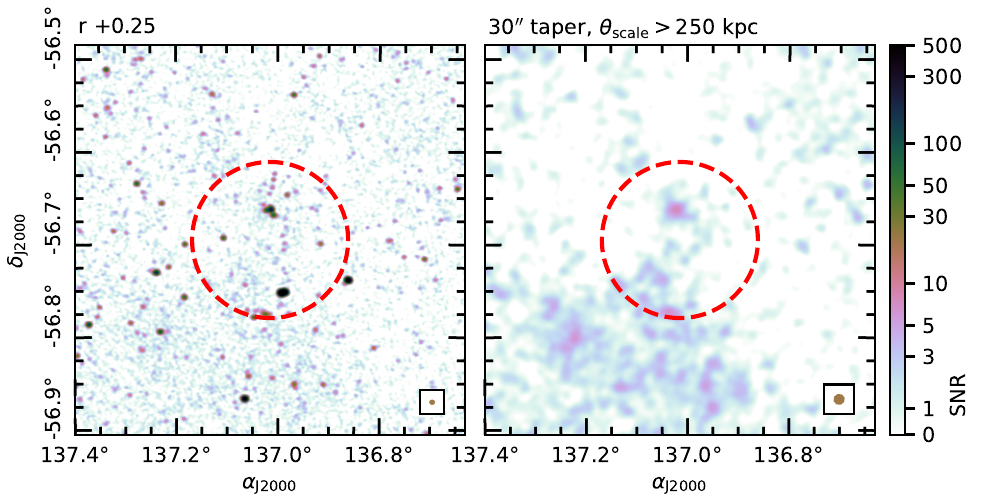}
\caption{\label{fig:app:PSZ2G275.73-06.12} PSZ2 G275.73$-$06.12.}
\end{subfigure}%
\begin{subfigure}[b]{0.5\linewidth}
\includegraphics[width=\linewidth]{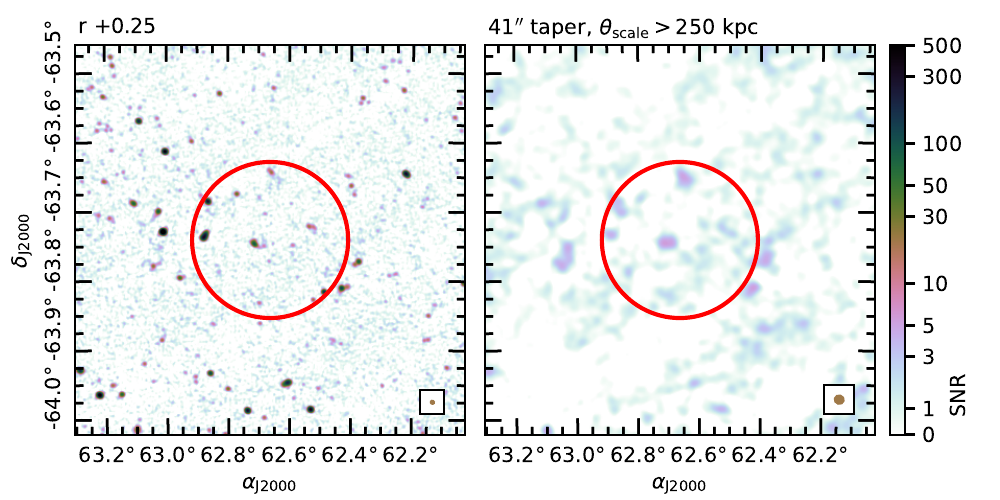}
\caption{\label{fig:app:PSZ2G276.09-41.53} PSZ2 G276.09$-$41.53.}
\end{subfigure}\\%
\begin{subfigure}[b]{0.5\linewidth}
\includegraphics[width=\linewidth]{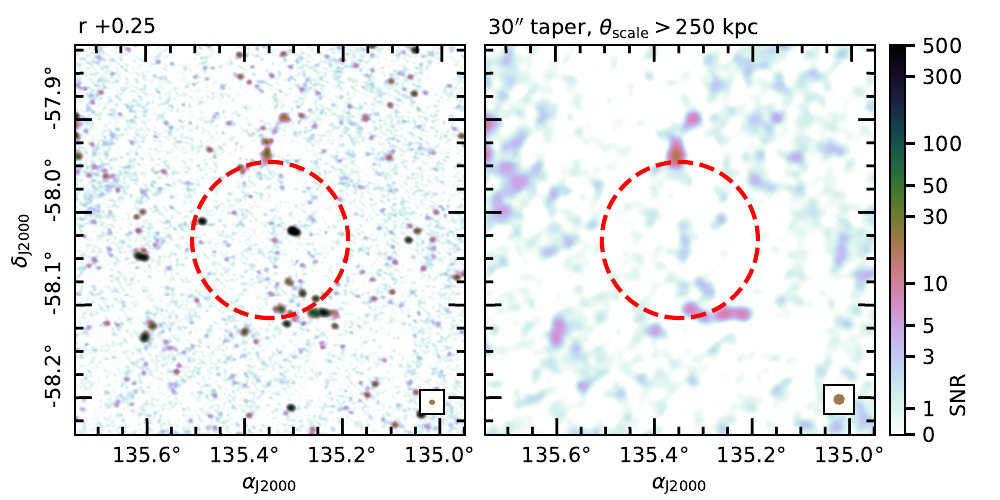}
\caption{\label{fig:app:PSZ2G276.14-07.68} PSZ2 G276.14$-$07.68.}
\end{subfigure}%
\begin{subfigure}[b]{0.5\linewidth}
\includegraphics[width=\linewidth]{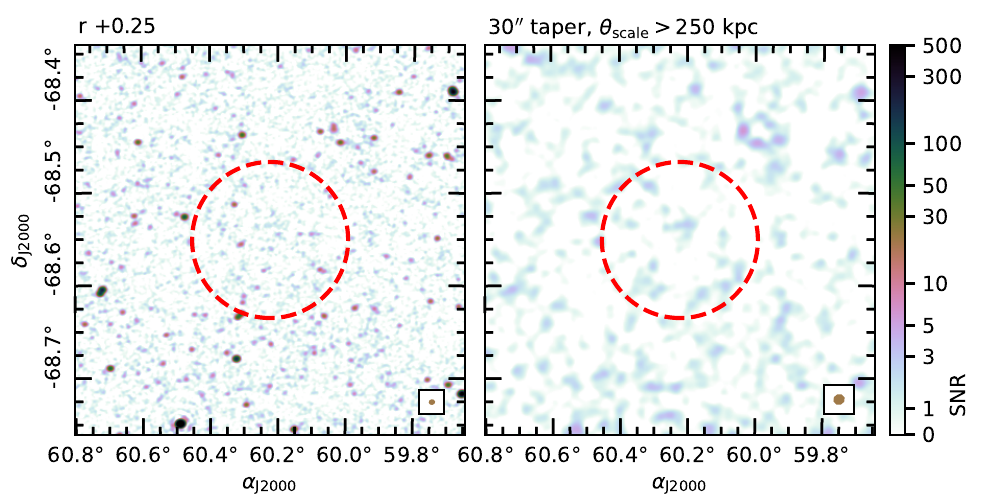}
\caption{\label{fig:app:PSZ2G282.32-40.15} PSZ2 G282.32$-$40.15.}
\end{subfigure}\\%
\caption{(Continued).}
\end{figure*}

\begin{figure*}
\centering
\ContinuedFloat
\begin{subfigure}[b]{0.5\linewidth}
\includegraphics[width=\linewidth]{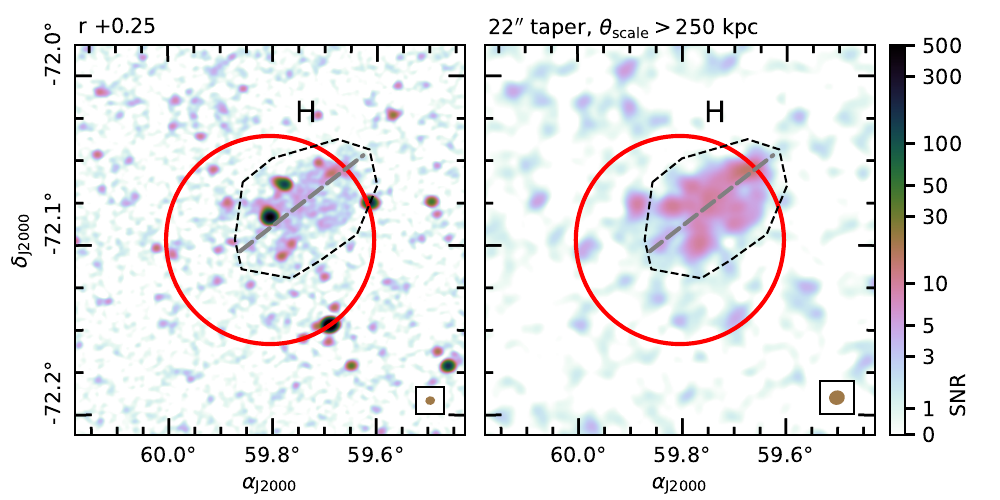}
\caption{\label{fig:app:PSZ2G286.28-38.36} \hyperref[sec:PSZ2G286.28-38.36]{PSZ2 G286.28$-$38.36}.}
\end{subfigure}%
\begin{subfigure}[b]{0.5\linewidth}
\includegraphics[width=\linewidth]{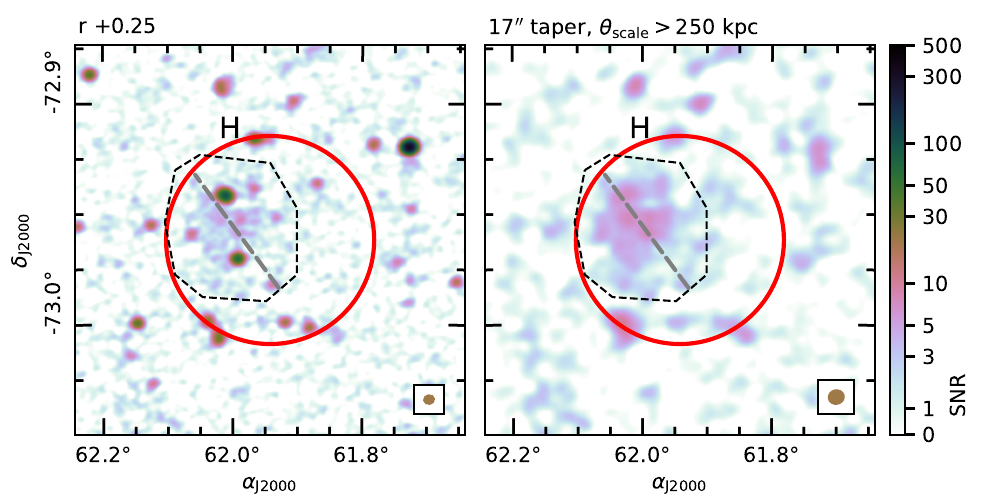}
\caption{\label{fig:app:PSZ2G286.75-37.35} \hyperref[sec:PSZ2G286.75-37.35]{PSZ2 G286.75$-$37.35}.}
\end{subfigure}\\%
\begin{subfigure}[b]{0.5\linewidth}
\includegraphics[width=\linewidth]{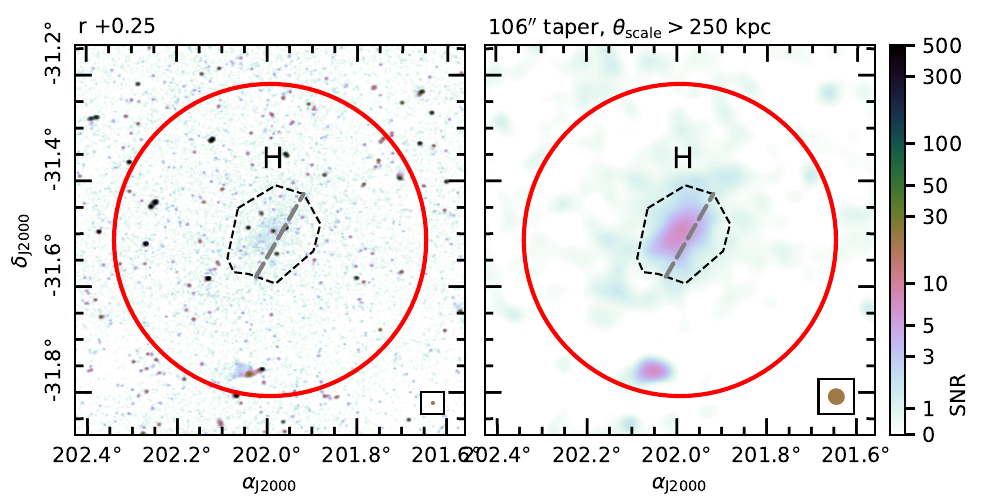}
\caption{\label{fig:app:PSZ2G311.98+30.71} \hyperref[sec:PSZ2G311.98+30.71]{PSZ2 G311.98$+$30.71}.}
\end{subfigure}%
\begin{subfigure}[b]{0.5\linewidth}
\includegraphics[width=\linewidth]{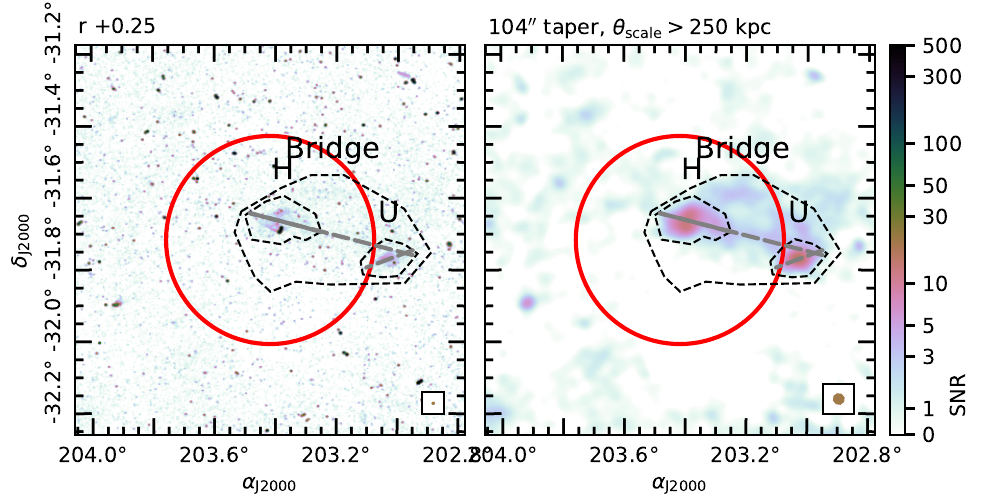}
\caption{\label{fig:app:PSZ2G313.33+30.29} \hyperref[sec:PSZ2G313.33+30.29]{PSZ2 G313.33$+$30.29}.}
\end{subfigure}\\%
\begin{subfigure}[b]{0.5\linewidth}
\includegraphics[width=\linewidth]{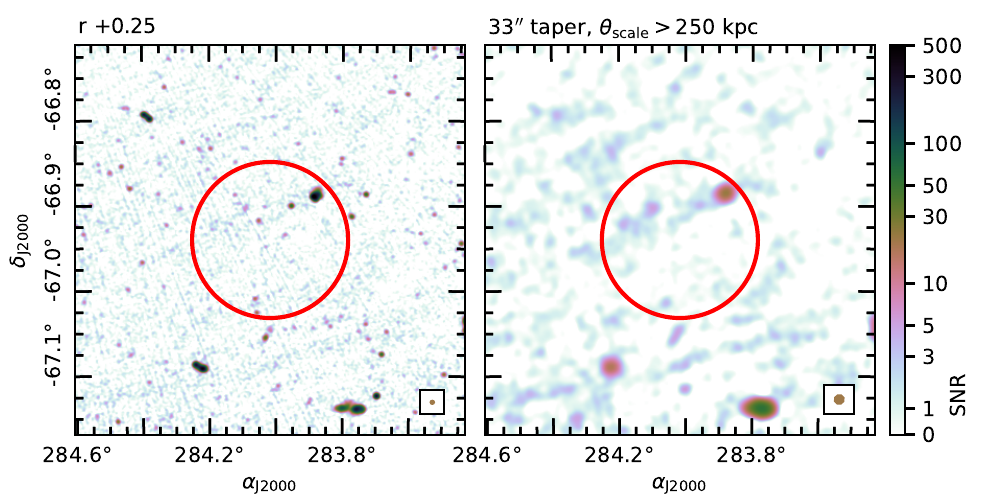}
\caption{\label{fig:app:PSZ2G328.58-25.25} PSZ2 G328.58$-$25.25.}
\end{subfigure}%
\begin{subfigure}[b]{0.5\linewidth}
\includegraphics[width=\linewidth]{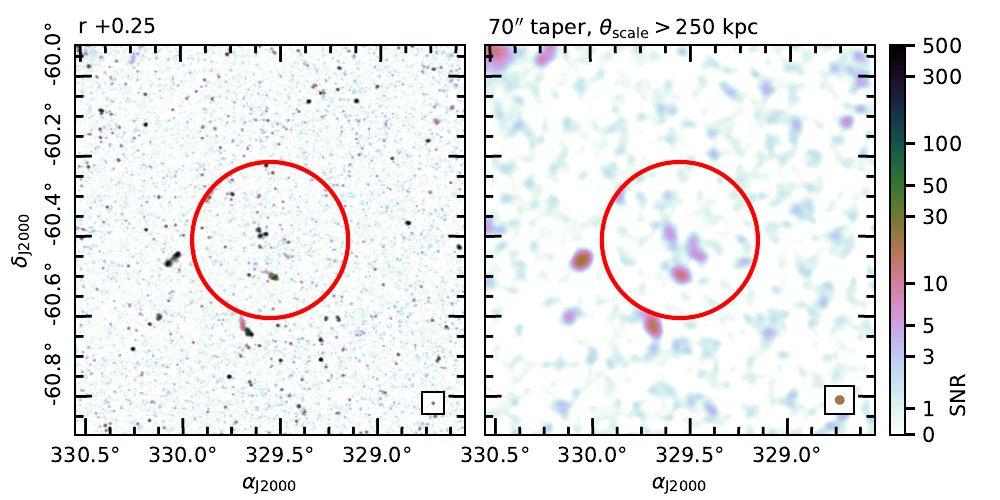}
\caption{\label{fig:app:PSZ2G331.96-45.74} PSZ2 G331.96$-$45.74.}
\end{subfigure}\\%
\begin{subfigure}[b]{0.5\linewidth}
\includegraphics[width=\linewidth]{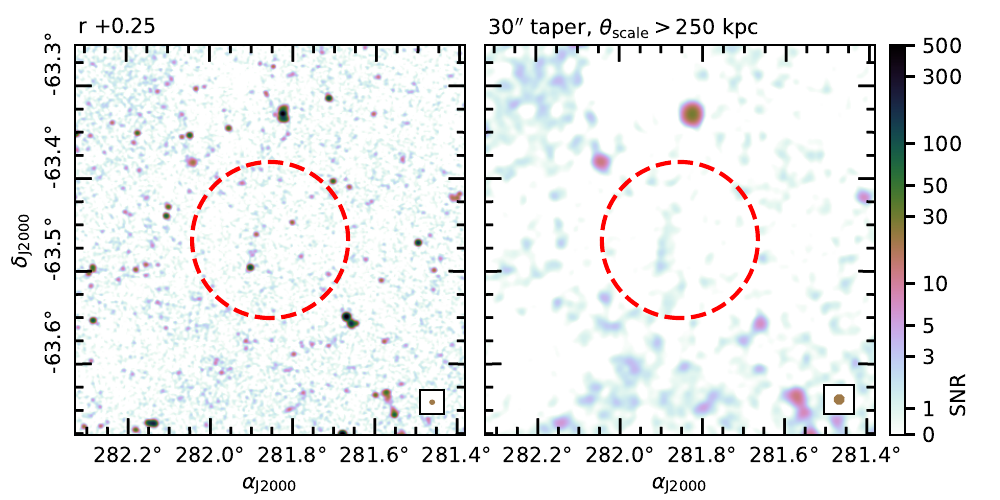}
\caption{\label{fig:app:PSZ2G332.11-23.63} PSZ2 G332.11$-$23.63.}
\end{subfigure}%
\begin{subfigure}[b]{0.5\linewidth}
\includegraphics[width=\linewidth]{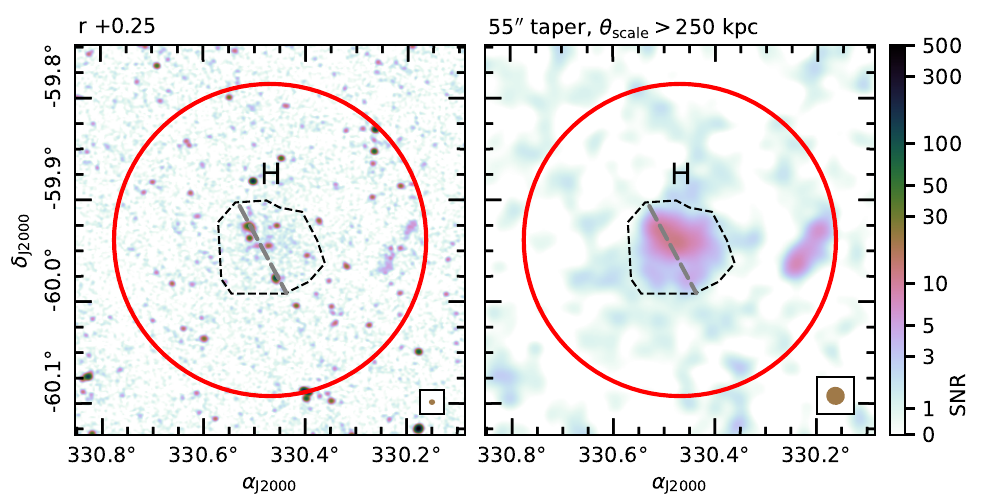}
\caption{\label{fig:app:PSZ2G332.23-46.37} \hyperref[sec:G332.23-46.37]{PSZ2 G332.23$-$46.37}.}
\end{subfigure}\\%
\caption{(Continued).}
\end{figure*}

\begin{figure*}
\centering
\ContinuedFloat
\begin{subfigure}[b]{0.5\linewidth}
\includegraphics[width=\linewidth]{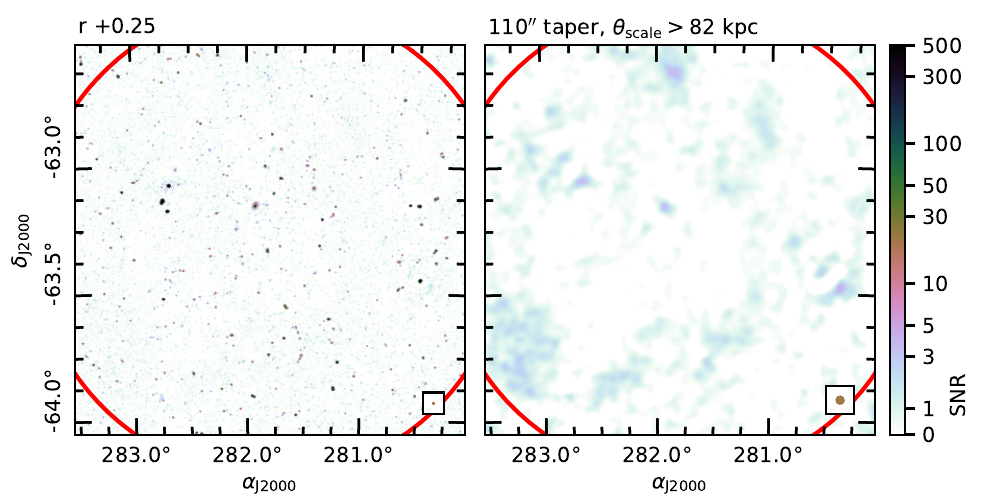}
\caption{\label{fig:app:PSZ2G332.29-23.57} PSZ2 G332.29$-$23.57.}
\end{subfigure}%
\begin{subfigure}[b]{0.5\linewidth}
\includegraphics[width=\linewidth]{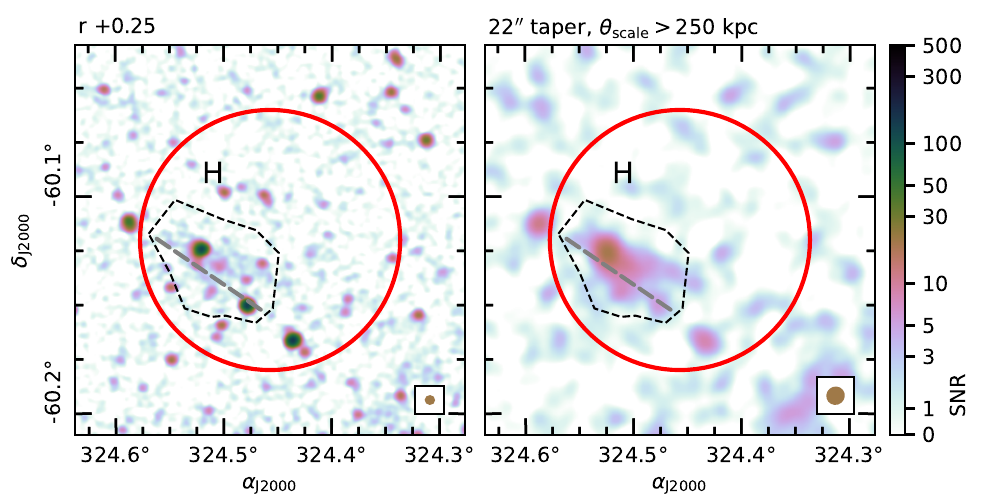}
\caption{\label{fig:app:PSZ2G333.89-43.60} \hyperref[sec:PSZ2G333.89-43.60]{PSZ2 G333.89$-$43.60}.}
\end{subfigure}\\%
\begin{subfigure}[b]{0.5\linewidth}
\includegraphics[width=\linewidth]{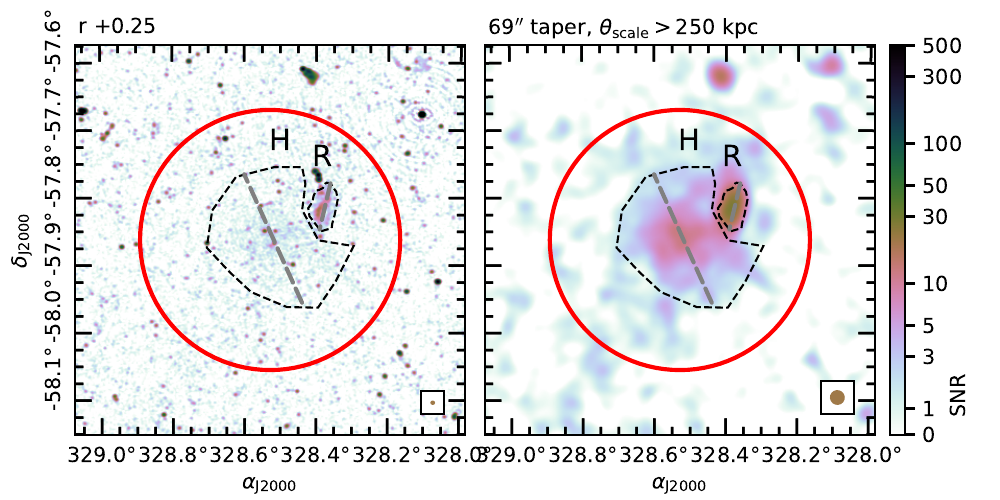}
\caption{\label{fig:app:PSZ2G335.58-46.44} \hyperref[sec:PSZ2G335.58-46.44]{PSZ2 G335.58$-$46.44}.}
\end{subfigure}%
\begin{subfigure}[b]{0.5\linewidth}
\includegraphics[width=\linewidth]{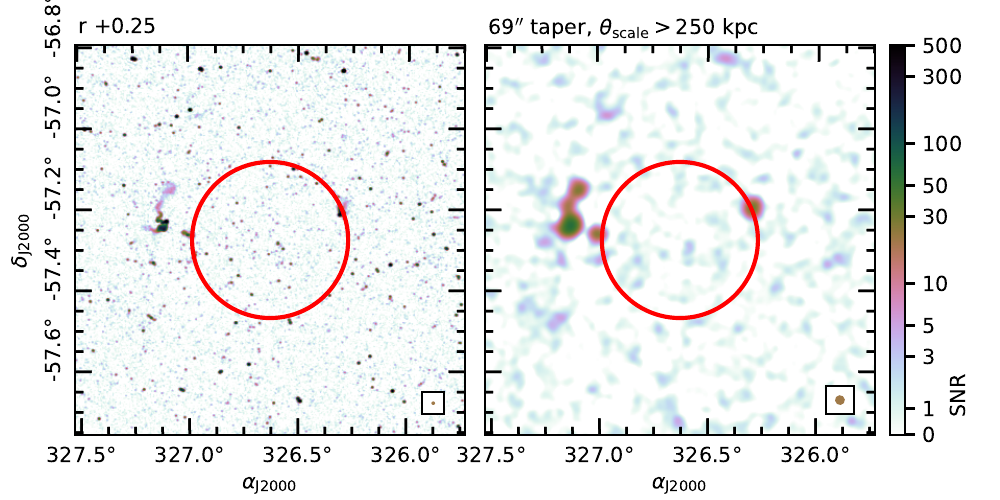}
\caption{\label{fig:app:PSZ2G336.95-45.75} PSZ2 G336.95$-$45.75.}
\end{subfigure}\\%
\begin{subfigure}[b]{0.5\linewidth}
\includegraphics[width=\linewidth]{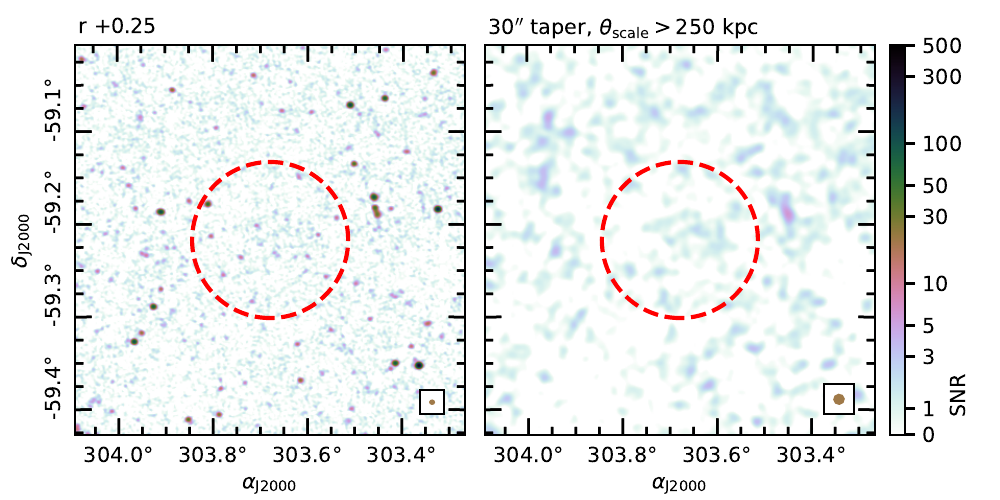}
\caption{\label{fig:app:PSZ2G337.99-33.61} PSZ2 G337.99$-$33.61.}
\end{subfigure}%
\begin{subfigure}[b]{0.5\linewidth}
\includegraphics[width=\linewidth]{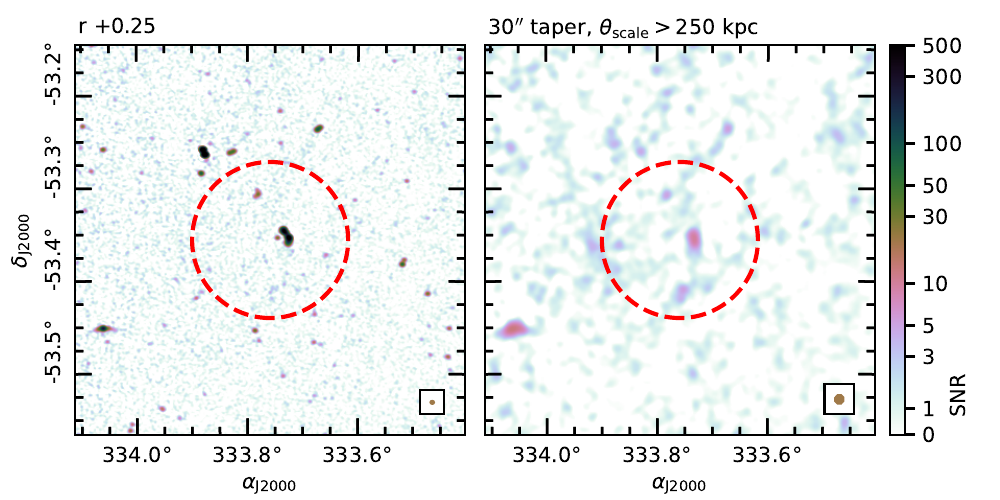}
\caption{\label{fig:app:PSZ2G339.74-51.08} PSZ2 G339.74$-$51.08.}
\end{subfigure}\\%
\begin{subfigure}[b]{0.5\linewidth}
\includegraphics[width=\linewidth]{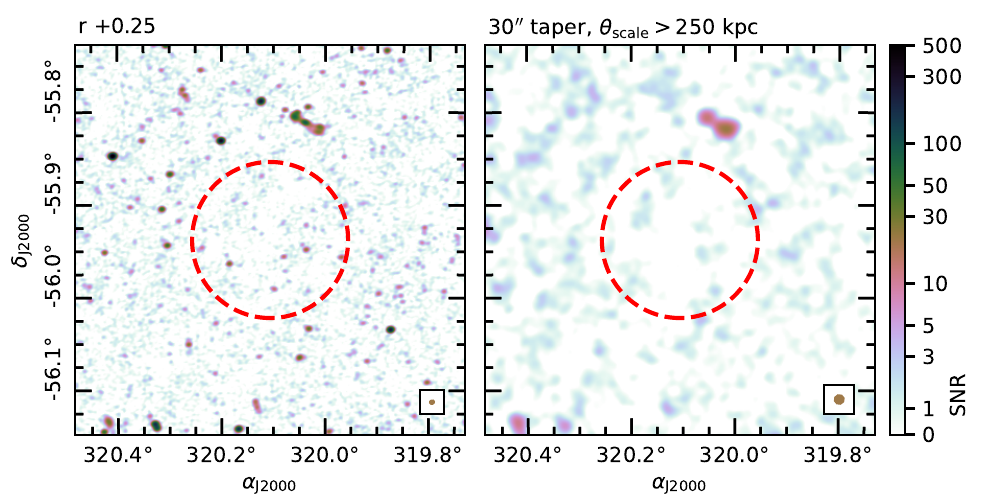}
\caption{\label{fig:app:PSZ2G340.35-42.80} PSZ2 G340.35$-$42.80.}
\end{subfigure}%
\begin{subfigure}[b]{0.5\linewidth}
\includegraphics[width=\linewidth]{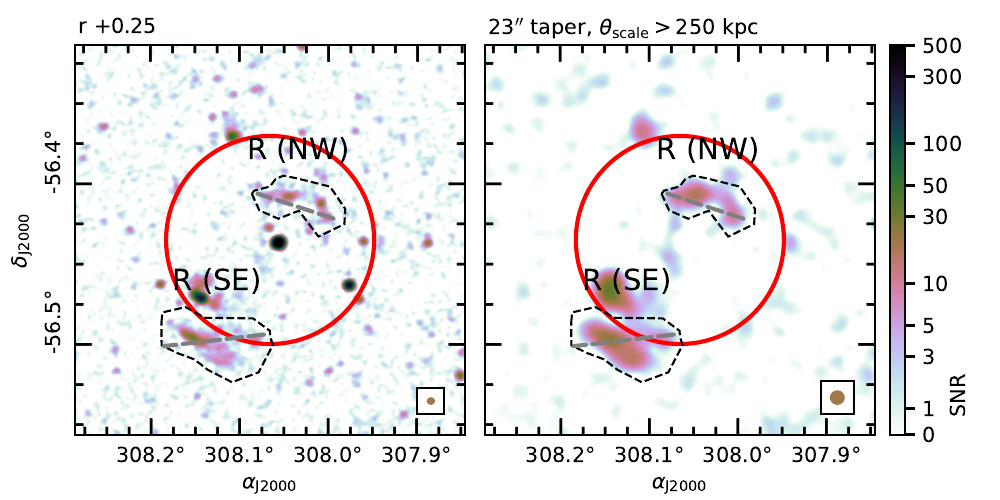}
\caption{\label{fig:app:PSZ2G341.19-36.12} \hyperref[sec:PSZ2 G341.19-36.12]{PSZ2 G341.19$-$36.12}.}
\end{subfigure}\\%
\caption{(Continued).}
\end{figure*}

\begin{figure*}
\centering
\ContinuedFloat
\begin{subfigure}[b]{0.5\linewidth}
\includegraphics[width=\linewidth]{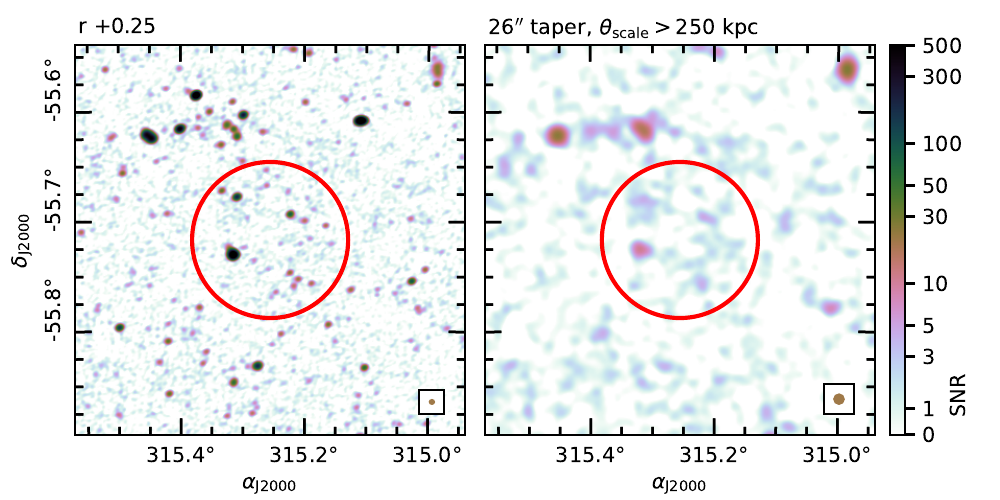}
\caption{\label{fig:app:PSZ2G341.44-40.19} PSZ2 G341.44$-$40.19.}
\end{subfigure}%
\begin{subfigure}[b]{0.5\linewidth}
\includegraphics[width=\linewidth]{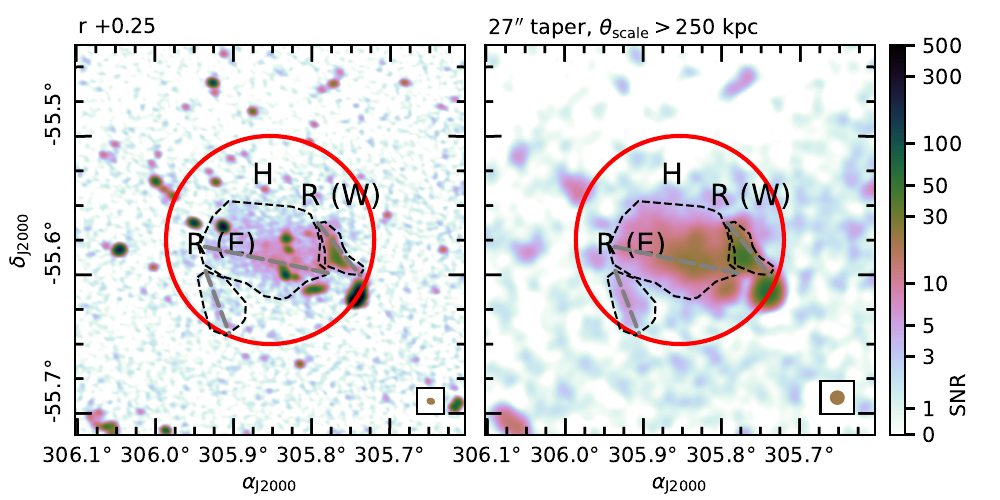}
\caption{\label{fig:app:PSZ2G342.33-34.93} \hyperref[sec:PSZ2G342.33-34.93]{PSZ2 G342.33$-$34.93}.}
\end{subfigure}\\%
\begin{subfigure}[b]{0.5\linewidth}
\includegraphics[width=\linewidth]{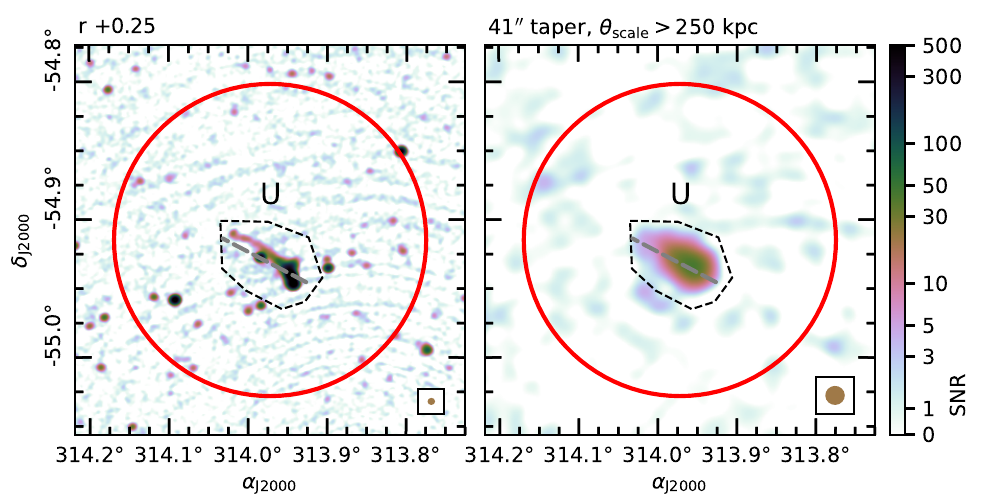}
\caption{\label{fig:app:PSZ2G342.62-39.60} \hyperref[sec:PSZ2G342.62-39.60]{PSZ2 G342.62$-$39.60}.}
\end{subfigure}%
\begin{subfigure}[b]{0.5\linewidth}
\includegraphics[width=\linewidth]{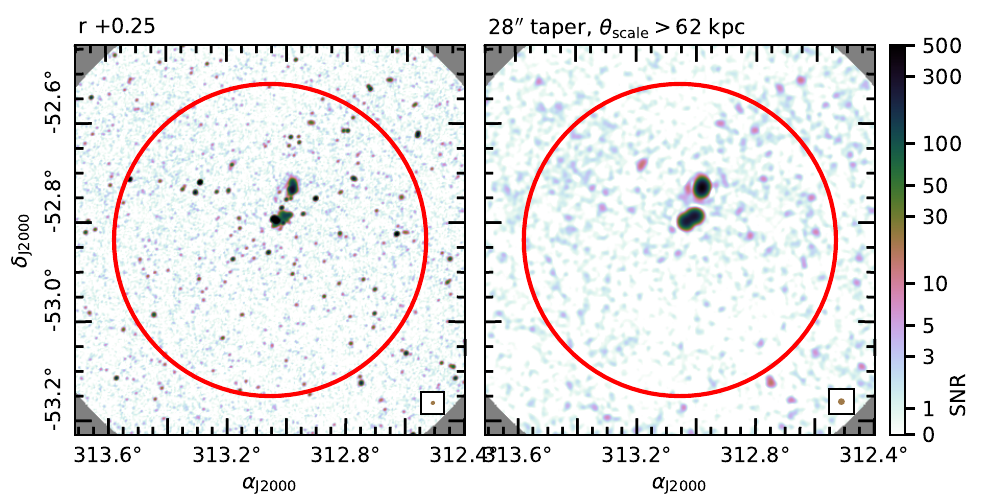}
\caption{\label{fig:app:PSZ2G345.38-39.32} PSZ2 G345.38$-$39.32.}
\end{subfigure}\\%
\begin{subfigure}[b]{0.5\linewidth}
\includegraphics[width=\linewidth]{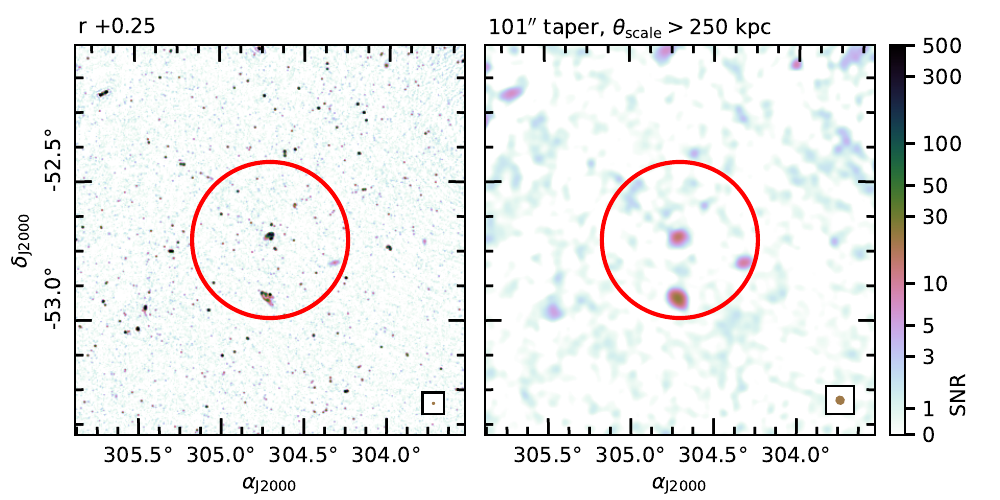}
\caption{\label{fig:app:PSZ2G345.82-34.29} PSZ2 G345.82$-$34.29.}
\end{subfigure}%
\begin{subfigure}[b]{0.5\linewidth}
\includegraphics[width=\linewidth]{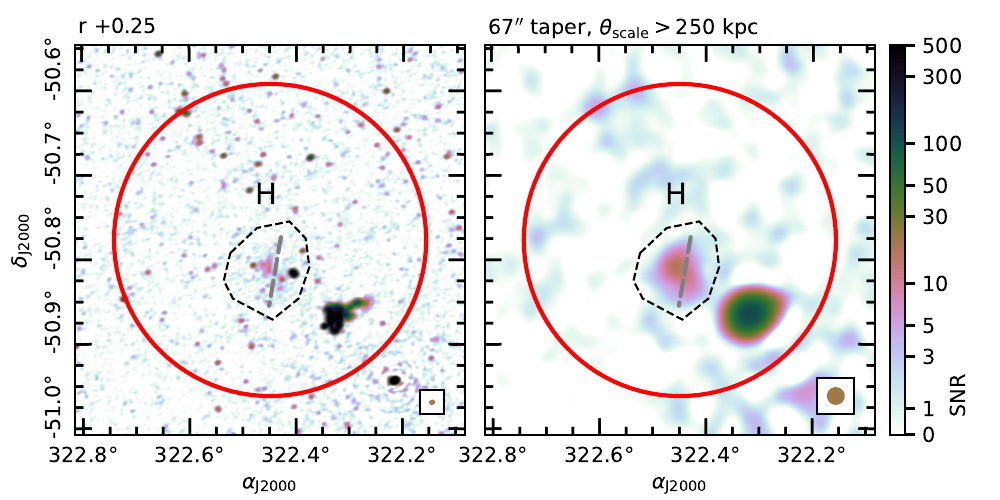}
\caption{\label{fig:app:PSZ2G346.86-45.38} \hyperref[sec:PSZ2G346.86-45.38]{PSZ2 G346.86$-$45.38}.}
\end{subfigure}\\%
\begin{subfigure}[b]{0.5\linewidth}
\includegraphics[width=\linewidth]{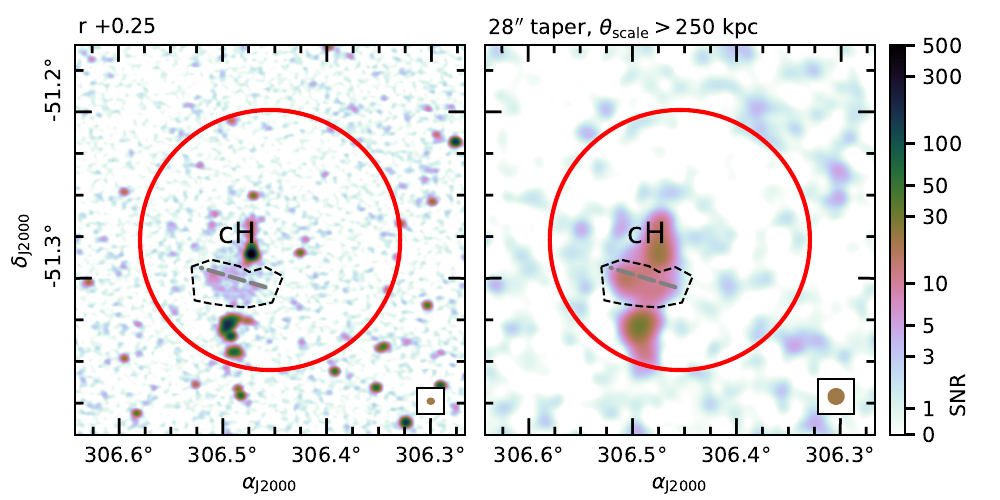}
\subcaption{ \hyperref[sec:PSZ2G347.58-35.35]{PSZ2 G347.58$-$35.35}.}
\label{fig:app:PSZ2G347.58-35.35}
\end{subfigure}%
\caption{(Continued).}
\end{figure*}

\section{Halo models}\label{sec:app:models}
\setcounter{figure}{0}
Figure~\ref{fig:app:models} shows the halo models output from \texttt{Halo-FDCA} (middle panels) along with the images used for model fitting (left panels) and the residuals after subtraction of the model from the data (right panels). The black, dashed contours on the left panels show the halo model, and regions that are masked are shown in grey. Only the best-fit model is shown (i.e. with the lowest reduced $\chi^2$), though other model fits are provided in the PASA Datastore along with other data products. In addition to the model images and residual images, log files from \texttt{Halo-FDCA} are also provided so interested users can obtain the model parameters from any of the models.

\begin{figure*}[t!]
    \centering
\begin{subfigure}[b]{0.5\linewidth}
\includegraphics[width=1\linewidth]{model_figures/PSZ2G008.31-64.74_circle_model.pdf}
\caption{\label{fig:app:model:PSZ2G008.31-64.74} PSZ2 G008.31$-$64.74.}
\end{subfigure}%
\begin{subfigure}[b]{0.5\linewidth}
\includegraphics[width=1\linewidth]{model_figures/PSZ2G011.06-63.84_skewed_model.pdf}
\caption{\label{fig:app:model:PSZ2G011.06-63.84} PSZ2 G011.06$-$63.84.}
\end{subfigure}\\%
\begin{subfigure}[b]{0.5\linewidth}
\includegraphics[width=1\linewidth]{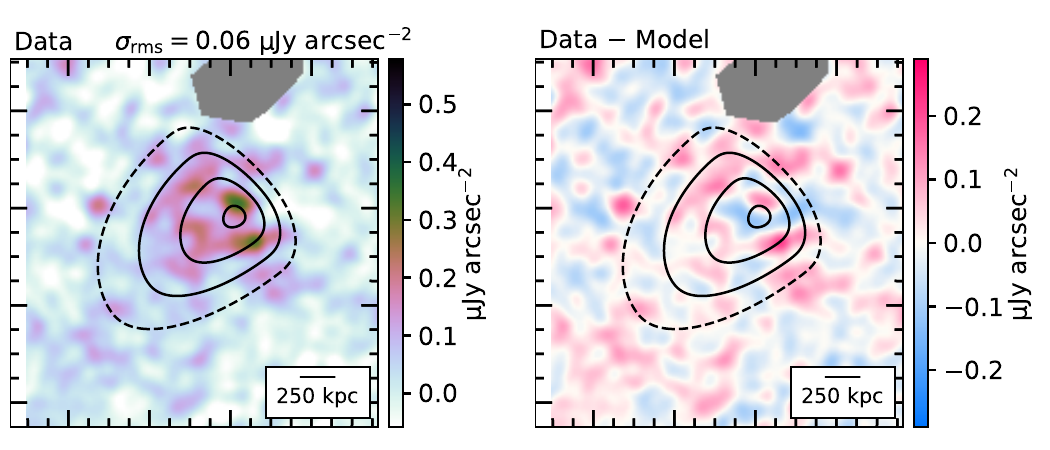}
\caption{\label{fig:app:model:PSZ2G011.92-63.53} PSZ2 G011.92$-$63.53.}
\end{subfigure}%
\begin{subfigure}[b]{0.5\linewidth}
\includegraphics[width=1\linewidth]{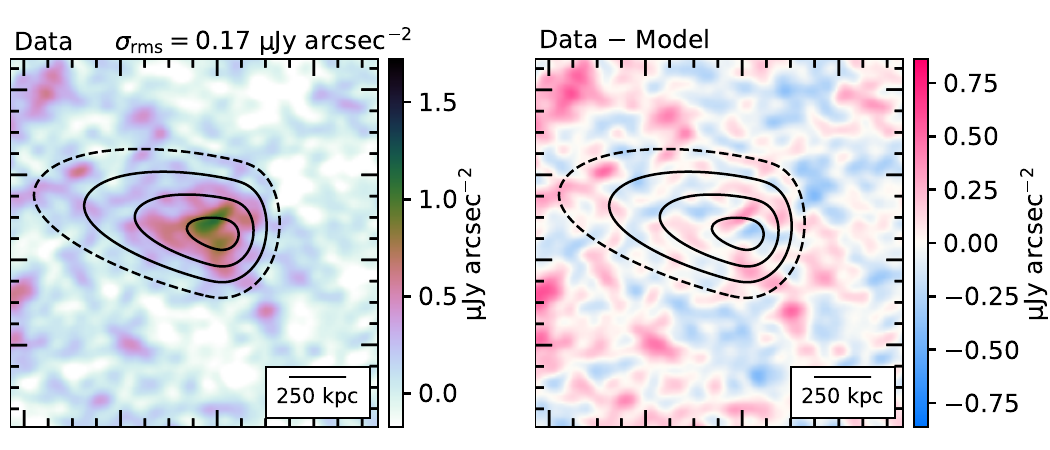}
\caption{\label{fig:app:model:PSZ2G110.28-87.48} PSZ2 G110.28$-$87.48.}
\end{subfigure}\\%
\begin{subfigure}[b]{0.5\linewidth}
\includegraphics[width=1\linewidth]{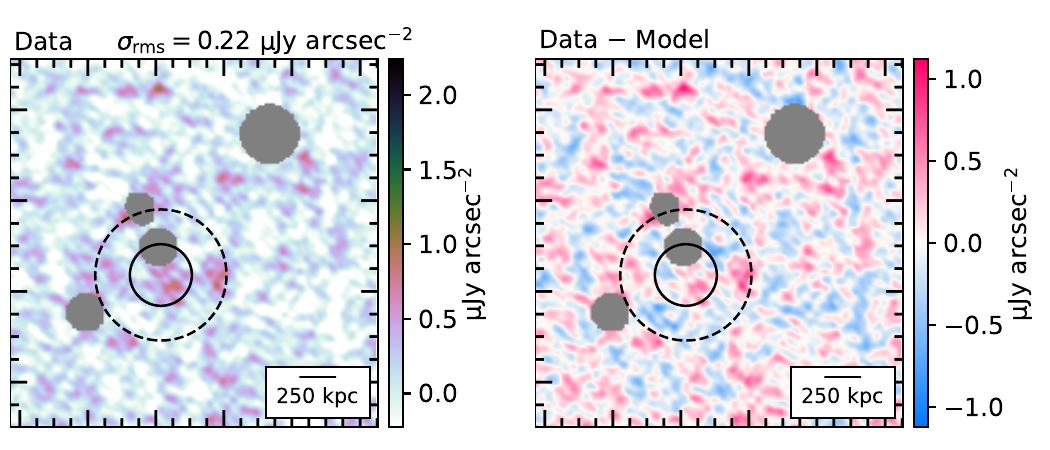}
\caption{\label{fig:app:model:PSZ2G172.98-53.55} PSZ2 G172.98$-$53.55.}
\end{subfigure}%
\begin{subfigure}[b]{0.5\linewidth}
\includegraphics[width=1\linewidth]{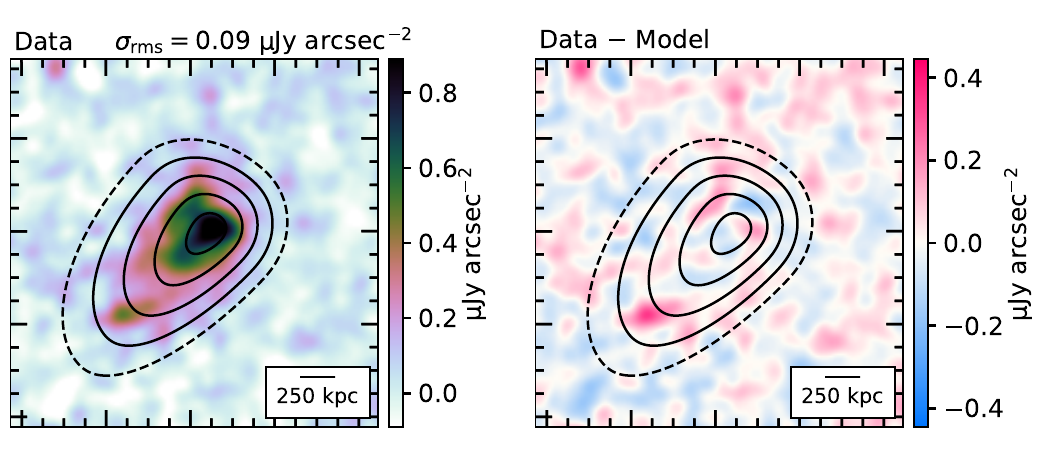}
\caption{\label{fig:app:model:PSZ2G175.69-85.98} PSZ2 G175.69$-$85.98.}
\end{subfigure}\\%
\begin{subfigure}[b]{0.5\linewidth}
\includegraphics[width=1\linewidth]{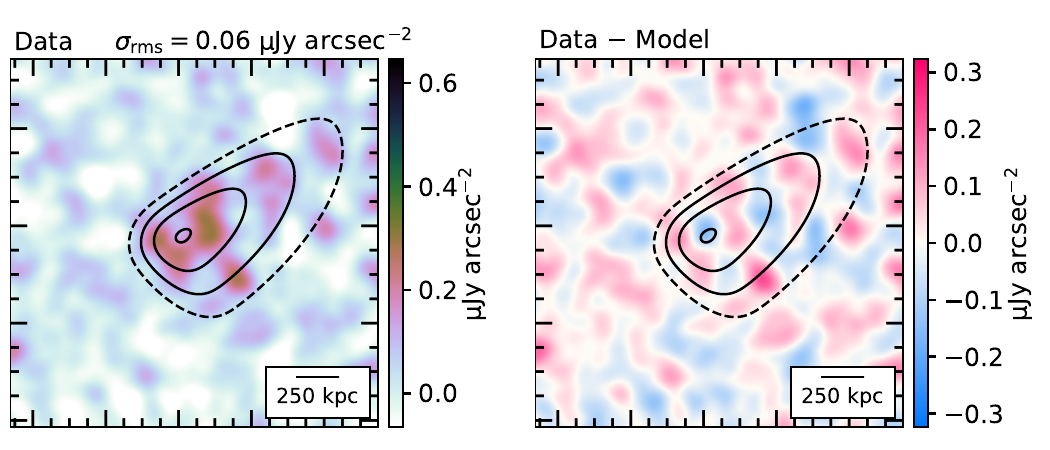}
\caption{\label{fig:app:model:PSZ2G223.47+26.85} PSZ2 G223.47$+$26.85.}
\end{subfigure}%
\begin{subfigure}[b]{0.5\linewidth}
\includegraphics[width=1\linewidth]{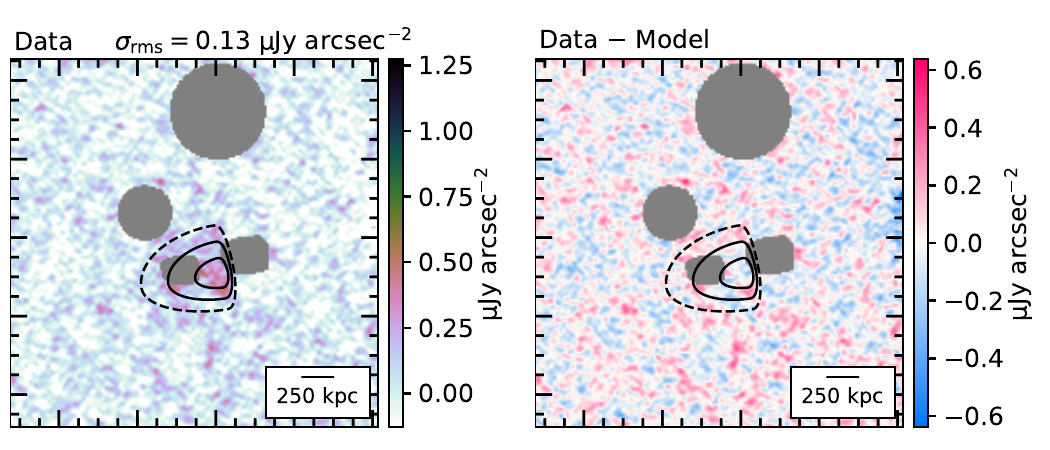}
\caption{\label{fig:app:model:PSZ2G225.48+29.41} PSZ2 G225.48$+$29.41.}
\end{subfigure}%
\caption{\label{fig:app:models} Radio halo models fit using \texttt{Halo-FDCA}. \emph{Left.} Compact source-subtracted image used for modelling the halo (and flux density measurements). \emph{Right.} Residual image after subtraction of the model. The model is shown as black contours in both panels (solid: $[1, 2, 4, 8, 16, 32]\times\sigma_\text{rms}$, dashed: $0.5\sigma_\text{rms}$). The left panel colourscale is linear between $[-1, 10]\times\sigma_\text{rms}$ and the right panel colourscale is linear between $[-5, 5]\times\sigma_\text{rms}$. Grey regions correspond to regions that are masked during fitting. Note the units are in \textmu Jy\,arcsec$^{-2}$ for consistency with the literature.}
\end{figure*}

\begin{figure*}[t!]
\centering
\ContinuedFloat
\begin{subfigure}[b]{0.5\linewidth}
\includegraphics[width=1\linewidth]{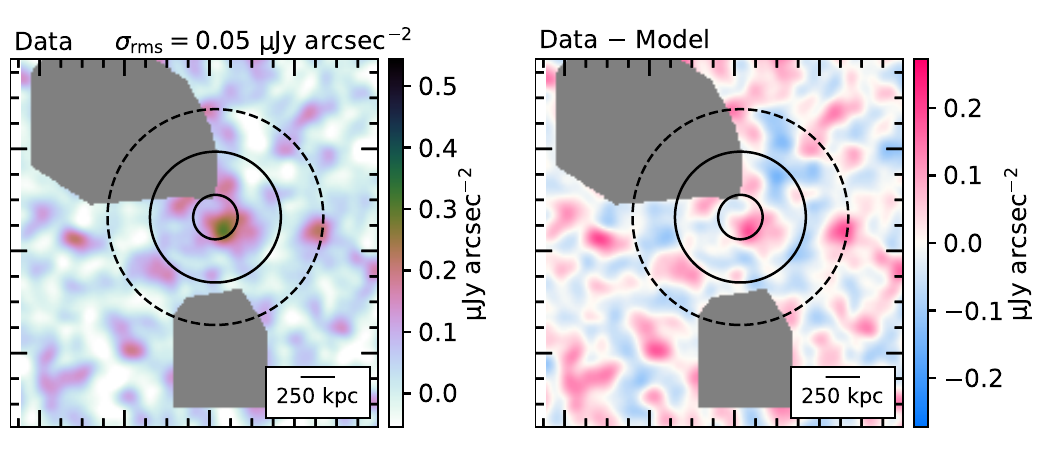}
\caption{\label{fig:app:model:PSZ2G228.50+34.95} PSZ2 G228.50$+$34.95.}
\end{subfigure}%
\begin{subfigure}[b]{0.5\linewidth}
\includegraphics[width=1\linewidth]{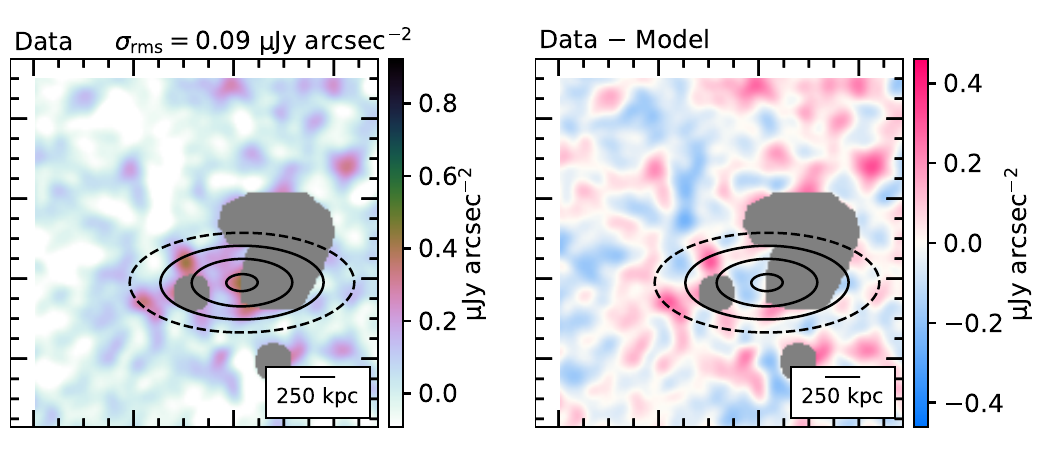}
\caption{\label{fig:app:model:PSZ2G231.79+31.48} PSZ2 G231.79$+$31.48.}
\end{subfigure}\\%
\begin{subfigure}[b]{0.5\linewidth}
\includegraphics[width=1\linewidth]{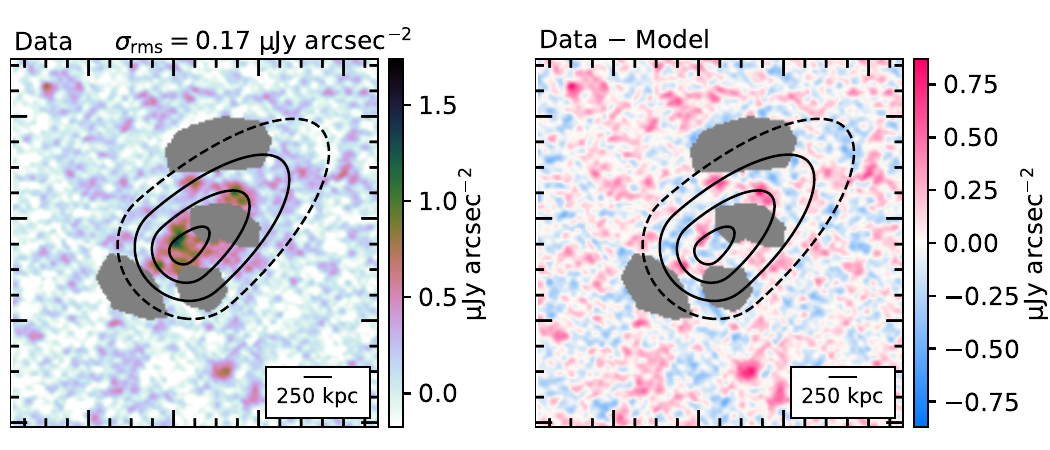}
\caption{\label{fig:app:model:PSZ2G233.68+36.14} PSZ2 G233.68$+$36.14.}
\end{subfigure}%
\begin{subfigure}[b]{0.5\linewidth}
\includegraphics[width=1\linewidth]{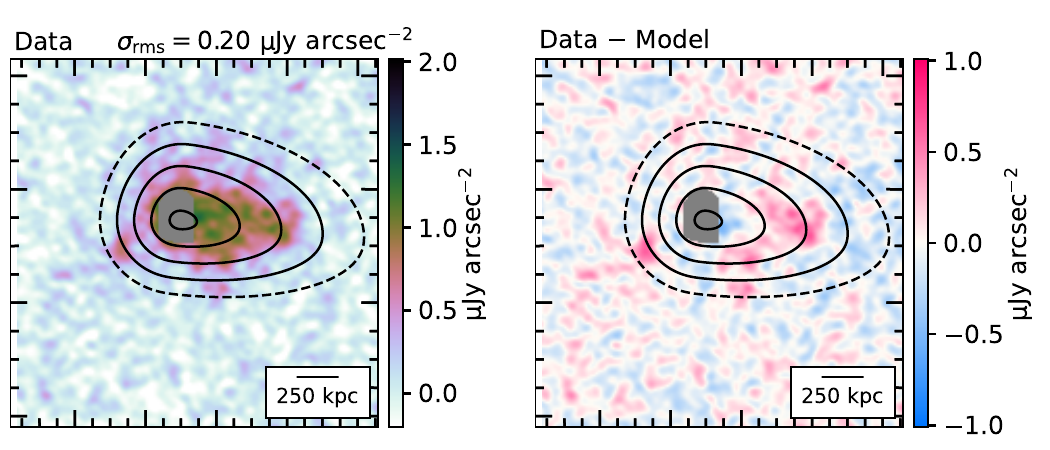}
\caption{\label{fig:app:model:PSZ2G239.27-26.01} PSZ2 G239.27$-$26.01.}
\end{subfigure}\\%
\begin{subfigure}[b]{0.5\linewidth}
\includegraphics[width=1\linewidth]{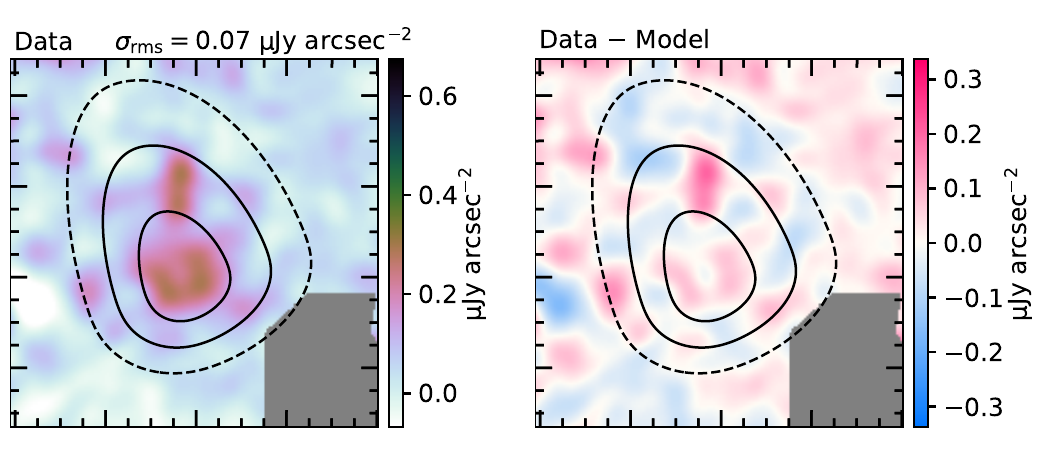}
\caption{\label{fig:app:model:PSZ2G260.80+06.71} PSZ2 G260.80$+$06.71.}
\end{subfigure}%
\begin{subfigure}[b]{0.5\linewidth}
\includegraphics[width=1\linewidth]{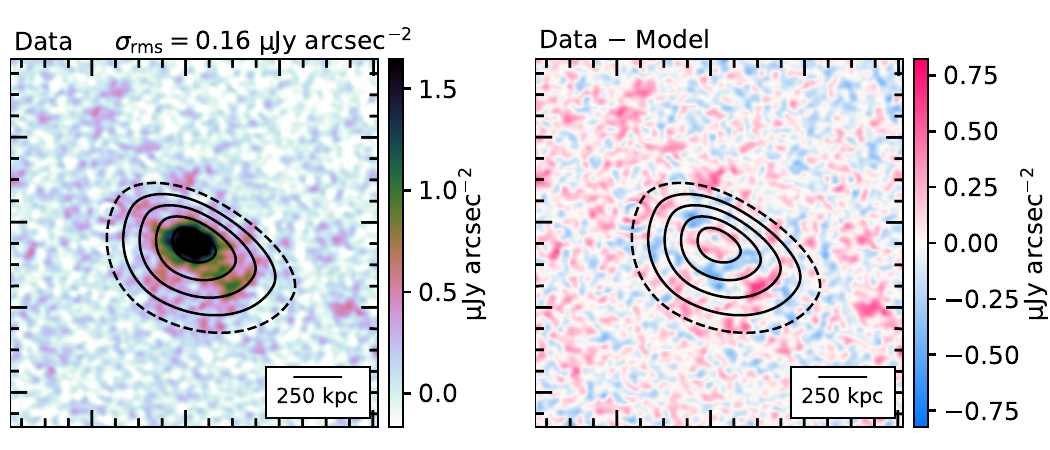}
\caption{\label{fig:app:model:PSZ2G263.14-23.41} PSZ2 G263.14$-$23.41.}
\end{subfigure}\\%
\begin{subfigure}[b]{0.5\linewidth}
\includegraphics[width=1\linewidth]{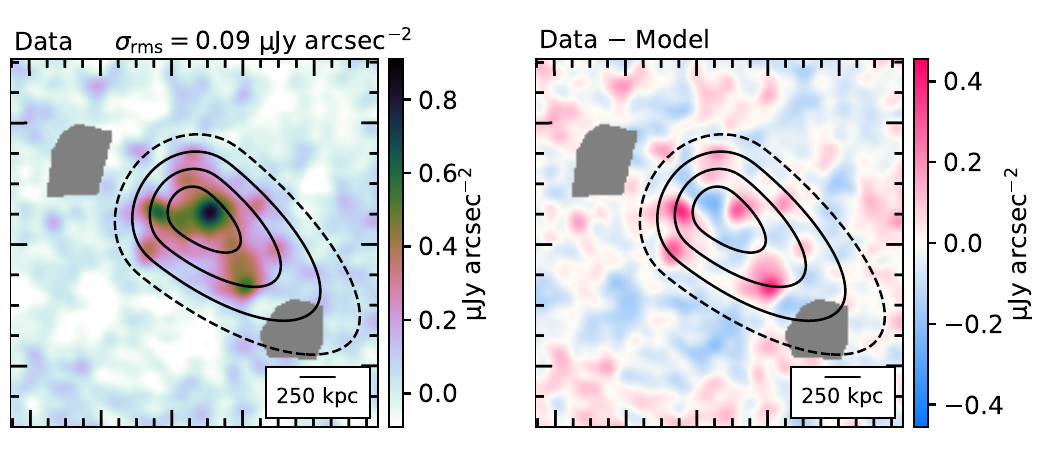}
\caption{\label{fig:app:model:PSZ2G263.68-22.55} PSZ2 G263.68$-$22.55.}
\end{subfigure}%
\begin{subfigure}[b]{0.5\linewidth}
\includegraphics[width=1\linewidth]{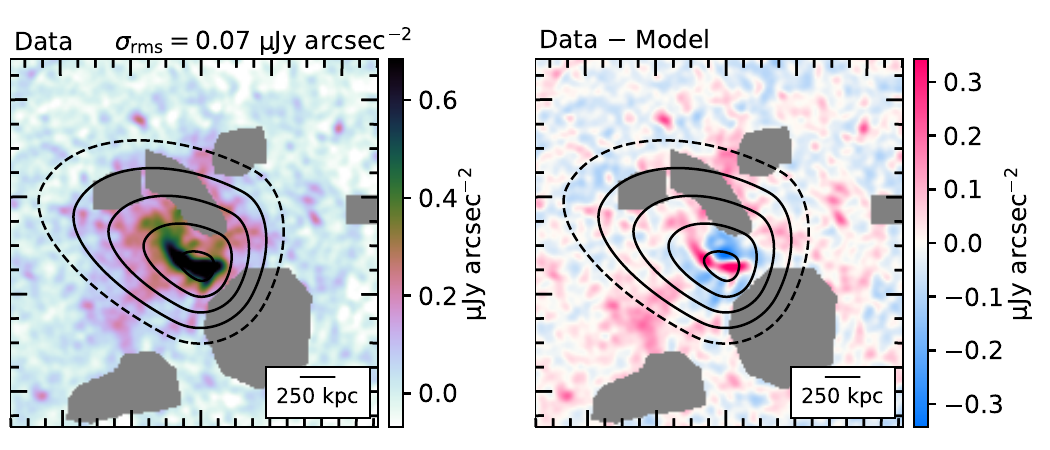}
\caption{\label{fig:app:model:PSZ2G272.08-40.16} PSZ2 G272.08$-$40.16.}
\end{subfigure}\\%
\begin{subfigure}[b]{0.5\linewidth}
\includegraphics[width=1\linewidth]{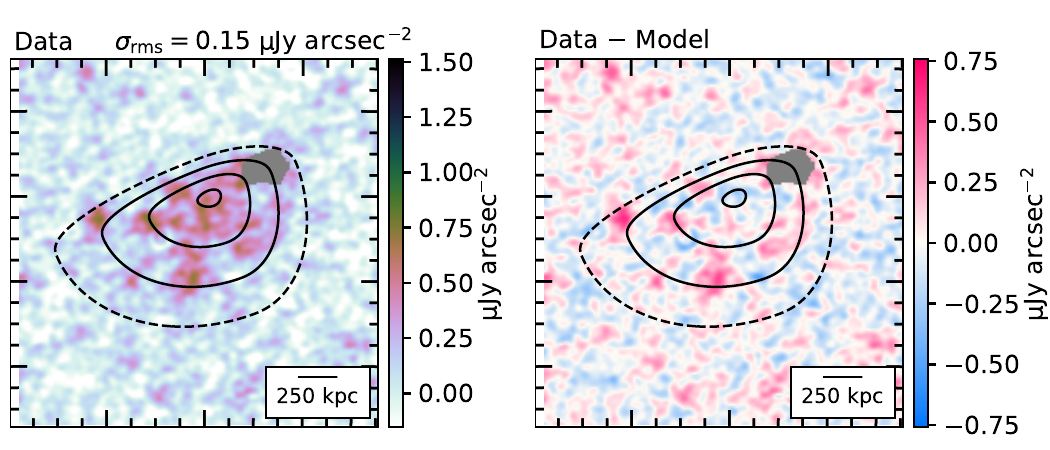}
\caption{\label{fig:app:model:PSZ2G286.28-38.36} PSZ2 G286.28$-$38.36.}
\end{subfigure}%
\begin{subfigure}[b]{0.5\linewidth}
\includegraphics[width=1\linewidth]{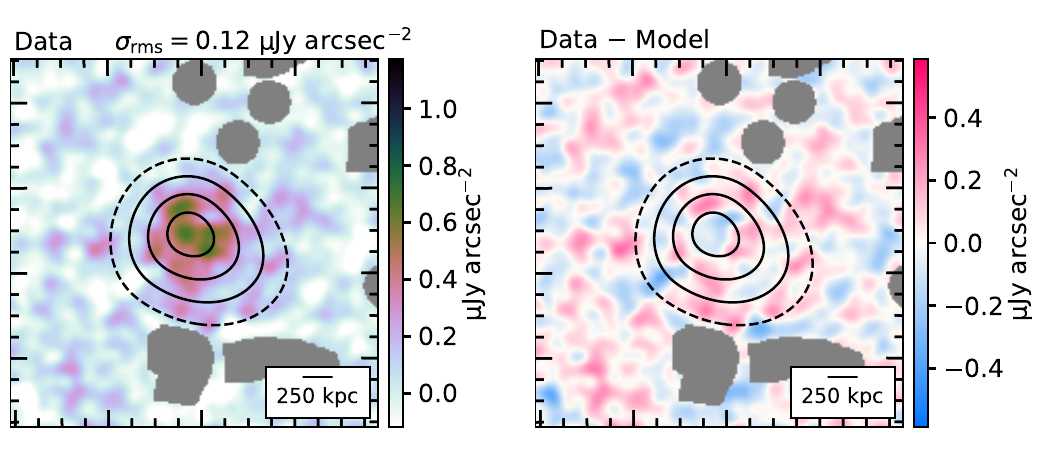}
\caption{\label{fig:app:model:PSZ2G286.75-37.35} PSZ2 G286.75$-$37.35.}
\end{subfigure}\\%

\caption{(Continued).}
\end{figure*}

\begin{figure*}[t!]
\centering
\ContinuedFloat
\begin{subfigure}[b]{0.5\linewidth}
\includegraphics[width=1\linewidth]{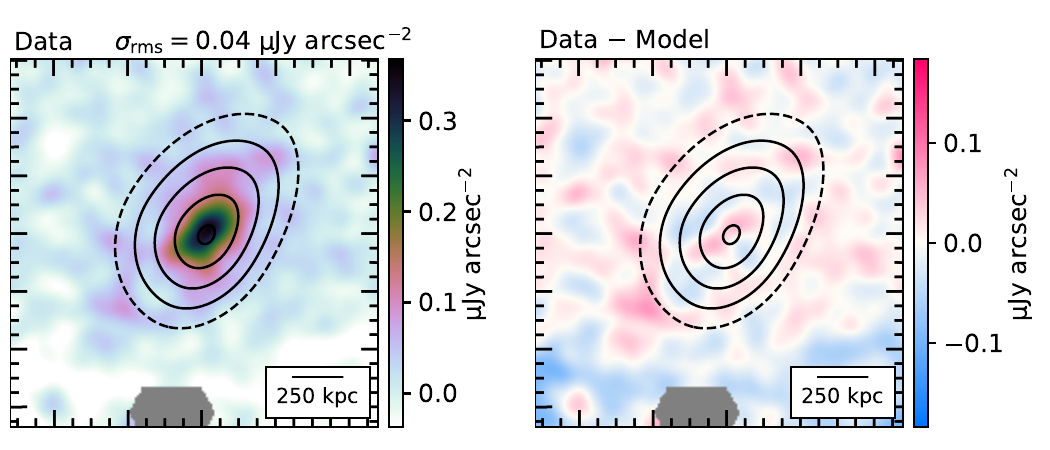}
\caption{\label{fig:app:model:PSZ2G311.98+30.71} PSZ2 G311.98$+$30.71.}
\end{subfigure}%
\begin{subfigure}[b]{0.5\linewidth}
\includegraphics[width=1\linewidth]{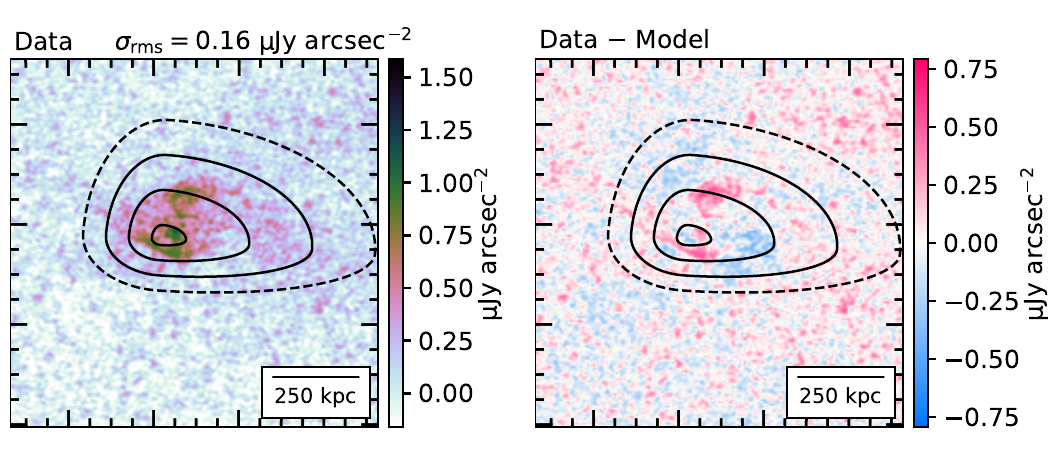}
\caption{\label{fig:app:model:PSZ2G313.33+30.29} PSZ2 G313.33$+$30.29.}
\end{subfigure}\\%
\begin{subfigure}[b]{0.5\linewidth}
\includegraphics[width=1\linewidth]{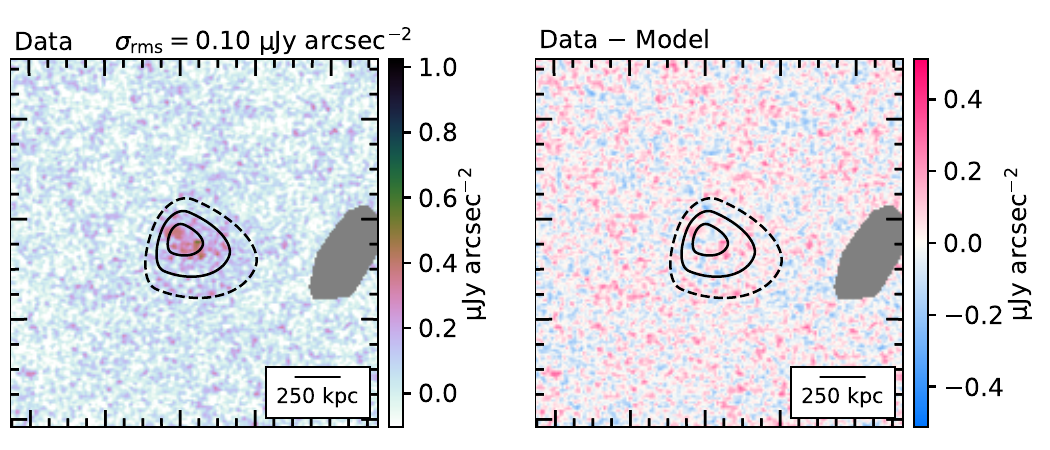}
\caption{\label{fig:app:model:PSZ2G332.23-46.37} PSZ2 G332.23$-$46.37.}
\end{subfigure}%
\begin{subfigure}[b]{0.5\linewidth}
\includegraphics[width=1\linewidth]{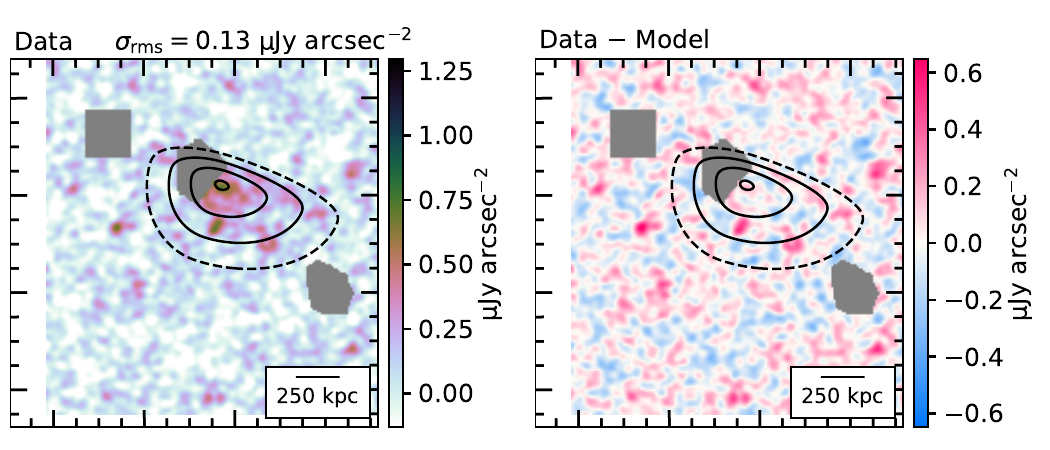}
\caption{\label{fig:app:model:PSZ2G333.89-43.60} PSZ2 G333.89$-$43.60.}
\end{subfigure}\\%
\begin{subfigure}[b]{0.5\linewidth}
\includegraphics[width=1\linewidth]{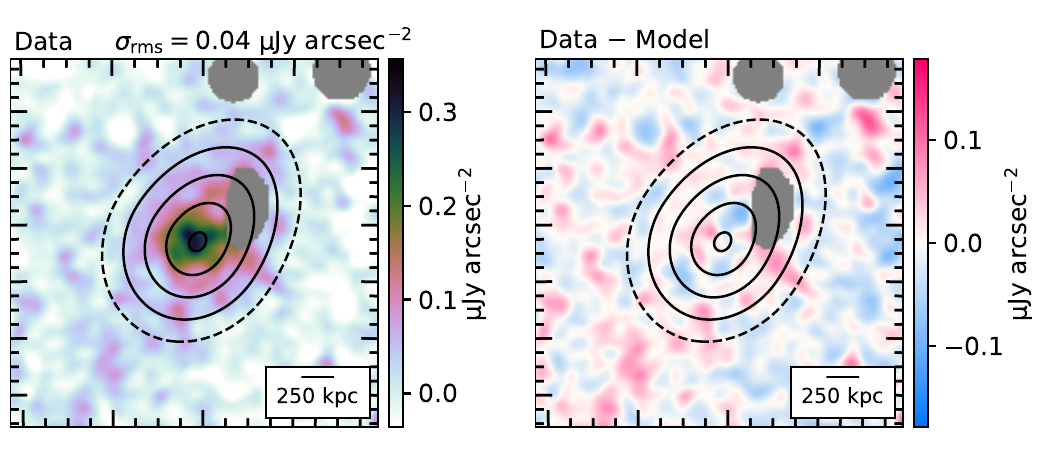}
\caption{\label{fig:app:model:PSZ2G335.58-46.44} PSZ2 G335.58$-$46.44.}
\end{subfigure}%
\begin{subfigure}[b]{0.5\linewidth}
\includegraphics[width=1\linewidth]{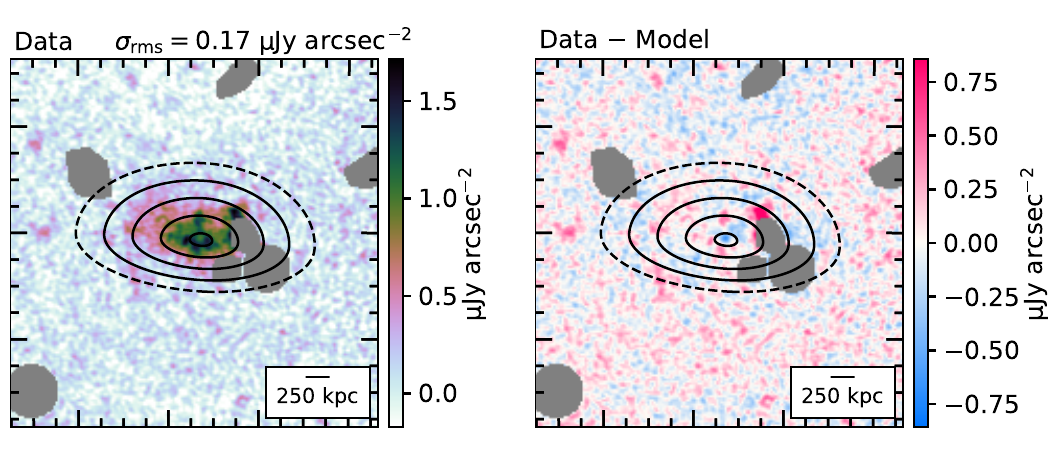}
\caption{\label{fig:app:model:PSZ2G342.33-34.93} PSZ2 G342.33$-$34.93.}
\end{subfigure}\\%
\begin{subfigure}[b]{0.5\linewidth}
\includegraphics[width=1\linewidth]{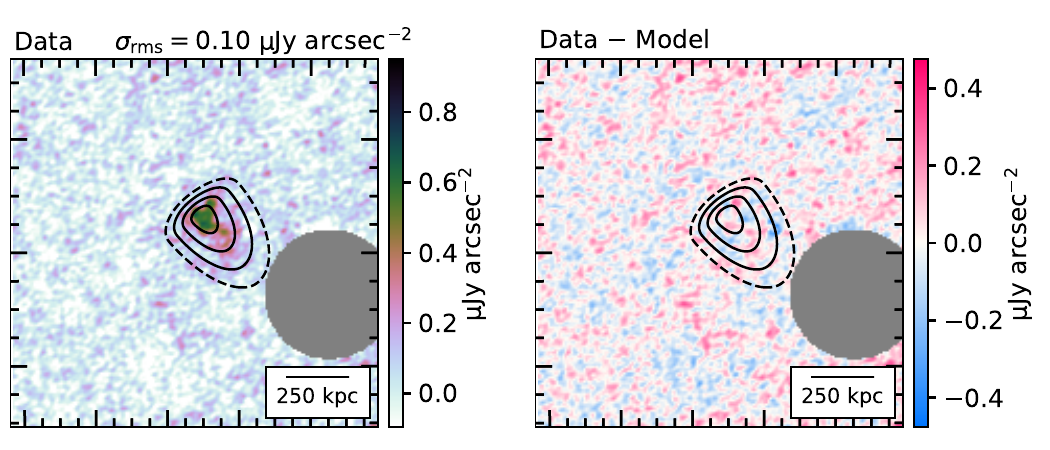}
\caption{\label{fig:app:model:PSZ2G346.86-45.38} PSZ2 G346.86$-$45.38.}
\end{subfigure}%
\begin{subfigure}[b]{0.5\linewidth}
\includegraphics[width=1\linewidth]{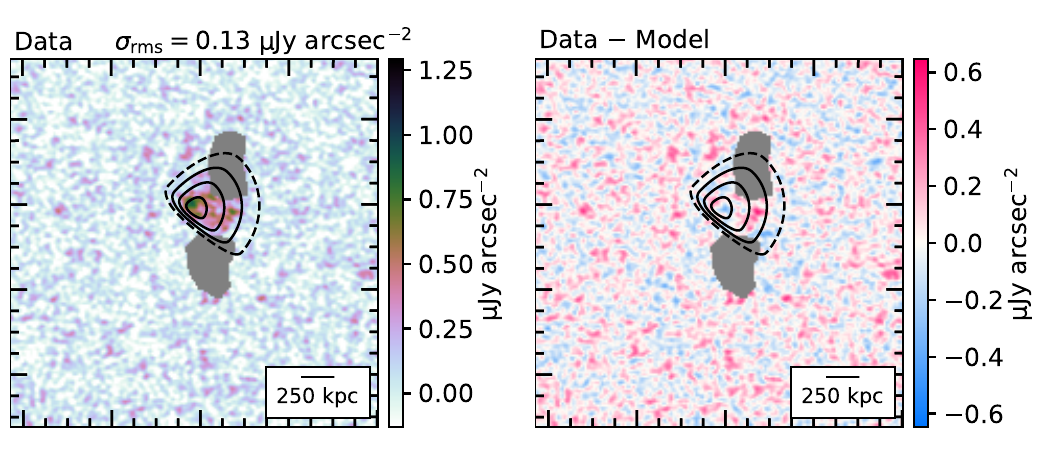}
\caption{\label{fig:app:model:PSZ2G347.58-35.35} PSZ2 G347.58$-$35.35.}
\end{subfigure}%
\caption{(Continued).}
\end{figure*}

\end{document}